\documentclass[12pt]{article}
\usepackage{amsmath,amssymb}
\usepackage{epsfig}
\usepackage{pstricks,pst-plot,pst-node}
\psset{unit=1cm,linewidth=1pt,arrowscale=1.5}
\newrgbcolor{purple}{0.8 0 1}
\newrgbcolor{pink}{1 0.6 0.6}
\usepackage[left=1in,right=1in,top=1in,bottom=1.25in]{geometry}
\usepackage{setspace}
\numberwithin{equation}{section}

\newcommand\be{\begin{equation}}
\newcommand\ee{\end{equation}}
\newcommand\half{\tfrac{1}{2}}
\newcommand\tr{\mathop{\rm tr}\nolimits}
\newcommand\etc{{\it etc.}}
\newcommand\E{{\cal E}}
\newcommand\bfsquare{%
	\hbox{%
		\vrule width 1.5pt
		\vbox to 1.5ex{\hrule height 1.5pt\vfil\hrule height 1.5pt width 0.9ex}%
		\vrule width 1.5pt
		}
	}
\newcommand\Z{{\mathbb Z}}
\newcommand\R{{\mathbb R}}

\def\cline#1#2{%
	\setbox0=\hbox{$\black #2$}%
	$#1\underline{\box0}$%
	}
\def\vev#1{\left\langle#1\right\rangle}

\def\CF{\mathcal{F}}

\def\bx{{\bf x}}

\begin{document}
\thispagestyle{empty}


\begin{flushright}
UTTG--07--13 \\
TAUP--2961/13
\end{flushright}

\vspace{10pt}
\begin{center}
\Large\bf
    Dimension Changing Phase Transitions\\
    in Instanton Crystals\\[10pt]
\rm
    Vadim Kaplunovsky$\strut^{a}$ and Jacob Sonnenschein$\strut^b$ \\[10pt]
\small\it
    a: Physics Theory Group and Texas Cosmology Center\\
    		University of Texas, Austin, TX 78712, USA\\[5pt]
 	b: The Raymond and Beverly Sackler School of Physics and Astronomy,\\
	Tel Aviv University, Ramat Aviv 69978, Israel\\
\end{center}

\vspace{15pt}
\centerline{ABSTRACT}
We investigate lattices of instantons and the dimension-changing transitions between them.
Our ultimate goal is the $\rm 3D\to 4D$ transition, which is holographically dual to the
phase transition between the baryonic and the quarkyonic  phases of cold nuclear matter.
However, in this paper (just as in \cite{Kaplunovsky:2012gb}) we focus on lower dimensions
--- the 1D lattice of instantons in a harmonic potential
$V\propto M_2^2x_2^2+M_3^2x_2^2+M_4^2x_4^2$,\space\space
and the zigzag-shaped lattice as a first stage of the $\rm 1D\to 2D$ transition.
We prove that in the low- and moderate-density regimes, interactions between the instantons
are dominated by  two-body forces.
This drastically simplifies
finding the ground state of the instantons' orientations, so we made a numeric scan
of the whole orientation space instead of assuming any particular ansatz.
\par
We find that depending on the $M_2/M_3/M_4$ ratios, the ground state of instanton
orientations can follow a wide variety of patterns.
For the straight 1D lattices, we found orientations periodically running over elements of
a $\Z_2$, Klein, prismatic, or dihedral subgroup of the $SU(2)/\Z_2$, as well as
irrational but link-periodic patterns.
For the zigzag-shaped lattices, we detected 4 distinct orientation phases --- the anti-ferromagnet,
another abelian phase, and two non-abelian phases.
Allowing the zigzag amplitude to vary as a function of increasing compression force, we obtained
the phase diagrams for the straight and zigzag-shaped lattices in the $({\rm force},M_3/M_4)$,
$({\rm chemical\ potential},M_3/M_4)$, and $({\rm density},M_3/M_4)$ planes.
Some of the transitions between these phases are second-order while others are first-order.

Our techniques  can be applied to other types of non-abelian crystals.

\newpage

\tableofcontents

%
\section{Introduction}

In the ordinary $N_c=3$ QCD cold nuclear matter forms a quantum liquid,
but for large $N_c$ it becomes a crystalline solid~\cite{Kaplunovsky:2010eh}.
In many holographic models of QCD \cite{SakaiSugimoto2004,Aharony:2006da}  baryons
are represented by  instantons of the $U(N_f)$ gauge theory
living on the flavor branes~\cite{Hata:2007mb}, so  cold nuclear matter corresponds
to a whole crystalline lattice of instantons~\cite{Rho:2009ym}.
The geometry and even the dimensionality of this lattice depend on the baryon density:
at medium densities, the instantons form a 3D lattice in the $x_4=0$ hyperplane of
the holographic 4D space (not counting the time); this corresponds to the baryonic phase
of  nuclear matter.
At high densities, the baryons spread out into the $x_4$ dimension and form a 4D lattice;
this is dual to the quark liquid phase of  nuclear matter, or rather to the
high-density {\it quarkyonic} phase \cite{McLerran:2007qj}
in which the quarks fill the Fermi sea, but the excitations near the Fermi surface
are baryon-like bound states of $N_c$ quarks rather than free quarks by themselves.
According to Rozali {\it et~al} \cite{Rozali:2007rx}, the chemical potential of
the quarks in this phase is related to the thickness of the lattice in the $x_4$ direction.

Most investigations of the low-temperature holographic nuclear matter focus on
the extreme high-density limit where the $x_4$ dimension is thick
and can be studied macroscopically (for example, see
\cite{FiniteDensity,Yamada:2007,Rozali:2007rx,deBoer:2012ij}).
But our main interest is in the  microscopic structures of the instanton lattices,
and especially the phase transition  between the 3D and 4D lattices.
In our previous paper \cite{Kaplunovsky:2012gb} (with Dmitry Melnikov) we
found that there is a whole sequence of such transitions: from a single 3D layer in the $x_4=0$ hyperplane,
to two 3D layers in two parallel hyperplanes, to 3 layers, to 4 layers, {\it etc.,}
until the number of layers becomes too large to count individually.
Alas, the 3D and 4D instanton lattices turned out to be too hard to analyze, so we resorted
to  oversimplified toy models.
In the first toy model, we approximated the instantons as point-like charges repelling each
other with 4D Coulomb forces; this approximation is very crude --- it ignores
the instantons' orientations, never mind the interference between the instantons ---
but it makes for a simple model of transitions between lattices of different dimensions.
Indeed, in this model we found that increasing density makes a 3D lattice of instantons in $x_4=0$ hyperplane
suddenly split into two sublattices in the $x_4=\pm\epsilon$ hyperplanes;
pictorially, this looked like the popcorn suddenly jumping up in the popper, so we dubbed
the dimension-changing transitions the {\it popcorn transitions.}

In our second toy model, we used actual instantons and the exact ADHM\cite{ADHM} construction for
the multi-instanton system.
However, we had confined the instanton centers to a 2D plane $x_1=x_2=0$ by making the
inverse 5D gauge coupling rise steeply in the $x_1$ and $x_2$ directions.
We also had the $1/g_5^2$ slowly rising in the $x_3$ direction,  altogether
\be
{8\pi^2\over g_5^2(x)}\ =\ N_c\lambda M\Bigl(1\,
	+\,M^2x_3^2\,+\,M^{\prime2}\bigl(x_1^2+x_2^2\bigr)\,
	+\,O\bigl(M^4x_{1,2,3}^4\bigr)\Bigr)\qquad
{\rm for}\quad M'\gg M.
\ee
Consequently, at low densities the instantons formed a 1D lattice in the $x_4$ dimension
(which acted as the model's only flat space dimension), but for higher densities
the instantons moved into the $x_3$ dimension (which acted as the holographic
dimension of the model).
In that model, we saw two phase transitions in response to increasing instanton density.
The first transition --- from a straight 1D chain of instantons to a zigzag-shaped chain
\be
\begin{pspicture}(-7,-0.9)(+7.4,+1.3)
\psline[linewidth=0.5pt]{>->}(-7,0)(+7,0)
\uput[r](7,0){$x_4$}
\psline[linewidth=0.5pt]{>->}(0,-1)(0,+1.2)
\uput[r](0,1.2){$x_3$}
\psline[linestyle=dotted](-6.5,+0.4)(+6.5,+0.4)
\psline[linestyle=dotted](-6.5,-0.4)(+6.5,-0.4)
\multido{\n=-6+2}{7}{\pscircle*[linecolor=blue](\n,+0.4){0.1}}
\multido{\n=-5+2}{6}{\rput(\n,0){%
	\psline[linecolor=magenta,linewidth=0.5pt](-1,+0.4)(0,-0.4)(+1,+0.4)
	\pscircle*[linecolor=blue](0,-0.4){0.1}
	}}
\end{pspicture}
\nonumber
\ee
--- is second-order, so the zigzag amplitude behaves as $\epsilon\propto\sqrt{\rho-\rho_c}$.
On both sides of this transition, the instantons' orientations form an anti-ferromagnetic
pattern: two alternating orientations related by a $180^\circ$ rotation.
At higher densities there is another transition, but this time it's first-order:
the zigzag amplitude $\epsilon$ jumps discontinuously, and the instantons' orientations
also change their pattern: the orientations of the nearest neighbors are now related by
a rotation through angle $\phi<110^\circ$ instead of $180^\circ$.

Presumably, at still higher densities we would have seen popcorn transitions from the zigzag
--- which is a kind of two-layer lattice --- to a 3-layer lattice, then to 4 layers, {\it etc., etc.}
However, calculating the net energies of such multi-layer lattices turned out to be too hard,
so we  stopped at the zigzag.

Technically, the main difficulty in working with more complicated instanton lattices is
setting up the ADHM construction \cite{ADHM} for the infinite number of instantons.
Indeed, the ADHM construction for $A$ instantons involves 4 non-commuting $A\times A$ matrices
$\Gamma^\mu_{mn}$ 
whose off-diagonal elements follow
from non-trivial constraint equations;
for $A\to\infty$ solving these equations becomes very hard.
For a simple 1D lattice of instantons whose orientations are related by {\it commuting} $SU(2)$
symmetries, the ADHM constraints were solved by Kraan and Baal \cite{Kraan:1998pm}, and we have
re-derived and used their explicit formula for the instanton number density profile $I(x)$ in
\cite{Kaplunovsky:2012gb}.
For a zigzag-shaped instanton chain with link-periodic instanton orientations, we found
an exact solution of the ADHM constraints in terms of Fourier transforms of some rather messy
functions, but we could not use it to derive an exact formula for the $I(x)$;
instead, we had to use the small-instanton-size approximation.
And it gets worse for more complicated instanton lattices and/or orientation patterns:
For the {\it simple} 2D and 3D lattices and purely abelian orientation patters
the ADHM constraints have known formal solutions\footnote{%
	A Nahm transform can map the ADHM constraints to an electrostatic-like problem in 2D or 3D.
	}
--- which alas are  way too complicated for any practical use ---
while for the non-abelian orientation patterns there are no known solutions at all.
This is particularly unfortunate since the lowest-energy 3D configuration for low densities
is likely to be an FCC lattice with non-commuting orientations of the instantons
\cite{Manton:1994rf}.

In this paper we follow a shortcut around the ADHM construction --- and also around
the more difficult part of the energy calculation.
We assume from the beginning that the instanton density is not too high, so the distances
between instantons are much larger than the instantons' radii.
In \S3 we show that for all such multi-instanton systems, interactions
between the instantons are dominated by the two-body forces.
The irreducibly-three-body forces, 4-body forces, {\it etc.}\ also exist, but they are
suppressed by powers of the small $\rm(radius/distance)^2$ ratio compared to the two-body forces.
Consequently, we obtain a manageably simple approximate formula for the net energy
of a multi-instanton system in terms of the instantons' positions and orientations.

Thanks to this formula, we do not need to assume (as we did in \cite{Kaplunovsky:2012gb})
that in a lattice of instantons, their relative orientations have the same symmetries
as their relative positions.
Instead, we may use numerical methods to minimize the net energy with respect to all
the orientations treated as independent parameters, without any symmetry assumptions.
The algorithm we use in this paper is fairly simple:
We set up a large lattice (200 sites is more than enough for 1 dimension) of fixed geometry,
with a free $SU(2)$ matrix $y_n$ at each site encoding the $n^{\rm th}$ instanton's orientation
(for $N_f=2$).
We start with a completely random set of the initial $y_n$ and make them evolve along the steepest
descent of the net energy function until they converge to a local minimum.
Then we repeat the process many times with different random sets of initial $y_n$.
This gives us a short list of the local minima, from which we select the deepest
--- which is presumably the global minimum of the net energy function.
This algorithm has convergence problems near the phase boundaries
--- where two or more local minima are almost degenerate ---
but it is good at finding all the orientation patterns that {\it might} be ground states
for some combinations of background parameters and lattice geometries.
And once we know all these patterns, we fit them all with common ansatz, calculate the
net energy analytically as a function of the ansatz's parameters, and then we  minimize
it with a much higher precision to get the accurate phase diagram of the instanton lattice.

In this paper we put the instantons in a harmonic potential that rises at different rates along
three out of four space coordinates;
in terms of the 5D gauge coupling,
\be
{8\pi^2\over g_5^2({\red\not}x_1,x_2,x_3,x_4)}\ =\ N_c\lambda M\times\Bigl(
	1\,+\,M_2^2x_2^2\,+\,M_3^2x_3^2\,+\,M_4^2x_4^2\,+\,O(M^4x_{2,3,4}^4)\Bigr)
\label{Ipot}
\ee
with  3 independent parameters $M_2,M_3,M_4$.
Here we use different coordinate axes than in  \cite{Kaplunovsky:2012gb}:
the 1D straight chain --- or the long axis of the zigzag --- runs along the $x_1$ axis instead of the $x_4$.
But the  important difference is allowing for $M_3\neq M_4$ and $M_2\sim M_3\sim M_4$
instead of $M_3=M_4=M'\gg M=M_2$;
it is this difference which leads to the wide variety of instanton orientation patterns
we shall see in this paper.

Specifically, we shall see that for the 1D instanton lattice in the almost-holographic background
$M_2,M_3\ll M_4$ (where $M_2$ and $M_3$ help to line up the instanton centers along the $x_1$
axis but do not affect the two-body forces between the instantons),
there is a large degenerate family of ground states for the instanton orientations $y_n\in SU(2)/\Z_2$
(for $N_f=2$).
This family includes many periodic patterns in which all the $y_n$ belong to a finite
subgroup of the $SU(2)/\Z_2$: the anti-ferromagnetic pattern in which $y_{{\rm odd}\,n}$ and
$y_{{\rm even}\,n}$ belong to 2 elements of a $\Z_2$ subgroup, the $\rm period=4$ pattern in which
the $y_n$ span the Klein group $\Z_2\times\Z_2$, and two series of ${\rm period}=2k$ patterns
in which the orientations span a prismatic group $\Z_k\times\Z_2$ or a dihedral group $D_{2k}$.
There are also many link-periodic patterns in which the relative orientations $y_n^\dagger y_{n+1}^{}$
of neighboring instantons are periodic but the orientations $y_n$ themselves are not periodic.
But the vast majority of the degenerate ground states are not periodic at all.

Increasing  the $M_2$ and $M_3$ parameters lifts the degeneracy.
For $M_2\ll M_3\sim M_4$, the ground state of the 1D lattice is anti-ferromagnetic, while for
$M_2\sim M_3\sim M_4$ the ground state is periodic with a longer period or link-periodic,
depending on the $M_2/M_3$ ratio.
Specifically, for $M_2=M_3$ the instanton's orientation in the ground state span the Klein group,
for other rational $M_2^2/M_3^2$ they span a dihedral group, while the irrational
$M_2^2/M_3^2$ favor the link-periodic patterns.

For the zigzag-shaped chains of instantons we need $M_2\ll M_3\sim M_4$ to make sure
the transition from a straight chain to a zigzag happens when
$\rm lattice~spacing\ \gg\ instanton~radius$
--- otherwise the two-body-force approximation would not work ---
but we may vary the $M_3/M_4$.
In this setting, the ground state of instanton orientations may follow one of
four different patterns, depending
on the zigzag amplitude and on the $M_3/M_4$ ratio.
Besides the anti-ferromagnetic and the abelian $\phi<\pi$ patterns seen  in \cite{Kaplunovsky:2012gb},
there are two non-abelian link-periodic pattern in which the $y_n^\dagger y_{n+1}^{}$ ``twists''
for even and odd links of the lattice do not commute with each other.
The non-abelian patterns appear only for $M_3<M_4$, that's why we did not see them
in  \cite{Kaplunovsky:2012gb}.

Focusing on the ``popcorn transition'' from a straight chain of instantons to
a zigzag-shaped chain (which is a two-layer lattice), we shall see that the thermodynamic order
of this transition depends on the $M_3/M_4$ ratio:
the transition is first-order for $M_3<0.725 M_4$ but second order for $M_3>0.725 M_4$.
Likewise, when we further increase the instanton density
--- or rather, the 1D compression force which maintains it ---
the number and the orders of the subsequent phase transitions between
different orientation phases of the zigzag-shaped chain also depend on the $M_3/M_4$ ratio.

This paper is organized as follows:
For the  reader's convenience, in \S2 we review the basic features of holographic nuclear matter:
effects of the large $N_c$ and large $\lambda$ limits,
the generalized  Sakai--Sugimoto model of holographic QCD,  realization of holographic baryons as
instantons of the flavor gauge symmetry, basic features of the holographic nuclear forces,
and our basic approach to the multi-baryon systems and the bulk nuclear matter.
In \S3, we argue that the interactions between multiple instantons include 2-body forces
as well as irreducibly-3-body force, 4-body forces, {\it etc.,}
but when the distances between the baryons are much larger that the instantons' radii,
the two-body forces dominate the net energy, while the 3-body,\dots, forces are suppressed
by  powers of $\rm(radius/distance)^2\ll1$.
First we give a heuristic argument, and then we back it up with a technical proof for $N_f=2$;
we also calculate the precise form of the net two-body force and its dependence on the
two instantons' orientations.

In \S4 we use the two-body-force approximation to study  1D lattices of instantons.
We start with the $M_2,M_3\ll M_4$ background and show that there is a very large degenerate family
of ground states of the instantons' orientations.
Then we move on to the $M_2\sim M_3\sim M_4$ backgrounds; this changes the precise form of the two-body
force in a way that lifts the degeneracy between the orientation states.
Consequently, the ground state of the lattice has a specific periodic or link-periodic
pattern of orientations, depending on the $M_2/M_3$ ratio.

In \S5 we study  zigzag-shaped instanton chains in the $M_2\ll M_3\sim M_4$ backgrounds.
After a few preliminaries, we use numerical methods to find the lowest-energy state of
instanton orientations without presuming any particular symmetries.
We repeat the calculation for 50 different lattice geometries ($\epsilon/D=0.01,\ldots0.99$ by 0.02)
and 50 values of the $M_3/M_4$ ratio; the results are presented on the phase
diagram~\ref{ZigzagRoughDiagram} on page~\pageref{ZigzagRoughDiagram}.
Our numeric code is not very accurate and has convergence problems near the phase boundaries,
but it is good at identifying the symmetries of the lowest-energy patterns.
Using these symmetries, we construct a 3-parameter ansatz that includes all the preferred patterns,
then calculate the net energy as analytic function of the parameters and minimize it.
For the reader's convenience, the energy calculation and minimization is described in the
Appendix, while the resulting phase diagram is presented in figure~\ref{ZigzagDiagram}
(page \pageref{ZigzagDiagram}) in the main body of \S5.

In the last part of \S5 we treat the zigzag amplitude and the lattice spacing
as dynamic parameters, vary the  compression force of the 1D instanton lattice,
and study the popcorn transitions from the straight chain to the zigzag
and between different orientations phases of the zigzag.
We show that the net number of such transitions --- 1, 2, or 3 --- and their orders depend
on the $M_3/M_4$ ratio.
The overall phase diagram in the compression versus $M_3/M_4$ plane is presented in
figure~\ref{BigPhaseDiagramFM}(a) on page \pageref{BigPhaseDiagramFM}; we also present
the phase diagrams in the chemical potential versus $M_3/M_4$ plane
(figure~\ref{BigPhaseDiagramFM}(b) on page \pageref{BigPhaseDiagramFM})
and in the instanton density versus $M_3/M_4$ plane
(figure~\ref{BigPhaseDiagramRho} on page \pageref{BigPhaseDiagramRho}).

Finally, in section 6 we summarize our results and give a short list of open questions
we hope to address in the near future.

%
\section{Review of Holographic Nuclear Physics}

In this section we review the basics of holographic nuclear physics that we shall need
later in this paper.
The {\it holography} here means gauge-gravity duality, and the semiclassical description
of the gravity side of the duality requires the limits of large $N_c$ and large
't~Hooft coupling $\lambda=g_{\rm YM}^2\times N_c$~\cite{tHooftLargeN}.
In the first two subsections \S2.1--2 we discuss the general consequences of these limits for the
nuclear physics.
In \S2.3 we get more specific and review the Sakai--Sugimoto model of holographic QCD.
There are many other models which differ in countless details --- and none of them
is exactly dual to the QCD, --- but their basic features are rather similar, and the
Sakai--Sugimoto model serves as a prototype.
In \S2.4 we introduce the baryons and explain their holographic duality to the
instantons of the flavor gauge symmetry.
We also review the interactions and the self-interactions of such instantons in the large $\lambda$
limit and explain why they have small radii $a\propto\lambda^{-1/2}$.
Finally, in \S2.5 we address the multi-instanton systems dual to nuclear matter and explain
how the interactions between the baryons appear in at the first order in the $1/\lambda$ expansion.
We also explain the technical difficulties in handling infinite crystals of baryons
and preview a shortcut around those difficulties that we shall take in \S3.

\subsection{Nuclear matter in the large $N_c$ limit}
\label{secLargeNQCD}
QCD perturbation theory in the large $N_c$ limit  is dominated by planar gluon diagrams
while contributions of the non-planar diagrams and of the quark loops are suppressed.
At zero temperature and chemical potential  below the mass of the baryon,
the large $N_c$ QCD is a theory of free mesons and glueballs whose interactions are suppressed
by powers of $N_c$ (except for interactions with the baryons).
The masses of mesons and glueballs are $O(\Lambda_{\rm QCD})$, and they scale as $N_c^0$.
On the other hand, the  baryons --- which are made out of $N_c$ quarks --- have  masses of order
$N_c\times\Lambda_{\rm QCD}$, so their relative abundance in thermal equilibrium is exponentially suppressed.
The interaction energy between baryons  also  scales as $N_c$.

\par\goodbreak
The $T-\mu$ phase diagram of the large--$N_c$ QCD is believed to look like figure~\ref{figLargeNQCD}
below.
(For a review see \cite{McLerran:2007qj} and references therein,  also \cite{Blake:2012dp}.)
\begin{figure}[htb]
\begin{center}
\vspace{3ex}
\includegraphics[width= 100mm]{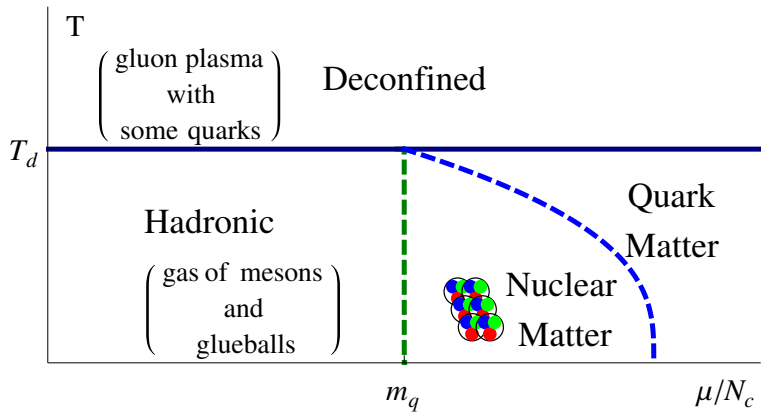}
\end{center}
\caption{Conjectured phase diagram for QCD at large $N_c$~\cite{McLerran:2007qj}}
\label{figLargeNQCD}
\end{figure}
We assume the 't~Hooft's $N_c\to\infty$ limit in which the number of flavors remains finite.
Consequently, the dynamics of the theory is dominated by the gluons, the quarks are sensitive
to the gluonic background, but the backreaction from the quarks to the gluons is suppressed
by $N_f/N_c\ll1$.
Thus, at lower temperatures there is confinement but for increasing temperature there is
a first order transition to the deconfined phase.
The transition temperature $T_d$ is at the $\Lambda_{\rm QCD}$ scale and it's almost independent
of the quark chemical potential $\mu_q=\mu_b/N_c$ (as long as $\mu_q$ does not grow with the $N_c$).

It is not clear whether for $N_c\to\infty$ the deconfining phase transition coincides
with the chiral symmetry restoration for the light quarks.
Several field theory arguments --- for example \cite{Casher,McLerran:2007qj} --- suggest that
for $\mu_q=0$ the two transitions should happen at the same point.
However, these arguments do not work for $\mu_q>0$ \cite{McLerran:2007qj},
and there are other arguments for the existence of confined but chirally restored phases
and/or deconfined phases where the chiral symmetry remains broken.
Moreover, some holographic models --- for example \cite{Aharony:2006da} --- have
deconfined but chirally broken phases even at zero chemical potentials.

For temperatures below the deconfining transition $t_d$ and baryon chemical potentials
below the baryon mass --- or equivalently for $\mu_q\lesssim m_q\equiv M_b/N_c$ ---
the thermal state of the theory is a gas of glueballs and mesons with almost no baryons
or antibaryons.
But at $\mu_q\approx m_q$ there is an abrupt phase transition to the bulk nuclear matter
with finite baryon density.
But unlike the ordinary nuclear matter which is in the quantum liquid state,
the large $N_c$ nuclear matter is crystalline solid since the ratio
of kinetic energy to potential energy decreases with $N_c$.
Indeed, the potential energy of baryon-baryon forces scales like $N_c$;
more precisely \cite{Kaplan}, in the large $N_c$ limit the two-baryon potential becomes
\begin{multline}
\label{Potential}
V\ \sim \  N_c\times A_C(r) \ + \ N_c\times A_S(r) \left({\bf I}_1\cdot {\bf I}_2\right)\left({\bf J}_1\cdot {\bf J}_2\right) \ +
\\  + \  N_c\times A_T(r) \left({\bf I}_1\cdot {\bf I}_2\right)\left(3  \left({\bf n}\cdot {\bf J}_1\right) \left({\bf n}\cdot {\bf J}_2\right)- \left({\bf J}_1\cdot {\bf J}_2\right)\right) \ + \
 O\left(1/N_c \right),
\end{multline}
for some $N_c$--independent profiles $A_C(r)$, $A_S(r)$, and $A_T(r)$ for the
central, spin-spin, and tensor forces;
their overall magnitudes are $A\sim\Lambda_{\rm QCD}$ for $r\sim 1/\Lambda_{\rm QCD}$.
Classically, this potential tries to organize the baryons into some kind of a crystal
where the distances between neighboring baryons do not depend on the $N_c$
while the binding energy (per baryon) scales like $N_c\times O(\Lambda_{\rm QCD})$.
In quantum mechanics, the baryons in such a crystal behave  like  atoms in ordinary crystals
--- they oscillate in their potential wells with zero-point kinetic energies
\be
\label{Kinetic}
{K}\  \sim \ \frac{\pi}{2m_B d^2}\sim \frac{\Lambda_{QCD}}{N_c}\,\frac{1}{d^2}\,,
\ee
where $d\sim 1/\Lambda_{\rm QCD}$ is the $N_c$-independent diameter of the potential well.
Therefore, at zero temperature the ratio of kinetic energy to the potential energy scales like
\be
\frac{K}{V}\ \sim \  \frac{1}{N_c^2}
\label{KVratio}
\ee
and becomes very small for large $N_c$.
At higher temperatures the kinetic energies of baryons become larger, $K\sim T$, but in the
confined phase we are limited to $T<T_d\sim\Lambda_{\rm QCD}$ and hence $K<O(\Lambda_{\rm QCD})$.
Consequently, the kinetic to potential energy ratio scales like
\be
\frac{K}{V}\ \sim\ \frac{1}{N_c}
\ee
--- which is larger than \eqref{KVratio} but still becomes small in the large $N_c$ limit.
Consequently, for large $N_c$ neither zero-point quantum motion nor thermal motion of baryons
can melt the baryon crystal, so nuclear matter remains solid all the way to the
deconfining temperature.

Before holography, the best models for the large--$N_c$ nuclear crystals were
lattices of skyrmions.
In this framework, Igor Klebanov \cite{Klebanov} had found a curious phase transition
from a lattice of whole skyrmions at low chemical potentials to a denser lattice of
half-skyrmions at higher potentials.
(According to \cite{KuglerShtrikman}, the whole-skyrmion lattice at low potentials
has FCC geometry while the half-skyrmion lattice at higher potentials is simple cubic.)
In the half-skyrmion-lattice phase, the order parameter for the chiral symmetry breaking
vanishes after space averaging, so in QCD terms
the transition is interpreted as chiral symmetry
restoration at high $\mu_q$.

In QCD with $N_c=3$ and two massless flavors there is a similar chirally-symmetric phase
at low $T$ and high $\mu_q$.
This phase is a quark liquid rather than a baryon liquid (the quarks are no longer confined
to individual baryons) and there is a condensate of quark pairs making this liquid
a color superconductor.
But for large $N_c$ the situation is more complicated: there is no color superconductivity,
and there is no deconfinement for $T<T_d$.
Instead, the dense cold nuclear matter forms a phase which combines the features of the
baryonic and quark phases: the quarks fill up a Fermi sea, but the interactions near the Fermi
surface are strong, so the excitations are not free quarks or holes but rather
meson-like quark-hole pairs or baryon-like states of $N_c$ quarks.
MacLerran {\it et~al} have dubbed this phase ``quarkyonic''.

For $\mu_q\gg\Lambda_{\rm QCD}$, the interior of the Fermi see is chirally symmetric,
but near the Fermi surface the symmetry is broken by the chiral density waves \cite{Rubakov}.
(Although Son and Shuster \cite{SonShuster} argue that such waves develop only
for very large $N_c>10^3$.)
To be precise, the chiral density waves mix the chiral symmetry of the quarkyonic phase
with the translational symmetry  rather than simply break it.
Averaging over space restores the chiral symmetry, just like it happens for the lattice
of half-skyrmions.

On the other hand, for $\mu_q$ just above $m_q=M_b/N_c$, the baryonic crystal has a completely
broken chiral symmetry.
Thus, at some critical $\mu_q^{(c)}=m_q\times\rm a$ few, there should be a chiral symmetry restoring
phase transition from the baryonic crystal to a distinct quarkyonic phase.
In holography, this transition is believed to be dual to the ``popcorn transition''
from a 3D instanton lattice to a 4D lattice.

%
\subsection{Effect of  the large $\lambda$ limit on the holographic nuclear matter}
In holography, the semiclassical description of the gravity side of the gauge-gravity duality
in terms of metric, fluxes, and branes requires the limits of large $N_c$ and also large
't~Hooft coupling $\lambda=N_c g_{\rm YM}^2$.
In the large $\lambda$ limit, the baryons become very heavy: in units of the mesonic mass scale
$M\sim\Lambda_{\rm QCD}$, the baryon mass is $M_b\sim\lambda N_c M$.
However, the interactions between the baryons do not grow with $\lambda$:
even for two baryons right on top of each other, the repulsive potential between them is only
$V\sim N_c M\sim M_b/\lambda$.
At larger distances, the forces are even weaker since the hard-core radius of a holographic
baryon shrinks with $\lambda$ as $R_b\sim M^{-1}\lambda^{-1/2}$.
Outside this radius, the repulsive potential decreases as $1/r^2$ until $r\sim M^{-1}$,
at which point it becomes dependent on the meson mass spectrum of a specific holographic model:
In some models, the potential becomes attractive for $r\gtrsim M^{-1}$ while in others
it remains repulsive at all distances.
The overall picture is shown in figure~\ref{figPotential}:

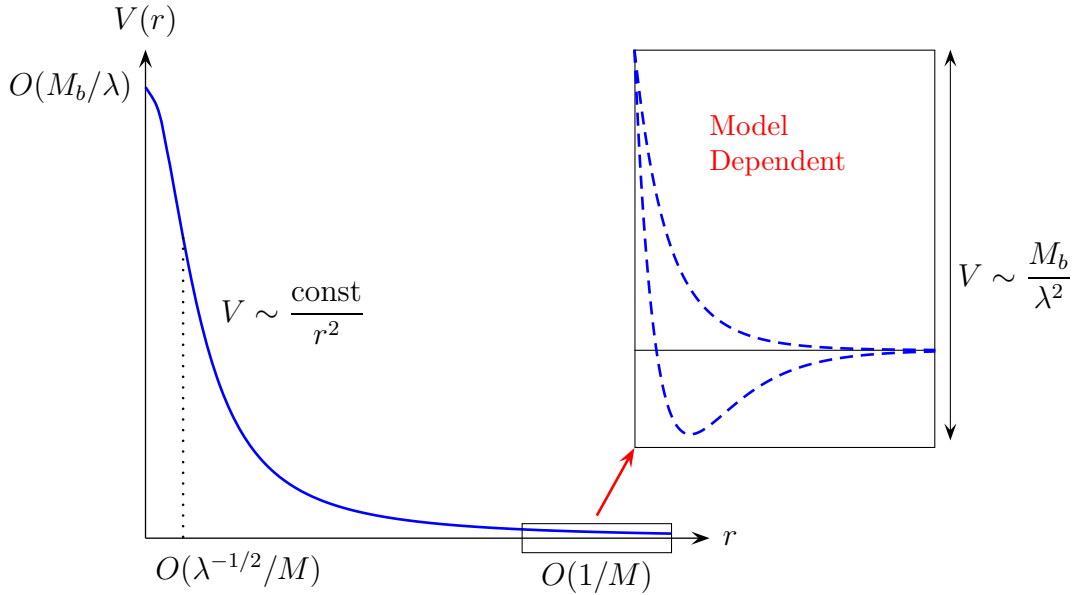
\begin{figure}[ht]
\begin{center}
\begin{pspicture}(-2,-0.5)(11,7.5)
\psline[linewidth=0.5pt,arrowscale=2]{->}(0,0)(7.5,0)
\uput[0](7.5,0){$r$}
\psline[linewidth=0.5pt,arrowscale=2]{->}(0,0)(0,6.5)
\uput[90](0,6.5){$V(r)$}
\psplot[plotstyle=curve,linecolor=blue]{0}{7}{%
	x dup mul 0.5 add  3 exch div %
	}
\uput[180](0,6){$O(M_b/\lambda)$}
\psline[linestyle=dotted,linewidth=1pt](0.5,0)(0.5,4)
\uput[300](0.5,0){$O(\lambda^{-1/2}/M)$}
\psframe[linewidth=0.4pt](5,-0.2)(7,+0.2)
\rput[t](6,-0.3){$O(1/M)$}
\psline[linecolor=red]{->}(6,0.3)(6.5,1.2)
\rput(6.5,2.5){%
	\psframe[linewidth=0.4pt](0,-1.3)(4,4)
	\psline[linewidth=0.4pt](0,0)(4,0)
	\psplot[plotstyle=curve,linecolor=blue,linestyle=dashed]{0}{4}{%
		x 0.25 exch exp dup 3 mul 2 sub mul x 4 add div 16 mul %
		}
	\psplot[plotstyle=curve,linecolor=blue,linestyle=dashed]{0}{4}{%
		x 0.25 exch exp dup 0.5 mul 0.5 add mul x 4 add div 16 mul %
		}
	\rput[l](1,3){\red\small Model}
	\rput[l](1,2.5){\red\small Dependent}
	\psline[linewidth=0.4pt,arrowscale=2]{<->}(4.2,-1.2)(4.2,4)
	\rput[l](4.3,1){$\displaystyle{V\sim{M_b\over\lambda^2}}$}
	}
\rput[l](1,3){$\displaystyle{V\sim{{\rm const}\over r^2}}$}
\end{pspicture}
\end{center}
\vspace{-10pt}
\caption{Two-body nuclear potential in holographic QCD.}
\label{figPotential}
\vspace{10pt}
\end{figure}

Since the nuclear forces are so weak in the holographic QCD, all transitions between
different phases of the cold nuclear matter happen at chemical potentials $\mu_b$
very close to  the baryon mass:
just below $M_b$ we have glueball/meson gas (or vacuum for $T=0$), while
just above $M_b$ we have dense quark matter.
To see the baryonic matter phase (or any other intermediate phases) we need to zoom
into the $\mu_b\approx M_b$ region.
\begin{figure}[ht]
\tabskip=1pt plus 1em
\halign to \hsize{%
	\hfil #\quad &\hfil #\cr
	(a)\quad & (b)\quad\cr
	\includegraphics[width=0.48\hsize]{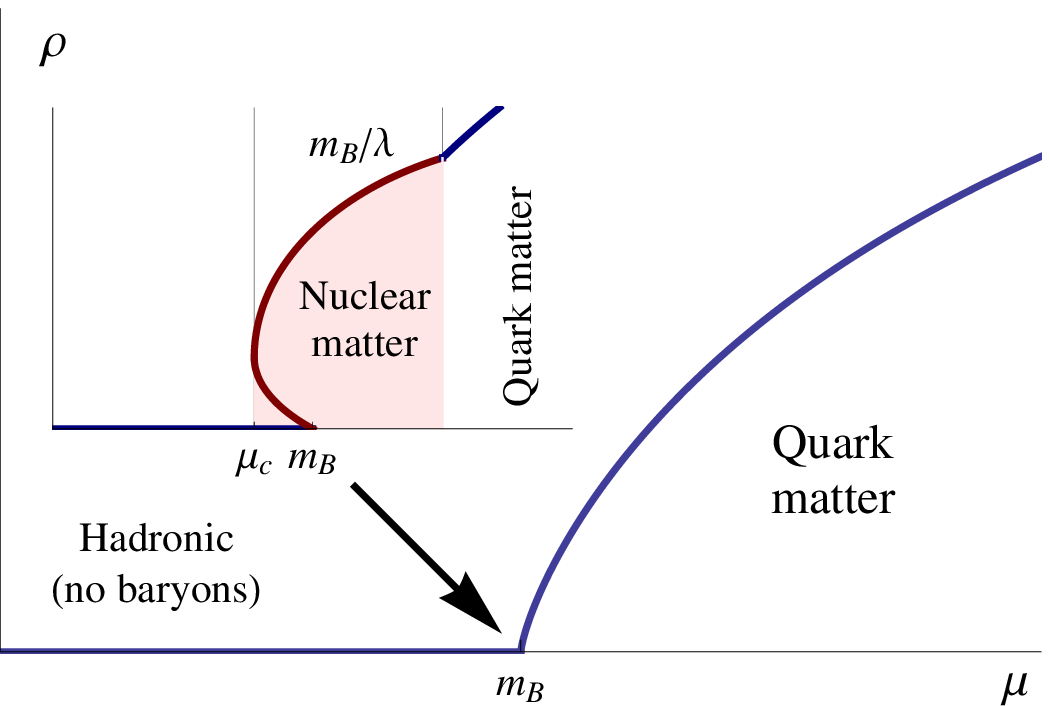}&
	\includegraphics[width=0.48\hsize]{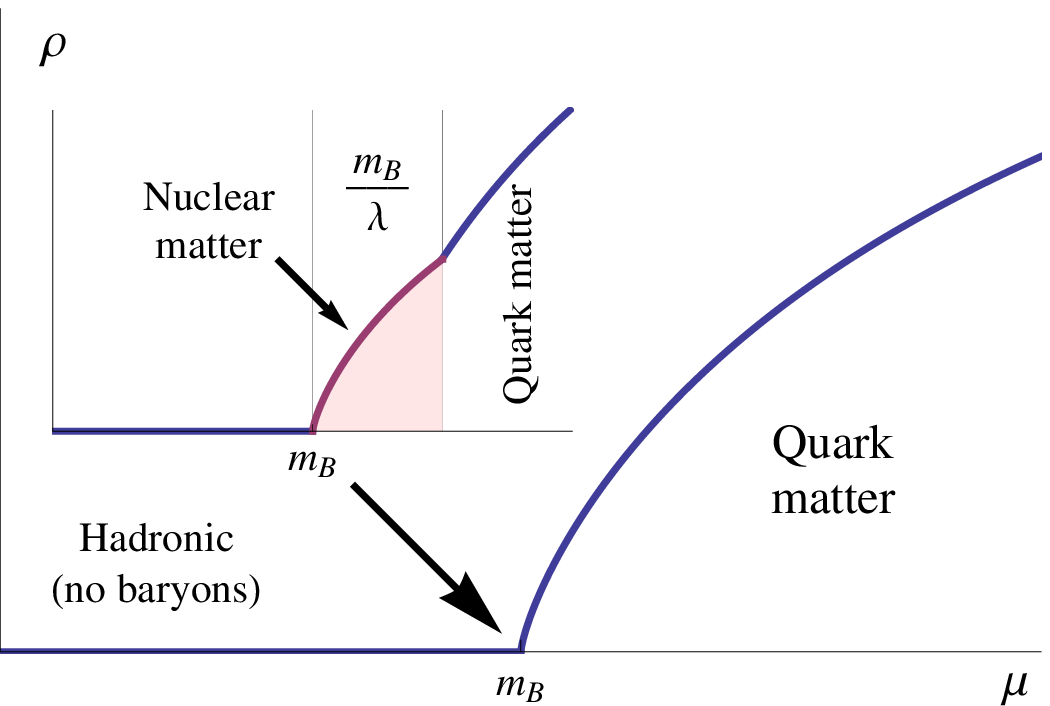}\cr
	}
\caption{Density as a function of the chemical potential in the large $\lambda$ regime
	for attracting~(a) and repelling~(b) baryons.
	The nuclear matter phase is confined to a narrow window of the order $\Delta\mu\sim B_B/\lambda$.
	In the naive diagram the transition occurs directly from the no-baryon to a quark phase.
	}
\label{figAttractiveRepulsive}
\end{figure}
Figure~\ref{figAttractiveRepulsive} (on the next page)
illustrates this point: to see the baryonic-matter phase
between the vacuum and the quark-matter phases on the plot of baryon density $\rho$
as a function of the chemical potential $\mu$, we need to zoom into the narrow range
$\mu-M_b=O(M_b/\lambda)$.
The figure also shows that the thermodynamic order of the phase transition between
the vacuum (or the meson/glueball gas for $T>0$) and the baryonic matter depends on the
sign of the long-distances nuclear force.
If the force becomes attractive at long distances, then bulk baryonic matter exists at zero
external pressure and has $\mu=M_b-\rm binding$ energy.
Consequently, the transition from the vacuum (or gas) to the nuclear matter (in the form of a
baryonic crystal) is  first-order as shown on figure~\ref{figAttractiveRepulsive}(a).
On the other hand, if the nuclear forces are repulsive at all distances, then the bulk nuclear
matter does not exists except at positive external pressures, and its chemical potential
must be $\mu>M_b$.
Moreover, $\mu$ raises monotonically with the pressure and the density, so the transition
from the vacuum to the bulk nuclear matter is  second-order
as shown on figure~\ref{figAttractiveRepulsive}(b).
\looseness=-1

In this article we do not wish to focus on a particular model of holographic QCD, so
we make no assumptions about the long-distance nuclear forces.
Consequently, we cannot say anything specific about the transition from the vacuum to the
baryonic phase.
Instead, we focus on the transition from the baryonic phase to the quarkyonic phase,
which correspond in the holographic picture to changing the lattice geometry
of the instanton crystal, from a 3D lattice, through a sequence of intermediate steps,
to a 4D lattice.
Or rather, that's the goal of our program; in this article we shall focus on a simplified problem,
namely the transitions between 1D and 2D instanton lattices.

%
\subsection{Sakai--Sugimoto model as a prototype of holographic QCD}

Some gauge theories --- especially the ${\cal N}=4$ SYM --- have exact holographic duals,
where both sides of the duality follow as IR limits of the same string-theoretical construction,
and all the extra degrees of freedom are superheavy.
Alas, the Quantum Chromo Dynamics --- or even the large $N_c$ limit of QCD --- is not so lucky:
it either does not have an exact holographic limit, or we have not found it yet.
Instead, there is a large number of ``holographic QCD'' {\it models}, which are dual
not to the QCD as such but rather to some QCD-like theories with a lot of extra junk
that the real QCD does not have.
Such models make for ``spherical horse in a vacuum'' approximations --- good enough to
get qualitative understanding, but not too accurate to make useful numerical predictions.

In a moment, we shall review the Sakai--Sugimoto model~\cite{SakaiSugimoto2004}
of holographic QCD.
That model should not be taken too seriously --- its predictions for the hadronic spectrum
are at best so-so, and for the nuclear forces are even worse, ---
but it serves as a type specimen for a large class of models.
Here are some general features of this class:
\begin{itemize}

\item
The construction starts with $N_c$ coincident D-branes, which span the Minkowski
$\rm space\times a$ compact cycle.
Typically, one uses D3 branes at a singularity, or D4-branes wrapping a circle,
although some constructions use D5 branes on a 2-cycle, {\it etc.}
For weak 't~Hooft couplings $\lambda=N_c\times g_{\rm YM}\sim N_c\times g_{\rm str}\ll1$,
the open strings between the branes give rise to the gluons of the $U(N_c)$ gauge theory
(plus a lot of extra junk).

\item
For $\lambda\gg1$, the D-branes merge into a big fat black brane.
All we see outside the horizon is a warped space-time geometry and the Ramond--Ramond fluxes
induced by the conserved charges of the D-branes.
Modes of the bulk-field excitations in this geometry are dual to the QCD {\it glueballs}.
Or rather, some modes are dual to the glueballs while other models are dual to the
extra junk that the real QCD does not have
\begin{pspicture}(0,0)(0.5,0.5)
	\pscircle*[linecolor=yellow](0.25,0.15){0.25}
	\psarc[linewidth=1pt](0.25,-0.25){0.3}{60}{120}
	\pscircle*(0.15,0.25){1pt}
	\pscircle*(0.35,0.25){1pt}
\end{pspicture}

\item
To make the hadrons (mesons and baryons), add $N_f$ flavor D-branes, usually
D5, D6, D7, or D8.
At weak $\lambda$, the open strings connecting the color branes to the flavor branes give
rise to the quarks and the anti-quarks.

\item
In the holographic limit $N_c\to\infty$ and $\lambda\to\infty$, but the $N_f$ stays finite
and the $N_f\times g_{\rm str}$ remains weak.
Consequently, the flavor branes remains D-branes (rather than merge into a black brane of their own),
but now they live in a warped geometry with fluxes induced by the $N_c$ color branes.

\item
The open strings between the flavor branes give rise to $N_f^2$ vector and scalar fields
living on those branes.
The 4D modes of these vector and scalar fields are dual to the QCD {\it mesons.}

\item
The YM instantons of the vector fields on the flavor branes are dual to the QCD {\it baryons,}
see \S2.4 for more details.

\end{itemize}

And now let us review the Sakai--Sugimoto model~\cite{SakaiSugimoto2004}
in a little more detail.
The color sector of the model is the Witten's  model~\cite{WittensModel}:
$N_c$ D4 branes spanning the Minkowski $\rm space\times a$ circle;
the circle has antiperiodic boundary conditions for the fermions, which breaks
the ${\cal N}=4$ SUSY down to ${\cal N}=0^*$.
In the field theory limit $\lambda\ll1\,\Longrightarrow\,\Lambda_{\rm QCD}\ll M_{\rm KK}$
the effective low-energy theory is pure $U(N_c)$ Yang---Mills,
but in the holographic limit $\lambda\gg1$ everything happens right at the Kaluza--Klein scale
$M_{\rm KK}=1/\rm circle$ radius, so the YM glueballs end up with similar $O(M_{\rm KK})$
masses to a lot of non-YM junk.

On the gravity side of the duality, the D4 branes merge into a black brane which warps the 10D metric.
Instead of flat 10D spacetime, we now have a warped product of $\R^{3,1}$ Minkowski space,
the $S^4$ sphere (originally surrounding the D4 branes), and a 2D cigar spanning the
radial direction ($\perp$ to the branes) and the $S^1$ circle.
The radial coordinate $u$ runs from $u_\Lambda>0$ to infinity; at $u_\Lambda$ the $S^1$ circle
shrinks to a point, hence the cigar.
Altogether, we have warped metric,  the 4--form flux, and the dilaton's gradient according to
\begin{align}
ds^2\ &
=\ \left( \frac{u}{R_{D4}}\right)^{3/2}
\Bigl[-d t^2+\delta_{ij}d x^i d x^j+f(u)d x_4^2\Bigr]\,
+\,\left( \frac{R_{D4}}{u}\right)^{3/2}
\left[\frac{d u^2}{f(u)}+u^2d\Omega_4^2\right]\nonumber\\[7pt]
F_4\ &
=\ 3\pi\ell_s^3 N_c\times\mbox{volume\_form}(S^4)\,,\qquad
e^{\phi}\ =\ g_s\left( \frac{u}{R_{D4}}\right)^{3/4},
\label{D4background}
\end{align}
where $x_4$ was the coordinate along the $S^1$ circle and now is the polar angle on the cigar,
\be
R_{D4}^3\ =\ \pi g_s\ell_s^3 N_c\,,\qquad
f(u)\ =\ 1\,-\,\left( \frac{u_{\Lambda}}{u}\right)^3\,,
\ee
$\ell_s=\sqrt{\alpha'}$ is the string length scale, and $g_s$ is the string coupling.
The $u_\Lambda$ --- the minimal value of the radial coordinate $u$ at the tip of the cigar ---
is related to the original radius $R$ of the $S^1$ circle as
\be
\label{KKscale}
2\pi R\ =\ \frac{4\pi}{3}\left( \frac{R_{D4}^3}{u_{\Lambda}}\right)^{1/2}.
\ee
The same radius $R$ also controls the 4D Yang--Mills coupling and hence the 't~Hooft's
coupling $\lambda$.
Analytically continuing from $\lambda\ll1$ to $\lambda\gg1$, we have
\be
\lambda\ =\ g^2_{\rm 4D}\times N_c\
=\ {g^2_{5D}\over2\pi R}\times N_c\
=\ {2\pi g_s\ell_s\over R}\times N_c\,.
\label{'tHooftconst}
\ee

To add the flavor degrees of freedom to the model, Sakai and Sugimoto added $N_f$ D8 branes
and $N_f$ anti--D8 branes.
On the field-theory side of the holographic duality ($i.\,e.$, for $\lambda\ll1$),
they span all space coordinates except the $x_4$ coordinate along the $S^1$ circle.
At the intersections of D8 and D4 branes, the open strings give rise to the massless quarks;
likewise, at the intersections of the anti-D8 branes and the D4 branes we get massless antiquarks.

\par
\begin{figure}[ht]
\begin{center}
\vspace{3ex}
\includegraphics[width= 100mm]{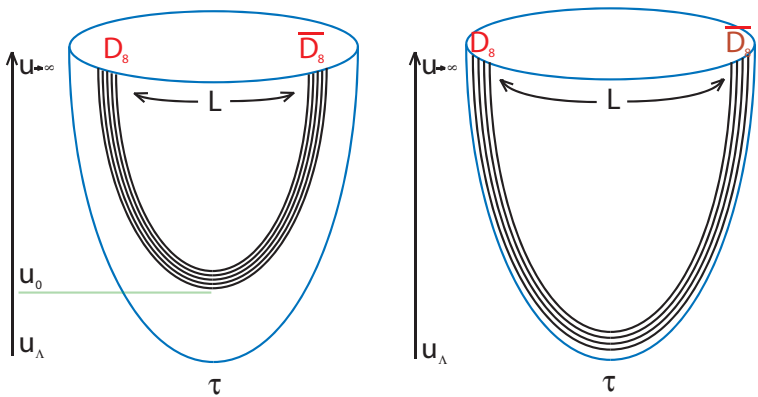}
\vspace*{-2cm}
\end{center}
\caption{The figure on the right is the generalized  non-antipodal configuration.
	The figure on the right describes the limiting antipodal case $L=\pi R$, where
	the branes connect at $u_0=u_{\Lambda}$.
	}
\label{Ushape}
\end{figure}

\goodbreak
On the holographic side $\lambda\gg1$ of the duality, the exact solution for the flavor
branes interacting with the warped metric and fluxes is not known, but for $N_f\ll N_c$
and $N_f\times g_s\ll1$ we may use the probe approximation:
The flavor branes seek the lowest-action configuration in the background metric~\eqref{D4background},
while their back-reaction upon the metric is neglected.
Consequently, at low temperatures (below the deconfinement transition)\footnote{%
   At higher temperatures --- above the deconfining transition ---
   the background metric has different topology, and the flavor branes
   also have different shapes, see Aharony {\it et~al} \cite{Aharony:2006da} for details.
   },
the flavor branes span the Minkowski ${\rm space}\times S^4\times\rm a$ line on the cigar;
the exact shape of this line follows from minimizing the branes' action, but its topology
follows from the cigar itself:
Since the  D8 and the anti-D8 branes
cannot reach all the way to the origin $u=0$, they must reconnect to each other and form
U-shaped configurations as shown on figure~\ref{Ushape}.%
The reconnection is the geometric realization of the spontaneous chiral symmetry breaking:
the separate stacks of $N_f$ D8 and $N_f$ anti-D8 probe branes give rise to the
$U(N_f)_L\times U(N_f)_R$ gauge symmetry, which corresponds to the $U(N_f)_L\times U(N_f)_R$
global chiral symmetry in 4D.
But when the D8 and anti-D8 branes reconnect, there is only a single stack of $N_f$ U-shaped branes
and hence only one unbroken $U(N_f)$ symmetry.
Thus, the chiral symmetry is spontaneously broken,
\be
U(N_f)_L\times U(N_f)_R\to U(N_f).
\ee

As shown on figure~\ref{Ushape}, the U-shaped profiles of reconnected branes depend
on one parameter --- the asymptotic distance $L$ between the D8 and the anti-D8 branes along the
$S^1$ circle for $u\to\infty$.
For $L=\pi R$ the branes form the {\it antipodal configuration} in which the branes
remain at opposite points on the circle for all $u$, all the way from $u\to\infty$ down to
$u=u_\Lambda$ where the branes reconnect;
this is the original configuration of Sakai and Sugimoto.
In the more general version of the model~\cite{Aharony:2006da} we allow for the $L<\pi R$
{\it non-antipodal configurations.}
In such configurations, the distance between the branes in the $x_4$ direction depends on $u$
--- it becomes smaller for smaller $u$ --- and eventually the branes reconnect at $u_0$
before they reach the bottom of the cigar.
The $u_0/u_\Lambda$ ratio may be used to parametrize the non-antipodal configurations
instead of the $L/R$.

The $u_0/u_\Lambda$ or $L/R$ parameter of the Sakai--Sugimoto model does not correspond
to any adjustable parameters of the real-life QCD.
Unfortunately, this parameter affects many physical properties of the model.
For example, for $(L/R)>0.97$ the deconfinement and the restoration of chiral symmetry
happen at the same temperature, but for $(L/R)<0.97$ they happen at different temperatures
and the model has an intermediate deconfined but chirally broken phase~\cite{Aharony:2006da}.
Also, in the antipodal model the central nuclear forces are purely repulsive, while
the non-antipodal models give rise to both repulsive and attractive nuclear forces
\cite{Kaplunovsky:2010eh}
(but unfortunately the net force remains repulsive at all distances).

The low-energy dynamics of the flavor degrees of freedom living on the D8 branes is governed by
the effective action comprising the Dirac--Born--Infeld (DBI) and Chern--Simons (CS) terms,
\be
S=S_{\rm DBI} + S_{\rm CS}\,.
\ee
The DBI action is
\be
\label{DBIaction}
S_{\rm DBI}\ =\ T_8\!\int\limits_{D8+\overline{D8}}\! d^9x\, e^{-\phi}
\mathop{\rm Str}\nolimits\left(\sqrt{-\det(g_{mn} + 2\pi\alpha' \CF_{mn})}\right),
\ee
where $T_8 = (2\pi)^{-p}\ell_s^{-p-1}$ is the D8-brane tension,
$g_{mn}$ is the nine-dimensional induced metric on the branes,
$\mathcal{F}_{mn}$ is the $U(N_f)$ gauge field strength, and Str denotes the symmetrized trace
over the flavor indices.
%
%
In the limit of fixed brane geometry and weak gauge fields, the DBI action reduces to Yang--Mills,
\be
S_{\rm DBI}[{\cal F}]\ =\ {\rm const}\ +\ S_{\rm YM}[{\cal F}]\
+\ O({\cal F}^4).
\ee
Furthermore, the low-energy field modes we are interested in are constant along the $S^4$ sphere
and the vector fields' directions are $\perp S^4$.
Consequently, we are going to dimensionally reduce the flavor gauge theory down to 5 dimensions:
the 4 Minkowski dimensions $x^{0,1,2,3}$, plus one coordinate $z$ along the U-shaped line on the cigar.
We find it convenient to choose a particular $z$ coordinate that makes the 5D metric conformal
\be
ds^2\ =\ A(z)\times\bigl( -dt^2\,+\,d\bx^2\,+\,dz^2\bigr),\qquad
A(z)\ =\ \left(\frac{u(z)}{R_{D4}}\right)^{3/2};
\ee
although its relations to the $u$ and $x^4$ coordinates of the cigar follow from the
differential equations
\be
\frac{du(z)}{dz}\ =\ \sqrt{\frac{u^8(z)f(u(z))\,-\,u_0^8f(u_0)}{u^5(z) R^3_{D4}}}\,,\qquad
\frac{dx^4(z)}{dz}\ =\ \frac{u_0^4\sqrt{f(u_0)}}{u^4(z)f(u(z))}\,.
\label{CigarCoordinates}
\ee
In the $(x^0,x^1,x^2,x^3,z)$ coordinates, the 5D YM action for the flavor gauge fields becomes
\be
S_{\rm YM}\ \approx\int\!\!d^4x\!\int\!\!dz\,{1\over 2g^2_{\rm YM}(z)}\,\tr\bigl({\cal F}_{MN}^2\bigr)
\label{actionz0}
\ee
for
\be
{1\over 2g^2_{\rm YM}(z)}\
=\ {N_c\lambda M_{\rm KK}\over 216\pi^3}\times{u(z)\over u_\Lambda}\,.
\ee
Near the bottom of the U-shaped flavor branes --- which is the only region we are going to
care about in this article ---
\be
{1\over 2g^2_{\rm YM}(z)}\ =\ {N_c\lambda M_{\rm KK}\over 216\pi^3}\times\left(
	\zeta\,+\,{8\zeta^3-5\over 9\zeta}\,M_{\rm KK}^2\times z^2\,
	+\,O(m_{\rm KK}^4 z^4)\right)\qquad
\mbox{where}\quad\ \zeta\,=\,{u_0\over u_\Lambda}\,.
\label{Uexpansion}
\ee

The Chern--Simons term arises from the couplings of the gauge fields on the D8 brane
to the bulk Ramond--Ramond field.
In 9 dimensions
\be
\label{CSterm0}
S_{\rm CS} = T_8\times\!\!\int\limits_{\rm D8+\overline{D8}} \!\! C_3\wedge \tr e^{2\pi\alpha' \CF}\,,
\qquad \text{where}\qquad F_4=d C_3.
\ee
After integrating over the $S^4$ and dimensionally reducing to 5D,
the Chern--Simons term becomes
\be
\label{CSterm1}
S_{\rm CS}\ =\ {N_c \over 24\pi^2} \int_{\rm 5D}\tr\left(
	\mathcal{A}\mathcal{F}^2\,
	-\,\tfrac{i}{2}\mathcal{A}^3\mathcal{F}\,
	-\,\tfrac{1}{10}\mathcal{A}^5
	\right) .
\ee
In a particularly interesting case of 2 flavors, it is convenient to separate the $U(2)$
gauge fields ${\cal A}_M$ into their $SU(2)$ components $A_M$ and the $U(1)$ components
$\hat A_M$.
In terms of these components, the Chern--Simons action becomes
\be
\label{CSterm2}
S_{\rm CS}\ =\ \frac{N_c}{16\pi^2}\int\! \hat A\wedge {\rm tr} F^2\
+\ \frac{N_c}{96\pi^2}\int\! \hat{A}\wedge \hat{F}^2\,.
\ee
We shall see in a moment that  the baryons and the multi-baryon systems have strong
self-dual $SU(2)$ magnetic fields $F_{\mu\nu}$.\footnote{%
	In our notations, the spacetime indices $0,1,2,3,z$ of the effective 5D theory are labeled
     $M,N,\ldots=0,1,2,3,z$ while the space indices $1,2,3,z$ of the same theory
     are labeled $\mu,\nu,\ldots$.
     When we need the 9D indices for the whole D8 brane, we use $m,n,\ldots$.
     }
Thanks to the first term in this Chern--Simons action, the instanton number density
\be
I(x,z)\ =\ {\epsilon^{\kappa\lambda\mu\nu} F^a_{\kappa\lambda}F^a_{\mu\nu}\over 32\pi^2}
\ee
acts as electric charge density for the abelian field $\hat A_0$;
the net electric charge of an instanton is $Q_{\rm el}=N_c/2$.

Besides the $U(N_f)$ gauge fields, the effective low-energy 5D theory also contains the
scalar fields $\Phi^a(x,z)$ describing the small fluctuations of the D8 branes in the transverse
direction.
For $N_f$ branes, the scalars form the adjoint multiplet of the $U(N_f)$ gauge symmetry.
The action for the scalar fields follow from the DBI action for the embedded metric $g_{mn}$
of the fluctuating branes.
For the $\Phi(x,z)$ fields normalized to have similar kinetic energies to the vector fields,
the scalar action looks like
\begin{align}
S_{\rm scalar}\ &
=\int\!\!d^4x\!\int\!\!dz\,{1\over 2g^2_{\rm YM}(z)}\,\tr\Bigl(
    (D_M\Phi)^2\,+\,V(\Phi)\Bigr)
\nonumber\\
&\quad+\ \frac{N_c}{16\pi^2}\int\!\!d^4x\!\int\!\!dz\, C(z)\times
	\tr\bigl(\Phi{\cal F}_{MN}{\cal F}^{MN}\bigr)\ +\ \cdots.
\label{ScalarAction}
\end{align}
The details of the scalar potential $V(\Phi)=m^2(z)\times\Phi^2+a(z)\times\Phi^4+\cdots$
need not concern us here, what's important is the second term describing the backreaction
of the gauge fields on the brane geometry.
In the antipodal Sakai--Sugimoto model $C(x)\equiv0$ and there is no
backreaction because of a geometric symmetry,
but in the non-antipodal models $C(z)\neq0$ and the scalar fields $\Phi$ induced
by the vector fields of the baryons lead to  attractive nuclear forces~\cite{Kaplunovsky:2010eh}.
The ratio of these attractive forces to the repulsive forces mediated by the abelian electric fields
depends of the $C(z)$ profile of the interaction term~\eqref{ScalarAction}.
For the Sakai--Sugimoto models
\be
{\rm attractive\over repulsive}\
=\ C^2(z)\ =\ {1-\zeta^{-3}\over 9}\times\left({u_0\over u(z)}\right)^8\ \le\ {1\over9}\ <\ 1,
\label{ScalarToVector}
\ee
so the net force is repulsive.

To see how that works, let us focus on the baryons in the Sakai--Sugimoto and other models
of the holographic QCD.

%
\subsection{Baryons  as flavor instantons}
\label{secSSBaryons}

In the old hadronic string, the baryons were made out of Y-shaped configurations
of 3 open strings connected to each other at one end;  the other end of each string
is connected to a quark.
To realize this picture in holographic QCD, we need a {\it baryon vertex} (BV)
--- some object connected to $N_c$ open strings.
The other ends of the strings are connected to the flavor branes and act like the quarks;
this gives the baryon its flavor quantum numbers.
Witten had constructed the baryon vertex for the ${\rm AdS}_5\times S^5$ model from a D5 brane
wrapping the $S^5$ sphere~\cite{WittenBaryons}; the generalized versions of
this construction~\cite{Brandhuber:1998xy} use ${\rm D}_p$ branes wrapping compact cycles
carrying $O(N_c)$ Ramond--Ramond fluxes.

In the Sakai--Sugimoto version of this construction,
the baryon vertex is realized as a D4 brane wrapped around
the $S^4$ sphere (but localized in all other dimensions except the time).
The $S^4$ carries $N_c$ units of the $F_4$ Ramond--Ramond flux,
\be
\frac{1}{(2\pi)^3l_s^3}\int_{S^4} F_4 = N_c\,,
\ee
so the Chern--Simons coupling of this flux to the $U(1)$ gauge field $\cal B$ living
on the D4 brane acts a $N_c$ units of the net electric charge for the $B_0$:
\be
T_4\int_{\rm D4} C_3\wedge e^{2\pi\alpha'd B} = N_c \int_{\rm time}\!\!B_0dt\,.
\label{D4BraneCharge}
\ee
In a compact space like  $S^4$, the net electric charge must vanish.
To cancel the charge~\eqref{D4BraneCharge} we need to connect the D4 brane to
open strings.
The back end of an oriented open string has electric charge $-1$, so we must connect the
D4 brane with $N_c$ such strings;
their front ends connect to the D8 flavor branes (since the strings do not have any other place to end)
and act as $N_c$ quarks.

We may put the D4 brane anywhere in space and anywhere on the cigar.
However, the $S^4$ volume increases with the $u$ coordinate, so the lowest-energy location
of the D4 is the cigar's tip $u=u_\Lambda$.
At other location, the brane feels a gravity-like force pulling it down to the tip.
However, the strings connected to the baryon vertex pull it towards the flavor D8 branes;
in the non-antipodal models the D8 branes do not reach the cigar's tip, so the strings
pull the baryonic vertex up from the tip towards the lowest point $u_0$ of the flavor branes.
The `tag of war' between the upward and downward forces on the baryon vertex determines its
ultimate location.
In some models, the forces reach equilibrium for the
baryon vertex hanging on strings below the flavor branes~\cite{Dymarsky:2010ci},
while in many other models --- including the non-antipodal Sakai--Sugimoto ---
the string forces win and pull the baryon vertex all the way to the lowest point $u_0$
of the flavor branes~\cite{Seki:2008mu}.

In all such models, the baryonic vertex is a ${\rm D}_p$ brane completely embedded in
a stack of ${\rm D}_{p+4}$ flavor branes, so it is equivalent to zero-radius Yang--Mills
instanton of the $U(N_f)$ gauge symmetry on the flavor branes, and for $N_f>1$ it may be smoothly
inflated to a finite-radius instanton~\cite{Dbrane-instanton}.
In $p+5$ dimensions of the flavor branes, this instanton is a fat ${\rm D}_p$--brane wrapping
some compact cycle,
but once we dimensionally reduce to 5 dimensions, the instanton becomes a finite-size particle.
Thus, {\it in the low-energy effective 5D theory of the holographic QCD, a baryon is realized
as a finite-size instanton of the $U(N_f)$ gauge theory.}

Our research program of studying the instanton crystals as holographic duals of the cold nuclear matter
stems from this realization of baryons as instantons in 5D, but it does not depend on any
specific details of the Sakai--Sugimoto model.
We use that model as a prototypical example, but any other model where the baryons are realized
as instantons would be just as good for our purposes, for example the model of \cite{DKS}
for some 7--brane geometries \cite{Dymarsky:2010ci}, or the $\rm AdS_5\times S^1$ model of
\cite{Kuperstein:2004yf} (the baryons of that model are studied in \S6 of \cite{Seki:2008mu}).

In the baryon-vertex picture, each of the $N_c$ strings connecting the vertex to the flavor branes
has electric charge $1/N_f$ under the abelian $U(1)$ subgroup of the $U(N_f)$, so the whole baryon
has abelian charge $N_c/N_f$.
In the instanton picture, the same electric charge obtains from the Chern--Simons coupling
between the abelian electric field and the non-abelian magnetic fields of the instanton.
\begin{align}
S_{\rm CS}\ =\ {N_c\over 24\pi^2}\int\!\!tr\Bigl(
	{\cal AF}^2\,-\,\tfrac{i}{2}{\cal A}^3{\cal F}\,-\,\tfrac{1}{10}{\cal A}^5\Bigr)\ &
\supset\ {N_c\over N_f}\int\!\hat A_{U(1)}\wedge{\tr\bigl(F\wedge F\bigr)_{SU(N_f)}\over 8\pi^2}\,,
\nonumber\\
&\supset\ {N_c\over N_f}\int\!\!d^5x\, \hat A_0(x)\times I(x)
\label{CSterm3}
\end{align}
where $I(x)$ is the instanton number density of the magnetic fields.
For $N_f\ge3$, the Chern--Simons couplings also endow instantons with non-abelian
electric charges.
Altogether,
\be
S_{\rm CS}\ \supset\int\!\!d^5x\,A_0^a(x)\times Q_{\rm el}^a I(x)
\ee
where the net electric charge $Q^a_{\rm el}$ has both abelian and non-abelian components;
in $N_f\times N_f$ matrix language,
\be
{\rm for}\ F^{\mu\nu}_{\rm magnetic}\,\in\,\begin{pmatrix}
	\, SU(2)\, &
	\omit\vrule  height 16pt depth 6pt &
	\quad\hbox{\Large\bf 0}\quad\\
	\noalign{\hrule}
	\hbox{\Large\bf 0} &
	\omit\vrule  height 20pt depth 6pt &
	\hbox{\Large\bf 0}
	\end{pmatrix},\qquad
Q_{\rm el}\ =\ {N_c\over2}\times\begin{pmatrix}
	\enspace\hbox{\Large\bf 1}\enspace &
	\omit\vrule  height 16pt depth 6pt &
	\enspace\hbox{\Large\bf 0}\enspace\\
	\noalign{\hrule}
	\hbox{\Large\bf 0} &
	\omit\vrule  height 20pt depth 6pt &
	\hbox{\Large\bf 0}
	\end{pmatrix},\qquad
\tr(Q^2_{\rm el})\,=\,{N_c^2\over2}\,.
\label{NAcharge}
\ee
Abelian or nonabelian, the like-sign electric charges repel each other;
it is this Coulomb repulsion between different parts of the same instantons that prevents it
from collapsing to a point-like D-brane.

However, the instantons do not grow large because the 5D gauge coupling decreases away from the
$z=0$ hyperplane: a large instanton would spread into regions of space where the coupling is weaker,
and that would increase the instanton's energy.
Instead, the equilibrium radius of the instanton scales like
\be
a_{\rm instanton}\ \sim\ {1\over M_{\rm KK}\times\sqrt{\lambda}}\,.
\label{RBscaling}
\ee
For a holographic model of a baryon, this radius is unrealistically small.
Indeed, using the $\rho$ meson's mass as a unit, the real-life baryon radius
$R_b\sim 3.4 M^{-1}_\rho$, while in holography $a\ll M_\rho^{-1}\sim M_{\rm KK}^{-1}$.
Moreover, it raises the question of whether we may adequately describe so small
an instanton using the $\rm DBI+CS$ action, or perhaps we need to include the higher-order
stringy corrections.\footnote{%
    In string theory, the  $\rm DBI+CS$ action for the gauge fields on a D-brane
    is exact for {\sl constant} tensions fields ${\cal F}_{mn}$, however strong.
    But for the variable tension fields, the DBI action includes all powers of the
    ${\cal F}_{mn}$ but neglects their derivatives $D_k{\cal F}_{mn}$,
    $D_kD_p{\cal F}_{mn}$, \dots.
    It is not clear what effect (if any) such higher-derivative terms would have on
    a small instanton.
    In a supersymmetric background, the instanton is BPS and its net mass is protected
    against stringy corrections, so the DBI action --- or even the Yang--Mills action ---
    gives the exact value.
    But what happens to small instanton in non-supersymmetric backgrounds is an open question.
    }
On the other hand, assuming the $\rm DBI+CS$ description is OK, the radius
as in eq.~\eqref{RBscaling}
allows for consistent expansion of the instanton's parameters in powers of
$\lambda^{-1}$~\cite{Hata:2007mb}.
In particular, the leading contribution to the instanton's mass is $O(\lambda\times N_cM_{\rm KK})$
while the corrections  due to $z$-dependent gauge coupling and
due to Coulomb self-repulsion are both $O(N_cM_{\rm KK})\sim M_I/\lambda$.

To see how that works, consider a static instanton --- a time-independent configuration
of $SU(N_f)$ magnetic fields, plus the electric fields induced by the CS couplings, and the
scalar fields induced by the $\tr(\Phi FF)$ coupling to the magnetic fields.
Since the canonically normalized couplings in 5D are $O(\lambda^{-1/2})$, the leading
contribution to the instanton's energy comes from the magnetic field.
In the DBI approximation\footnote{%
	In string theory, the complete action for scalar and vector fields living on a D-brane
	starts with the DBI and CS terms but also includes an infinite number of terms involving
	$D_n{\cal F}_{mn}$ and the higher derivatives.
	The $\rm DBI+CS$ action is merely a long-distance approximation that allows for
	strong gauge field tensions ${\cal F}_{mn}$ as long as their gradients are small.
    It is not clear how well this approximation works for small instantons, but at present
    we do not have a better approximation.
    }
\be
E_{\rm DBI}\ =\int\!\!d^3\bx\!\int\!dz\,{1\over 2 g_{\rm YM}^2(z)}
\times\mathop{\rm Str}\nolimits\left(\sqrt{\det\bigl(K(z)\delta_{\mu\nu}+F_{\mu\nu}\bigl)}\,-\,K^2(z)\right)
\ee
where $K(z)=(2\pi\alpha')^{-1}\times g_{11}(z)\sim1/(\lambda M_{KK}^2\ell_s^4)$.
Both $g_{\rm YM}(z)$ and $K(z)$ depend on the $z$ coordinate on the distance scale $O(1/M_{\rm KK})$,
so for instantons of much smaller size we may start with the approximation of constant $K$
and constant 5D gauge coupling.
And in this approximation, the DBI energy is minimized by the magnetic fields that are exactly
self-dual (in the 4D space of $(x^1,x^2,x^3,z)$);
moreover, the DBI energy of an instanton is equal to its Yang--Mills energy
\be
E_{\rm DBI}({\rm instanton})\ =\ E_{\rm YM}({\rm instanton})\ =\ {8\pi^2\over g^2_{\rm YM}}
\ee
regardless of its radius and of the $K$ parameter of the DBI action.

To determine the equilibrium radius of an instanton we need to calculate its energy
to the next order in $1/\lambda$ expansion.
To this order, we assume the magnetic fields to be
exactly self-dual --- which allows us to use the YM action
instead of DBI even for small instantons --- but the gauge coupling is $z$--dependent,
and we also account for the electric and the scalar fields.
For small instantons, we may approximate the 5D gauge coupling as
\be
{8\pi^2\over g^2(z)}\ =\ N_c\lambda M_{\rm KK}\left(
	B\,+\,DM_{\rm KK}^2\times z^2\,+\,O(M_{\rm KK}^4z^4)\right)
\ee
for some numerical constants $B$ and $D$; for the Sakai--Sugimoto model
$B=\zeta/27\pi$ and $D=(8\zeta^3-5)/9\zeta^2$, while other models may have different values.
Consequently, the YM energy of the nonabelian magnetic fields evaluates to
\be
E_{\rm NA}\ =\ N_c\lambda M_{\rm KK}\times\left(
	B\,+\,DM_{\rm KK}^2\times Z^2\,+\,DM_{\rm KK}^2\times {a^2\over2}\right)
\label{ENAsingle}
\ee
where $a$ is the instanton's radius and $Z$ is is the $z$ coordinate of its center.

The electric potentials $A_0^a$ couple to the $F_{\mu\nu}\tilde F^{\mu\nu}$ products
of the magnetic fields while the scalar potentials $\Phi^a$ couple to the $F_{\mu\nu}F^{\mu\nu}$.
For the self-dual magnetic fields, both potentials couple to the same source
$I(\bx,z)\times Q^a_{\rm el}$,
and the only difference is the coupling strength ratio $C(z)$
({\it cf.}\ eqs.~\eqref{ScalarAction} and \eqref{ScalarToVector}).
For small instantons we may neglect the $z$--dependence of this ratio and let $C(z)\approx C(0)\equiv C$,
and as long as we do not go very far from the instanton (for distances $r\ll R_{\rm KK}$),
we may also neglect the $z$--dependence of the gauge coupling.
Consequently, both electric and scalar potentials become the 4D Coulomb potentials
\be
\Phi(\bx,z)\ =\ C\times A_0(\bx,z)\
=\ {2Q_{\rm electric}\over B N_c\lambda M_{\rm KK}}\int\!\!d^3\bx'\!\int\!\!dz'\,
	{I(\bx',z')\over (\bx-\bx')^2+(z-z')^2}\,.
\ee
The electric fields lead to repulsive forces between the charges while the scalar forces lead
to the attractive forces.
Altogether, the net Coulomb energy amounts to
\begin{align}
E_C\ &
=\ {B\over8\pi^2}\,N_c\lambda M_{\rm KK}\times
\int\!\!d^3\bx\,dz\,\tr\bigl( (\nabla A_0)^2\,-\,(\nabla\Phi)^2\bigr)\\
&=\ (1-C^2)\times{N_c\over4B \lambda M}\int\!\!d^3\bx\,d^3\bx'\,dz\,dz'\,
	{I(\bx,z)\times I(\bx',z')\over (\bx-\bx')^2+(z-z')^2}\,,\nonumber
\end{align}
and for a single instanton of radius $a$ it  evaluates to
\be
E_C\ =\ {1-C^2\over B}\times {N_c\over5\lambda M_{\rm KK} a^2}\,.
\label{ECsingle}
\ee
Note: for $C<1$ the electric fields are stronger than the scalar fields and the net
Coulomb energy of the instanton is positive --- which makes for a net self-repulsive force
that prevents the instanton from shrinking to zero radius.
In models with $C>1$ (assuming they exist), the scalar fields would be stronger,
the net Coulomb energy would be negative, which means a self-attractive force rather than self-repulsive.
In such a model, the instanton would shrink to zero radius and our approximations would not be valid.

Altogether, the net energy of an instanton amounts to
\be
E(\mbox{instanton})\ =\ N_c\lambda M_{\rm KK}\times\left(
	B\,+\,DM_{\rm KK}^2\times Z^2\,+\,DM_{\rm KK}^2\times {a^2\over2}\,
	+\,{1-C^2\over 5B}\times{1\over \lambda^2 M_{\rm KK}^2 a^2}
	\right)
\ee
Note that for $a^2\sim1/(\lambda M_{\rm KK}^2)$ both radius-dependent terms here
are  $O(1/\lambda)$ corrections to the leading term.
Minimizing the net energy, we find the equilibrium value of the instanton radius
\be
a\ =\ {1\over M_{\rm KK}\sqrt{\lambda}}\times
\root 4\of{2(1-C^2)\over 5BD}\,,
\label{GenRadius}
\ee
the instanton center is in equilibrium at $Z=0$ (the bottom of the U-shaped flavor branes),
and the instanton's mass is
\be
M_I\ =\ N_c M_{\rm KK}\times\left(
	\lambda B\,+\,\sqrt{2D(1-C^2)\over 5B}\,+\,O(1/\lambda)\right).
\label{GenMass}
\ee
For the antipodal Sakai--Sugimoto model
\be
a\ =\ {\root 4\of{162\pi/5}\approx 3.2\over M_{\rm KK}\sqrt{\lambda}}\quad
{\rm and}\quad M_I\ =\ N_c M_{\rm KK}\times\left(
	{\lambda\over 27\pi}\,+\,\sqrt{18\pi\over5}\,+\,O(1/\lambda)\right),
\ee
while other models should have different $O(1)$ numeric factors.

We may absorb 2 out of 3 model-dependent parameters $B,C,D$ into a redefinition of
the $\lambda$ and $M_{\rm KK}$ parameters of the effective 5D theory,
for example
\begin{align}
\lambda\ &\to\ {\lambda\over\sqrt{DB^3}}\,,\quad
M_{\rm KK}\ \to M\,=\,M_{\rm KK}\times\sqrt{D\over B}\,,\\
\intertext{thus}
{8\pi^2\over g^2(z)}\ &
=\ N_c\lambda M\Bigl(1\,+\,M^2\times z^2\,+\,\cdots\Bigr).
\label{HoloGZ}
\end{align}
For the {\it static} instantons or multi-instanton systems we may also get rid of the third
model-dependent parameter $C$.
Indeed, for the static systems $\Phi(\bx,z)=C\times A_0(\bx,z)$, so the only effect of the scalar
fields is to reduce the net Coulomb force by a constant factor $(1-C^2)$.
We may simulate this effect without any scalar fields by using different gauge couplings
for the electric and magnetic fields, $g^2_{\rm el}=(1-C^2)\times g^2_{\rm mag}$,
or equivalently by rescaling the time dimension $x^0$ relative to the space dimensions
$(x^1,x^2,x^3,z)$,
\be
t\ \to\ {t\over\sqrt{1-C^2}}\quad{\rm but}\quad \bx\ \to\ \bx,\quad z\ \to\ z.
\ee
Consequently, the static instanton's energy becomes
\begin{align}
E_{\rm net}\ &
=\ N_c\lambda M\ +\ N_c\lambda M^3\times\!\int\!\!d^3\bx\,dz\,I(\bx,z)\times z^2
\nonumber\\
&\qquad+\ {N_c\over 4\lambda M}\times\!\int\!\!d^3\bx\,dz\,d^3\bx'\,dz'\,
{I(\bx,z)\times I(\bx',z')\over (\bx-\bx')^2+(z-z')^2}\
+\ {\rm subleading}\\
&=\ N_c\lambda M\ +\ N_c\lambda M^3\times\left(Z^2+{a^2\over2}\right)\
+\ {N_c\over 5\lambda M a^2}\ +\ {\rm subleading,}\nonumber
\end{align}
hence in equilibrium $Z=0$ and
\be
a\ =\ {\root 4\of{2/5}\over M\sqrt{\lambda}}\,.
\ee

Besides the radius $a$ and $Z$, the instanton has other moduli --- the $X^{1,2,3}$
coordinates of the center (which corresponds to the baryon's coordinates in 3D) and
$4N_f-5$ orientation moduli in the $SU(N_f)$ gauge algebra;
and the net energy is degenerate in  these moduli to all orders in $1/\lambda$.
For finite $N_c$ --- even if it's very large --- one should quantize the motion
of the instanton in those moduli directions.
Consequently, a holographic baryon acquires definite spin and isospin quantum numbers;
for $N_f=2$ the baryons have $I=J=0,1,2,\ldots,{N_c\over2}$ for even $N_c$ or
$I=J={1\over2},{3\over2},{5\over2},\ldots,{N_c\over2}$ for odd $N_c$
\cite{Hata:2007mb,Seki:2008mu},
and there are similar $\rm spin\leftrightarrow flavor$ multiplet relations for $N_f>2$.

However, in multi-baryon systems interactions between the baryons break the
rotational and flavor symmetries of individual baryons, and only the overall
$SO(3)$ and $SU(N_f)$ symmetries remain unbroken.
In holography, the multi-instanton systems suffer from the same problem:
the magnetic fields of multiple instantons interfere with each other, which spoils
the degeneracy of the net energy with respect to orientations moduli of the
individual instantons.
In the large $N_c$ limit, this effect becomes more important than the quantum motion
in the moduli space.

Consequently, in our research program into holographic nuclear matter and related issues,
we stick to classical static instantons with definite classical orientations in space and
in $SU(N_f)$.
From the quantum point of view, such instantons are superpositions of states
with different spins and isospins (or rather $SO(4)$ and $SU(N_f)$ quantum numbers).
We do not even care is the instantons are bosons or fermions --- at our level of analysis
it simply does not matter.
Minimizing the classical energy of a classical multi-instanton system with respect to
classical positions and orientations of all the instantons is already a very hard problem
--- see the next section for the highlights, --- and we do not wish to delve into
the quantization of moduli until we have the classical issues under control.

%
\subsection{Multi--Baryon Systems}
\label{subMBS}
In the large $N_c$ limit, nuclear forces between the baryons are dominated by the static
potentials.
Holographically, a static system of $A$ baryons corresponds to a time-independent
configuration of the non-abelian magnetic flavor fields $F^a_{\mu\nu}(\bx,z)$ ($\mu,\nu=1,2,3,z$)
of net instanton number $A$,
\be
\int\!\!d^3\bx\,dz\,{\epsilon^{\kappa\lambda\mu\nu}\over 16\pi^2}\,\tr(F_{\kappa\lambda}F_{\mu\nu})\
=\ A\ =\ \rm\#baryons,
\label{InstantonNumber}
\ee
accompanied by the Coulomb electric $A_0^a(\bx,z)$ and scalar $\Phi^a(\bx,z)$ potentials induced
by their Chern--Simons and $\Phi FF$ couplings to the magnetic fields \cite{Kaplunovsky:2010eh}.
The whole configuration should minimize the net $\rm DBI+CS$ energy of the system subject to
the constraint~\eqref{InstantonNumber}.

In the $\lambda\to\infty$ limit, the DBI energy of the magnetic fields is $O(\lambda)$
while the net effect of the electric and scalar fields is only $O(1)$.
Moreover, the magnetic fields are concentrated within $O(\lambda^{-1/2})$ distance from
the $X^4=0$ hyperplane, so to the leading order we may approximate the $4+1$ dimensional spacetime
as flat.
Thus,
\be
E_{\rm net}\ \approx\ E_{\rm DBI}\ \approx\ {1\over g^2_{YM}}
\int\!\!d^3\bx\,dz\,\mathop{\rm Str}\nolimits\left[
	\sqrt{\det\bigl(K\delta_{\mu\nu}+F_{\mu\nu}\bigr)}\,
	-\,K\right]
\label{DBIenergy}
\ee
where we neglect the $z$ dependence of the $g^2_{\rm YM}$ and $K=g_{11}(z)/(2\pi\alpha')$.
Similar to the Yang--Mills energy of an $A$ instanton system, this leading-order DBI energy
is minimized by the self-dual configurations of the magnetic fields $F^a_{\mu\nu}(\bx,z)$.
In fact, all such self-dual configurations (of the same instanton number $A$) have the same
leading-order energies
\be
E_{\rm leading\,order}\ =\ A\times\left({8\pi^2\over g_{YM}^2}\,=\,N_c\lambda M\right).
\label{LeadingOrder}
\ee
The self-dual configurations form a continuous family parametrized by $A\times 4N_f$ moduli,
which correspond to the locations, radii, and $SU(N_f)$ orientations of the $A$ instantons.
But the leading-order energy~\eqref{LeadingOrder} does not depend on any of these moduli.

Fortunately, the sub-leading corrections to the net energy lift the degeneracy of the leading
order, which provides for the $O(\lambda^0)$ interactions between the baryons.
To work out such interactions we need the degenerate perturbation theory for the
magnetic field configurations and their energies.
At  first order of this perturbation theory, we
(a) limit the $F^a_{\mu\nu}(\bx,z)$ configurations to the
degenerate minima of the leading-order energy function, $i.\,e.$, to the self-dual magnetic fields;
(b) calculate the $O(\lambda^0)$ corrections $\Delta E$ for the energies of these configurations;
(c) minimize the $\Delta E$ among the self-dual configurations.
At the next order, we would calculate the $O(\lambda^{-1})$ corrections
$\Delta F^a_{\mu\nu}(\bx,z)$ to the magnetic fields --- which would no longer be self-dual ---
and then use such corrections to calculate the net energy to the  order $O(\lambda^{-1})$.
But in this article, we  limit our calculations to the $O(\lambda^0)$ interaction
energies and zeroth-order magnetic fields, so we shall stop after step (c).

In more detail, to get $O(\lambda^0)$ interactions between $A$ baryons, we proceed as follows:
\begin{enumerate}
\item
General self-dual magnetic field configurations obtain via ADHM construction \cite{ADHM}
in terms of $A\times A$ and $A\times N_f$ matrices obeying certain quadratic constraints.
Our first task is to solve these constraints and write down the ADHM matrices
in terms of the instantons' locations, radii, and orientations.
\item
Given the ADHM matrices, we work out the instanton number density profile
\be
I(\bx,z)\ =\ {\epsilon^{\kappa\lambda\mu\nu}\over 16\pi^2}\,\tr(F_{\kappa\lambda}F_{\mu\nu}),
\ee
and for $N_f>2$ also the non-abelian adjoint density
\be
I^a(\bx,x)\ =\ {\epsilon^{\kappa\lambda\mu\nu}\over 16\pi^2}\times d^{abc} F^b_{\kappa\lambda}F^c_{\mu\nu}\,.
\ee
\item
Next, we calculate the $O(\lambda^0\times N_c M)$ corrections to the net energy of the system.
There are two sources of such corrections: the $z$--dependence of the 5D gauge coupling,
and the Coulomb electric and scalar potentials induced by the Chern--Simons
and the $\Phi FF$ couplings to the magnetic fields, thus
\be
\Delta E\ =\ \Delta E_{NA}\ +\ \Delta E_C\,.
\ee
The $z$--dependent 5D gauge coupling changes the DBI energy of the magnetic fields by
\be
\Delta E_{\rm NA}\ =\ N_c M\times\int\!\!d^3x\,dz\,\lambda M^2 z^2\times I(\bx,z),
\ee
while the Coulomb energy depends on the $N_f$.
For $N_f=2$, the  $U(2)$ Chern--Simons and  $\Phi FF$ terms couple the $SU(2)$ magnetic fields
to the $U(1)$ Coulomb fields only.
Consequently, the $A_0^a$ and the $\Phi^a$ fields are abelian and couple to the instanton density
$I(\bx,z)$, and their net energy is simply $4+1$ dimensional Coulomb energy
\be
\Delta E_C\ =\ {N_c M\over4}\times\int\!\!d^3\bx_1\,dx_1\!\int\!\!d^3\bx_2\,dz_2\,
{I(\bx_1,z_1)\times I(\bx_2,z_2)\over (\lambda M^2)\times( (\bx_1-\bx_2)^2+(z_1-z_2)^2)}\,.
\ee
For $N_f>2$, the $U(N_F)$ $U(2)$ Chern--Simons and  $\Phi FF$ terms couple the $SU(N_F)$
magnetic fields to both abelian and non-abelian electric and scalar fields;
the abelian Coulomb fields are sourced by the instanton density $I(\bx,z)$ while the
non-abelian fields are sourced by the adjoint density $I^a(\bx,z)$.
Altogether, the net energy of these Coulomb fields is
\be
\!\Delta E_C\,=\,{N_c M\over4}\times\int\!\!d^3\bx_1\,dz_1\!\!\int\!\!d^3\bx_2\,dz_2\,
{{2\over N_f}I(\bx_1,z_1)\times I(\bx_2,z_2)\,+\,4I^a(\bx_1,z_1)\times I^a(\bx_2,z_2)
	\over (\lambda M^2)\times( (\bx_1-\bx_2)^2+(z_1-z_2)^2)}\,.
\ee

\item
Steps 1, 2, and 3, give us $\Delta E$ as a function of baryons' locations, radii, and $SU(N_f)$
orientations.
Now we need to minimize the $\Delta E$ with respect to all these moduli.
\end{enumerate}

This 4-step procedure is fairly straightforward for a few baryons --- {\it cf.}\
calculations of the 2--body nuclear forces by Kim, Sin  and Zahed \cite{Kim:2006gp}
and by Hashimoto {\it et~al} \cite{SakaiSugimoto2inst}, --- but
it becomes prohibitively difficult for large numbers of baryons and outright impossible
for infinite baryon crystals.
At best, we can survey a small subspace of the $A$-instanton moduli space and try to
minimize the $\Delta E$ over that subspace.
For example, we may assume that all the instantons have the same radius, that their centers
form a periodic lattice of some particular symmetry,
and that the orientations of the instantons also form some kind of a periodic pattern;
this gives us an ansatz for all the $A\times 4N_f$ moduli in terms of just a few overall parameters,
and we can try to calculate and minimize the $\Delta E$ as a function of these parameters.
However, any such ansatz is likely to miss the true lowest-energy configuration of the system.
Indeed, in condensed matter guessing the crystalline symmetry of some substance from
the properties of the individual atoms is a game of chance with poor odds, and there is no reason
why the instanton crystals should be any simpler.
Moreover, even if we could somehow guess  all the symmetries of the instanton crystal,
actually working through steps 1--4 is impossible without additional approximations (besides $\lambda\gg1$).
For example, in our previous paper \cite{Kaplunovsky:2012gb} we needed the
$\rm (lattice\ spacing)\gg(instanton\ radius)$ approximation just to solve the ADHM constraints
for the zigzag-shaped chain of instantons, and even for a straight chain we needed
the $\rm spacing\gg radius$ approximation to calculate the Coulomb energy $\Delta E_C$.

In this article we explore a shortcut around steps 1, 2, and 3.
In section~3 we shall see that when the distances between the instantons are much larger
than their radii, the interactions between the baryons are dominated by the two-body forces,
\be
\E_{\rm interactions}^{\rm net}\ \approx\ {1\over2}\sum_{\textstyle{m,n=1,\ldots,A\atop m\neq n}}
F_2\bigl(X_m^\mu-X_n^\mu,{\rm orientation}_m,{\rm orientation}_n\bigr)
\ee
for a manageably simple function  $F_2$ of the two instanton's positions and orientations.
Consequently, we can minimize the net interaction energy over the entire moduli space of
the multi-instanton system using a simple numerical simulation:
Starting from a random set of instanton positions and orientations, use the steepest descent
algorithm to find the nearest local minimum of the net energy; repeat this procedure for  different
random starting points to find other local minima; eventually, we find all the local minima,
compare their energies, and identify the global minimum.
In our next paper \cite{OurNextPaper} we shall use this algorithm to construct the
2D and 3D instanton lattices starting from scratch, $i.\,e.$, from the $F_2$ function.
In the present article, we focus on  the 1D lattice (\S4) and on the first step in the transition
form 1D to 2D --- the zigzag (\S5).
In both cases, we assume the lattice geometry for the instanton positions, but we use the
numeric simulation to find the lowest-energy pattern of their orientations.
Once we know all the patterns that show up, we use them as ansatz's (with a few parameters)
for which we calculate the net energy as  analytic functions of the parameters.
In this way, we map the boundaries between different orientation patterns much more accurately
than we could do it in the numerical simulation.

To make the baryons form a 1D lattice instead of spreading out in  three dimensions,
we curve the $x_2$ and $x_3$ dimensions of the flavor brane similar to the curvature of the $z$
coordinate.
In terms of the effective $4+1$ dimensional theory, this corresponds to the 5D flavor gauge coupling
depending on the $x_2$ and the $x_3$ as well as the $x_4\equiv z$,
\be
{8\pi^2\over g_5^2(x)}\ =\ N_c\lambda M\times\Bigl(
	1\,+\,M_2^2x_2^2\,+\,M_3^2x_3^2\,+\,M_4^2x_4^2\,+\,O(M^4x^4)\Bigr).
\tag{\ref{Ipot}}
\ee
This gauge coupling acts as a harmonic potential for the instantons which pulls them towards
the $x_1$ axis, so at low densities the instantons form a 1D lattice along the $x_1$.
At higher densities, the instantons push each other away from the $x_1$ axis and form
more complicated 2D or 3D lattices, starting with the zigzag
\be
\begin{pspicture}(-7,-1.2)(+7.4,+1.6)
\psline[linewidth=0.5pt]{>->}(-7,0)(+7,0)
\uput[r](7,0){$x^1$}
\psline[linewidth=0.5pt]{>->}(0,-1.2)(0,+1.2)
\uput[u](0,1.2){$x^2$}
\psline[linestyle=dotted](-6.5,+0.4)(+6.5,+0.4)
\psline[linestyle=dotted](-6.5,-0.4)(+6.5,-0.4)
\multido{\n=-6+2}{7}{\pscircle*[linecolor=blue](\n,+0.4){0.1}}
\multido{\n=-5+2}{6}{\rput(\n,0){%
	\psline[linecolor=magenta,linewidth=0.5pt](-1,+0.4)(0,-0.4)(+1,+0.4)
	\pscircle*[linecolor=blue](0,-0.4){0.1}
	}}
\end{pspicture}
\nonumber
\ee
To make sure the transition from the straight chain to the zigzag happens for lattice
spacings much larger than the instanton radius (which is required by the two-body force approximation),
we assume $M_2\ll M_3,M_4$.

This setting is similar to what we have used in our previous paper \cite{Kaplunovsky:2012gb},
except for a change of notations\footnote{%
	Back in \cite{Kaplunovsky:2012gb}, we had
	$$
	{8\pi^2\over g_5^2(x)}\ =\ N_c\lambda M\times\Bigl(
	1\,+\,M^{\prime2}\bigl(x_1^2+x_2^2\bigr)\,+\,M^2x_3^2\,+\,O(M^4x^4)\Bigr),\qquad
	M'\gg M,
    $$
    so the zigzag was in the $(x_3,x_4)$ plane, with the long axis along the $x_4$.
    Translating between the old notations and those of this paper, we have
    $(x_1,x_2,x_3,x_4)^{\rm old}=(x_4,x_3,x_2,x_1)^{\rm new}$,
    $M^{\rm old}=M_2^{\rm new}$, $M^{\prime\,\rm old}=M_3^{\rm new}=M_4^{\rm new}$.
    }
and allowing for $M_3\neq M_4$.
In \S5 we shall see that $M_3\neq M_4$ makes a big difference for the instanton orientation
patterns in a zigzag:  the lowest-energy  patterns are abelian for $M_3\approx M_4$
but non-abelian for anisotropic $M_3<0.94 M_4$.
Even the order of the  phase transition from the straight
chain to the zigzag depends on the $M_3/M_4$ ratio: second-order for $\rm ratios>0.725$
but first-order for $\rm ratios<0.725$.

%
\section{Two-Baryon versus  Multi-Baryon Interactions}
\label{TBvsMB}
In real-life nuclear physics, besides the two-body nuclear forces due to
meson exchanges, there are significant three-body forces, and presumably
also four-body forces, \etc,
\be
\hat H_{\rm nucleus}\
=\,\sum_{n=1}^A\hat H^{\rm 1\,body}(n)\
+\,\tfrac12\sum_{\textstyle{{\rm different}\atop m,n=1,\ldots A }}\hat H^{\rm 2\,body}(m,n)\
+\,\tfrac16\sum_{\textstyle{{\rm different}\atop \ell,m,n=1,\ldots A }}
	\hat H^{\rm 3\,body}(\ell,m,n)\
+\ \cdots
\ee
where $n$ stands for  the quantum numbers of the $n^{\rm th}$ nucleon.
Likewise, in the holographic nuclear physics interactions between multiple baryons
include two-body forces and also three-body, four-body, \etc, forces.
Even in the classical infinite-mass limit where the holographic baryons
become static instantons of the $SU(N_f)$ gauge fields in $4+1$ dimensions,
the potential energy (due to $g_{\rm 5D}\neq\rm const$ and due to Chern-Simons interactions)
of an $A$--instanton system has form
\begin{align}
\E^{\rm net}(1,2,\ldots,A)\ &
\equiv\ E^{\rm total}\ -\ A\times N_c\lambda M\\
&=\,\sum_{n=1}^A \E^{(1)}(n)\
+\,\half\!\!\!\sum_{\textstyle{{\rm different}\atop m,n=1,\ldots A }}\E^{(2)}(m,n)\
+\,\tfrac{1}{6}\!\!\!\sum_{\textstyle{{\rm different}\atop \ell,m,n=1,\ldots A }}
	\E^{(3)}(\ell,m,n)\
+\ \cdots\nonumber
\end{align}
with significant three-body, \etc, terms.

What about the relative magnitudes of the 2-body, 3-body, 4-body, \etc, interaction terms?
When the baryons are packed cheek-by-jowl till their instanton cores overlap and merge,
we expect all the $n$-body forces to have comparable strengths.
{\it But in the opposite low-density regime of baryons separated by distances much larger
then their radii,
the two-body forces dominate the interactions,
while the multi-body forces are smaller by powers of $\it(radius/distance)^2$.}
The purpose of this section is to prove this fact.

Before we delve into the technical issues, let us briefly summarize our argument.
First,  consider  the simplest case of $N_f=2$ and intermediate-range distances between
the baryons,
\be
\left(a\sim{1\over M\sqrt\lambda}\right)\ \ll\ |X_m-X_n|\ \ll\ {1\over M}\,,
\ee
which allows us to treat the 5D holographic space as approximately flat.
For $N_f=2$ the holographic baryons are instantons of the $SU(2)$ magnetic fields,
which source the $U(1)$ electric and scalar fields via CS and
$\tr(\Phi F_{\mu\nu}F^{\mu\nu}$) couplings.
When those instantons are small and separated from each other by large distances
$|X_m-X_n|\sim D\gg a$, their interactions  come from two sources:

\noindent {\bf 1.}
Direct Coulomb repulsion ($\rm electric-scalar$)
between nearly-point-like abelian charges in $4+1$ dimensions,
\be
\E^{\rm direct}\ =\ {N_c\over 4\lambda M}\sum_{m\neq n}{1\over |X_m^\mu-X_n^\mu|^2}\,,
\label{Edirect}
\ee
which is a manifestly two-body interaction.

\newpage
\noindent {\bf 2.}
The interference between the instantons changes the distribution of the instanton
number density in space,
\begin{align}
\label{DensityDef}
I(x)\ &
=\ \frac{\epsilon^{\mu\nu\rho\sigma}}{32\pi^2}\,
\tr\bigl(F_{\mu\nu}(x)F_{\rho\sigma}(x)\bigr),\\[5pt]
I^{A\,\rm instantons}(x)\ &
=\,\sum_{i=n}^A I^{\rm standalone}_n(x)\
+\,\Delta I^{\rm interference}(x),
\end{align}
which in turn changes the self-interaction energy (Coulomb and non-abelian) of each
instanton by an amount comparable to~(\ref{Edirect}).
Within an instanton --- $i.\,e.$, at $O(a)$ distance from some instanton's center $X_n^\mu$
--- the interference from the other instantons should be relatively weak,
\be
\Delta I^{\rm interference}(x)\ \sim\ {a^2\over D^2}\times I_n^{\rm standalone}(x),
\ee
so there should be some kind of a perturbation theory for it.
At the first order of such perturbation theory, the $\Delta I$ arises from interference
between the un-perturbed standalone-like instantons, so we expect it to be a sum
of pair-wise interferences from the other instantons,
\be
\Delta I^{\rm 1st\,order}(x\,{\rm near}\,X_n)\
=\ I_n^{\rm standalone}(x)\times\sum_{m\neq n}{\cal F}^{(1)}_{n,m}(x)
\label{DeltaIFirstOrder}
\ee
where
\be
{\cal F}^{(1)}_{n,m}(x)\ \sim\ O\left({a^2\over|X_n-X_m|^2}\right)
\ee
and depends only on the instantons $\#n$ and $\#m$ --- $i.\,e.$, on their positions,
radii, and orientations --- but not on any other instantons.
At the second-order, we expect to include the interference between the first-order
$\Delta I$ and the additional instantons, so at this order we obtain 3-body effects,
\be
\Delta I^{\rm 2nd\,order}(x\,{\rm near}\,X_n)\
=\ I_i^{\rm standalone}(x)\times\sum_{\ell,m\neq n}{\cal F}^{(2)}_{n,\ell,m}(x)
\ee
but
\be
{\cal F}^{(2)}\ \sim\ {a^4\over D^4}\ \ll\ {\cal F}^{(1)}.
\label{DeltaLast}
\ee
Likewise, the higher orders may involve more and more instantons, but the magnitudes
of such high-order interference effects are suppressed by the higher powers of $(a^2/D^2)$.

Now consider the Coulomb self-interaction of the instanton $\#n$,
\be
\E_C^{\rm self}(n)\ =\ {N_c\over 4\lambda M}\times{1\over\rho_n^2}
\ee
where $\rho_n$ is the instanton's effective charge radius,
\be
{1\over\rho_n^2}\ =\! \mathop{\intop\!\!\!\!\intop}\limits_{x_1,x_2\,{\rm near}\,X_n}\!\!
d^4x_1\,d^4x_2\,{I(x_1)I(x_2)\over|x_1-x_2|^2}\,.
\ee
A standalone instanton has
\be
{1\over\rho_n^2}\ =\ {4/5\over a_n^2}
\ee
but interference from the other instantons should change this radius by small amount
of similar relative magnitude to $\Delta I/I_n^{\rm standalone}$, thus
\be
{1\over\rho_n^2}\ \to\ {4/5\over a_n^2}\ +\ \Delta_n\,,\quad
\Delta_n\ \sim\ {1\over D^2}\,,
\ee
which changes the instanton's Coulomb self-interaction energy by
\be
\Delta\E_C^{\rm self}(n)\ =\ {N_c\over 4\lambda M}\times\Delta_n\
\sim\ {N_c\over\lambda M D^2}\,.
\ee
Note that this effect has a similar magnitude to the direct Coulomb repulsion~(\ref{Edirect})
between the instances.

Moreover, the charge radius correction $\Delta_n$ is linear in the $\Delta I^{\rm interference}$
at $x$ near the $X_n$, hence in light of eq.~(\ref{DeltaIFirstOrder}), the leading-order
contribution to the $\Delta_n$ is a sum of pair-wise interferences from the other instantons,
thus
\be
\Delta_n\ =\,\sum_{m\neq n}\Delta^{(1)}_{n,m}\ +\ O(a^2/D^4)
\ee
where each $\Delta_{n,m}^{(1)}$ depends only on the instantons $\#n$ and $\#m$.
Consequently, the leading effect of the interference on the net Coulomb self-energy
of all the instantons has form
\be
\label{ECtwobody}
\begin{split}
\Delta^{\rm interference}\E_C^{\rm self}\ &
\equiv\ \E_C^{\rm self}[{\rm interfering}]\ -\ \E_C^{\rm self}[{\rm standalone}]\\
&=\ {N_c\over4\lambda M}\sum_{\textstyle{n,m=1,\ldots,A\atop n\neq m}}\Delta_{n,m}^{(1)}\
+\ O\left({N_c\over\lambda M}\times{a^2\over D^4}\right)
\end{split}
\ee
where the leading terms act as two-body interactions between the instantons.

Besides the Coulomb energy, the non-abelian energy
is also affected by the interference between the instantons,
\be
\Delta^{\rm interference}\E_{NA}\
=\ N_c\lambda M^3\times\int\!\!d^4 x\,(x^4)^2\times  \Delta I^{\rm interference}(x),
\label{DENA}
\ee
but this time the integral should be taken over the whole 4D space, including
both the instantons and the inter-instanton space.
Indeed, we shall see later in this section that
\be
\label{DImagnitude}
\begin{split}
\text{at $O(a)$ distance from an $X_n$,}\quad &
\Delta I\ \sim\ {1\over a^2 D^2}\,,\\
\text{at $O(D)$ distances from {\it all} the $X_m$,}\quad &
\Delta I\ \sim\ {a^4\over D^8}\,,
\end{split}
\ee
and both  kinds of places make $O(a^4/D^2)$ contributions to the integral~(\ref{DENA}).
Moreover, in both kinds of places, the leading terms in the $\Delta I(x)$ is a sum
of  independent two-instanton interference terms,
\be
\Delta I^{\rm interference}(x)\
=\ \half\sum_{n\neq m}{\cal I}^{(2)}_{n,m}(x)\ +\ \rm subleading.
\label{Idecomp}
\ee
Although our heuristic argument 
for such decomposition near instanton centers does not work in the inter-instanton space,
we shall prove later in this section that the decomposition~(\ref{Idecomp}) works
everywhere in the 4D space.
Therefore, the non-abelian interactions between the instantons due to interference
are also dominated by the two-body terms
\be
\Delta^{\rm interference}\E_{NA}\
=\ \half\sum_{n\neq m} N_c\lambda M^3\!\int\!\!d^4x (x^4)^2\times {\cal I}_{n,m}(x)\
+\ \rm subleading.
\label{ENAtwobody}
\ee

Thus far we have assumed $N_f=2$.
For larger numbers of flavors, interference between magnetic fields of the $SU(N_f)$
instantons is harder to describe mathematically, but its qualitative features
for $a\ll D$ remain exactly as in eqs.~(\ref{DensityDef}) through~(\ref{Idecomp}).
In particular, the effect of interference between small instantons separated by
large distances on the density profile $I(x)$ is dominated by the two-instanton
interference terms as in
\def\secref#1.#2.#3;{#2}%
\edef\jjj{\ref{DeltaLast}}%
\edef\qqq{\expandafter\secref\jjj..;}%
eqs.~(\ref{Idecomp}) and (\ref{DeltaIFirstOrder}--\qqq).
Consequently, the instantons' non-abelian and  Coulomb interaction energies suffer
small corrections from the interference that give rise to predominantly-two-body
interactions between the instantons as in eqs.~(\ref{ECtwobody}) and (\ref{ENAtwobody}).

\par\smallskip\goodbreak
The qualitatively new aspect of $N_f\ge3$ comes not from the $SU(N_f)$ self-dual magnetic fields
of the multiple instantons but from the electric and scalar Coulomb fields
induced by the magnetic fields via Chern--Simons and $\tr(\Phi F^{|mu\nu}F_{\mu\nu})$
interactions.
For $N_f=2$ such electric and scalar fields are purely abelian, $A^0,\Phi\in U(1)\subset U(2)$,
but for $N_f\ge3$ they have both $U(1)$ and $SU(N_f)$ components.
For a single instanton, all fields --- magnetic, electric, and scalar ---
belong to the same Cartesian $U(2)$ subgroup of the $U(N_f)$ ---
in $N_f\times N_f$ matrix language,
\be
{\bf for}\ A^\mu_{\rm magnetic}\,\in\,\begin{pmatrix}
	SU(2) &\omit\vrule  height 16pt depth 6pt & \quad\hbox{\Large\bf 0}\quad\\
	\noalign{\hrule}
	\hbox{\Large\bf 0} & \omit\vrule  height 20pt depth 6pt & \hbox{\Large\bf 0}
	\end{pmatrix},\qquad
A^0,\Phi\,\propto\,\begin{pmatrix}
	&\half &&&& \lower 5pt\hbox{\huge\bf 0}\\
	&& \half &&&&\\
	&&& 0 &&&\\
	&&&& 0 &&\\
	&\hbox{\huge\bf 0} &&&& 0
	\end{pmatrix}
\ee
--- so the Coulomb {\it self-interaction} of a standalone instanton works
exactly as for $N_f=2$.
On the other hand, Coulomb interactions {\it between} multiple instantons whose magnetic fields
span different $SU(2)$ subgroups of the $SU(N_f)$ become much more complicated.

Moreover, in a general multi-instanton background, the Coulomb fields $A^0(x)$
and $\Phi(x)$ do not behave as simply $1/r^2$.
Instead, they satisfy gauge-covariant equations for the matrix-valued fields,
\be
-\Bigl(\partial_\mu+i[A_\mu(x),\ ]\Bigr)^2 A^0(x)\
=\ {g^2 N_c\over 32\pi^2}\,\epsilon_{\alpha\beta\mu\nu}F^{\alpha\beta}F^{\mu\nu}
\ee
and ditto for the scalar  $\Phi(x)$.
Consequently, the Coulomb force between two sources (such as instantons)
depends on the magnetic fields in the space between them.
In other words, for $N_f\ge3$ the direct Coulomb interactions are no longer two-body;
instead, the force between two instantons depends on the other instantons in the system.

Fortunately, for small instantons separated by large distances,
the magnetic fields are concentrated in small volumes of 4--space, so their
effect on the Coulomb field propagation becomes perturbatively weak:
\be
\begin{split}
\Bigl\langle x^\mu,b\Bigm|{-1\over\rm covariant\,\bfsquare}\Bigm|y^\mu,c\Bigr\rangle\ &
=\ {\delta^{bc}\over 4\pi^2|x^\mu-y^\mu|^2}\
+\,\sum_{n=1}^A {O(a^2_n)\,F(b,c,{\rm orientation}_n)
	\over |x^\mu-X_n^\mu|^2\times|y^\mu-X_n^\mu|^2}\\
&\qquad+\ O(a^4/D^6)~\text{two-instanton terms}\ +\ \cdots
\end{split}
\ee
where $b$ and $c$ are adjoint indices of the $U(N_f)$.
The leading term here is the ordinary Coulomb propagator in 4D space,
thus {\it in the $a\ll D$ limit, the direct Coulomb interactions between
small instantons are predominantly two-body,}
\be
\E^{\rm direct}\
\approx\ {N_c\over4\lambda M}\sum_{n\neq m}{\mathop{\rm overlap}(n,m)\over|X^\mu_n-X^\mu_m|^2}
\label{Edirect3plus}
\ee
where overlap$(n,m)\ge0$ is the overlap between the $SU(2)$'s spanned by the magnetic
fields of the instantons $\#n$ and $\#m$.

The bottom line is, {\it for all $N_f\ge2$, all the interactions between small instantons
separated by large distances are predominantly two-body.}
However, for $N_f\ge3$ the dependence of such two-body interactions on the instantons'
orientations becomes much more complicated than for $N_f=2$.

Finally, let us relax our second un-necessary assumption of intermediate-range
distances between the baryons and consider what happens at longer distances $D\sim1/M$.
In this regime, the curvature of the fourth space dimension --- and especially
the $x^4$--dependence of the 5D gauge coupling --- can no longer be treated as
a perturbation.
Consequently, the magnetic fields of an instanton or a multi-instanton system
are no longer self dual in the inter-instanton space --- although inside
the instanton cores or at intermediate-range distances $r\ll 1/M$ from instantons
they remain approximately self-dual.
Therefore, the interference between very distant instantons is no longer governed by the
self-dual ADHM solutions.
Instead, we must work it out the hard way: figure out how the magnetic fields of an
instanton propagate through the curved $4+1$ dimensions towards the other instantons,
and then find out how such fields disturb the other instantons' cores.

Fortunately, we do not need a hard calculation to see that at large distances
from an instanton its magnetic fields are very weak.
Indeed, even in flat 5D space the fields weaken with distance as $A^a_\mu\sim a^2/r^3$
(in the IR-safe singular gauge),
so at $r\gg a$ they are so weak that the field equations become effectively linear.
In the curved space, we may decompose the weak 5D gauge fields into 4D mesonic fields,
hence at large distances $r\gtrsim 1/M$ from an instanton, its fields become Yukawa-like
\be
A^a_\mu(r,Z)\ \sim\ {a^2\over r^3}\times
\sum_k \Psi_k(z)\Psi_k(z_{\rm inst}) e^{-m_k r}\
\ll\ {a^2\over r^3}\,.
\label{WeakFields}
\ee
Consequently, at the location $X_n^\mu$ of any particular instanton, the background
fields from the {\it other} instantons are very weak, and their effect on the
instanton $\#n$ itself can be adequately accounted by the first-order perturbation
theory.
In other words, the effects of other instantons $\#m\neq n$ on the instanton $\#n$
are weak and add up linearly!
For the self-interaction energy of the $n^{\rm th}$ instanton, this means
\be
\E(n)\ \approx\ \E({\rm standalone})\ +\,\sum_{m\neq n}\Delta_m\E(n)
\ee
where the second term gives rise to two-body interaction energies
\be
\E^{(2)}_{\rm interference}(n,m)\ =\ \Delta_m\E(n)\ +\ \Delta_n\E(m).
\ee
We expect the two-body terms to be rather small --- in addition to the
usual 5D $1/|X_m-X_n|^2$ factors they should carry Yukawa-like exponentials
$e^{-mr}$ (or rather sums of such exponentials), but these are the leading
interactions due to interference.
To  three-body or multi-body interactions follow from higher-order perturbation
by very weak fields~(\ref{WeakFields}), so they are much smaller that the
two-body interactions.

As to the direct Coulomb interactions between the holographic baryons via
electric or scalar fields, at large distances $|X_m-X_n|\sim1/M$ they also
decompose into sums of Yukawa forces.
Moreover, since the scalar mesons generally have different masses from the vector mesons,
the attractive potential due to scalars may have a different $r$ dependence
from the repulsive potential due to vectors.
Thus, for a right model, the net two-body force between two holographic baryons
may become attractive at large enough distances between them.
But regardless of the model, the direct Coulomb interactions are always
manifestly two-body for $N_f=2$, while for $N_f\ge3$ the multi-body terms
exist but become very small at large distances between the baryons.
Thus,
\be
\E_{\rm direct}\ =\ \half\sum_{m\neq n} \mathop{\rm overlap}(m,n)\times
V^{(2)}(|X_m-X_n|)
\ee
where the precise form of the potential $V^{(2)}(r)$
is model-dependent, but the two-body form of the direct interactions
is quite universal.

\medskip\centerline{$\star\quad\star\quad\star$}\medskip

At this point we have finished our heuristic arguments and about to start
a more technical analysis of the interactions between the holographic baryons.
To keep our presentation relatively simple, we assume just two quark flavors
and intermediate-range distances between
the holographic baryons --- $D\gg a$ but $D\ll 1/M$.

For $N_f=2$ the non-abelian gauge symmetry is $SU(2)$, so the ADHM data for an
$A$-instanton configuration consists of a quaternionic array $Y_n$ ($n=1,\ldots,A)$ and
a symmetric quaternionic $A\times A$ matrix $\Gamma_{mn}$.
Equivalently, we may use $A$ real instanton radii $a_n=|Y_n|$, $A$ $SU(2)$ matrices
$y_n$ (equivalent to unimodular quaternions $Y_n/a_n$) parametrizing the instanton's
orientations, and 4 real symmetric matrices $\Gamma^\mu_{mn}$ whose diagonal
elements $\Gamma_{nn}^\mu=X_n^\mu$ are the locations of the instanton's centers.
Finally, the off-diagonal matrix elements $\Gamma^\mu_{m\neq n}=\alpha^\mu_{mn}$
follow from the other parameters by solving the ADHM equations
\be
\Im\left( \bigl(\Gamma^\dagger\Gamma\bigr)_{mn}\,+\,Y_m^\dagger Y_n^{}\right)\ =\ 0
\ee
or equivalently
\be
\eta^a_{\mu\nu}\bigl[\Gamma^\mu,\Gamma^\nu\bigr]_{mn}\
+\ a_ma_n\times\tr\bigl(y_m^\dagger y_n^{}(-i\tau^a)\bigr) =\ 0
\label{ADHM}
\ee
where  $\eta^a_{\mu\nu}$
is the 't~Hooft's tensor mapping between the $SU(2)_{\rm gauge}$ and the $SU(2)_L$
inside ${\rm Spin}(4)=SU(2)_L\times SU(2)_R$,
\be
\eta_{44}^a=0,\quad
\eta^a_{4i}=-\delta^a_i\,,\quad
\eta^a_{i4}=+\delta^a_i\,,\quad
\eta^a_{ij}=\epsilon^{aij},\quad
a,i,j=1,2,3.
\ee

The ADHM data are somewhat redundant --- an $O(A)$ symmetry acting on all the
$\Gamma^\mu_{mn}$ and $Y_n=a_ny_n$ does not change any physical properties of
the multi-instanton data.
This symmetry includes $\Z_2^A$ which flips the instanton orientations
$y_n\to-y_n$ (independently for each $n$).
It also includes small $SO(A)$ rotations that change the off-diagonal elements by
$\delta\alpha^\mu_{mn}=\epsilon_{mn}(X_m-X_n)^\mu+O(\epsilon^2)$.
To eliminate these rotations,  the ADHM equations~(\ref{ADHM}) for the off-diagonal elements
should be combined with additional constrains (one for each $m\neq n$),
for example
\be
\forall m\neq n:\
(X_m-X_n)^\mu \alpha^\mu_{mn}\ =\ 0.
\label{ADHMsols}
\ee

For large distances $|X_m-X_n|\sim D$ between the instantons, $D\gg a_n$, we may
solve the ADHM equations (and the constraints~(\ref{ADHMsols})
as a power series in $a^2/D^2$:
\begin{subequations}
\label{AlphaExpansion}
\begin{align}
\alpha^\mu_{mn}\ &
\equiv\ \Gamma^\mu_{m\neq n}\
=\ \alpha^{(1)}_{\mu mn}\ +\ \alpha^{(2)}_{\mu mn}\ +\ \alpha^{(3)}_{\mu mn}\
+\ \cdots,\\
\alpha^{(1)}_{\mu mn}\ &
=\ {\eta^a_{\mu\nu}\,(X_m-X_n)_\nu\over|X_m-X_n|^2}\times
	\half a_ma_n\,\tr\bigl(y_m^\dagger y_n^{}(-i\tau^a)\bigr)\
=\ O(a^2/D),\\
\alpha^{(2)}_{\mu mn}\ &
=\ -{\eta^a_{\mu\nu}\,(X_m-X_n)_\nu\over|X_m-X_n|^2}\times
	\sum_{\ell\neq m,n}\eta^a_{\kappa\lambda}
		\alpha^{(1)}_{\kappa \ell m}\alpha^{(1)}_{\lambda \ell n}\
=\ O(a^4/D^3),\\
\alpha^{(3)}_{\mu mn}\ &
=\ -2{\eta^a_{\mu\nu}\,(X_m-X_n)_\nu\over|X_m-X_n|^2}\times
	\sum_{\ell\neq m,n}\eta^a_{\kappa\lambda}
		\alpha^{(1)}_{\kappa \ell m}\alpha^{(2)}_{\lambda \ell n}\
=\ O(a^6/D^5),\\
\omit\dotfill &\omit\dotfill \nonumber
\end{align}
\end{subequations}
Note that {\it for each off-diagonal matrix element,
the leading term $\alpha^{(1)}_{\mu mn}$ in this expansion depends
only on the instantons $\#m$ and $\#n$}
($i.\,e.\,$, on the positions, radii, and orientations of only these two instantons)
while the subleading terms $\alpha^{(2)}_{\mu mn},\alpha^{(3)}_{\mu mn},\ldots$
involve additional instantons.

Given the ADHM data $\Gamma^\mu_{mn}$, $a_n$, and $y_n$ of a multi-instanton system, the
4D instanton number density (\ref{DensityDef})
can be obtained as
\be
I(x)\ =\ -\frac{1}{16\pi^2}\,\square\square\,\log\det\bigl(L(x)\bigr)
\label{Idensity}
\ee
where $L(x)$ is a real $A\times A$  symmetric matrix
\be
L_{mn}(x)\
=\,\sum_\ell\bigl(\Gamma^\mu_{\ell m}-x^\mu\delta_{\ell m}\bigr)
	\bigl(\Gamma^\mu_{\ell n}-x^\mu\delta_{\ell n}\bigr)\
+\ \half a_m a_n\tr\bigl(y_m^\dagger y_n^{}\bigr).
\label{Ldef}
\ee
Thanks to the double D'Alambertian in eq.~(\ref{Idensity}), several moments
of the instanton density may be obtained via integrating by parts:
\begin{subequations}
\label{Moments}
\begin{align}
\int\!\!d^4 x\,I(x)\ &
=\ A,\\
\int\!\!d^4 x\,I(x)\times x^\nu\ &
=\ \tr\bigl(\Gamma^\nu\bigr),\\
\int\!\!d^4 x\,I(x)\times x^\mu x^\nu\ &
=\ \tr\bigl(\Gamma^\mu\Gamma^\nu\bigr)\ +\ \half\delta^{\mu\nu}\tr\bigl(T\bigr)\\
\text{where}\ T_{mn}\ &
\equiv\ \half a_m a_n\,\tr\bigl(y_m^\dagger y_n\bigr),\\
\int\!\!d^4 x\,I(x)\times x^\lambda x^\mu x^\nu\ &
=\ \half\bigl(\Gamma^\lambda\{\Gamma^\mu,\Gamma^\nu\}\bigr)\
+\ \half\delta^{\lambda\mu}\tr\bigl(\Gamma^\nu T\bigr)\\
&\qquad+\ \half\delta^{\lambda\nu}\tr\bigl(\Gamma^\mu T\bigr)\
+\ \half\delta^{\mu\nu}\tr\bigl(\Gamma^\lambda T\bigr).\nonumber
\end{align}
\end{subequations}
The quadratic moment~(\ref{Moments}c) for $\mu=\nu=4$ is
particularly important since it gives us an exact
formula for the non-abelian energy of the system directly in terms of the
ADHM data,
\begin{align}
\E_{NA}\ &
=\ N_c\lambda M^3\times\!\int\!\!d^4 x\,I(x)\times \bigl(x^4\bigr)^2\ 
=\ N_c\lambda M^3\times\Bigl(\tr\bigl(\Gamma^4\Gamma^4\bigr)\,+\,\half\tr\bigl(T\bigr)\Bigr)
\nonumber\qquad\\
&=\ N_c\lambda M^3\times\left(
	\sum_{i=n}^A\Bigl(\bigl(\Gamma^4_{nn}\bigr)^2\,+\,\half T_{nn}\Bigr)\
	+\,\sum_{m\neq n}\bigl(\Gamma^4_{mn}\bigr)^2
	\right)
\label{ENAprelim}\\
&=\ N_c\lambda M^3\sum_n\Bigl(\bigl(X^4_n\bigr)^2\,+\,\half a_n^2\Bigr)\
	+\ N_c\lambda M^3\sum_{m\neq n}\bigl(\alpha^4_{mn}\bigr)^2.\nonumber
\end{align}
Obviously, the first sum on the last line here is the sum of individual instantons'
potential energies due to their radii and locations (relative to the $x^4=0$ hyperplane)
while the second sum comprises the interactions between the instantons,
\begin{align}
\E_{NA}^{\rm net\,interaction}\ &
=\ N_c\lambda M^3\sum_{m\neq n}\bigl(\alpha^4_{mn}\bigr)^2.
\label{ENAgen}
\end{align}
This formula is general and applies to any multi-instanton configuration, the
problem is calculating the off-diagonal matrix elements $\alpha^\mu_{mn}=\Gamma^\mu_{m\neq n}$.
Fortunately, for the small instantons
separated by large distances $O(D)\gg a$,
those off-diagonal matrix elements are given by the expansion (\ref{AlphaExpansion}.a--d)
in powers of $(a^2/D^2)$.
Moreover, for each $m\neq n$ the leading term $\alpha^{(1)}_{\mu mn}$ in this expansion
depends only on the positions and orientations of the instantons $\#m$ and $\#n$
but does not depend on any other instantons.
Thus, {\it to the leading order in $a^2/D^2$, the non-abelian interactions~(\ref{ENAgen})
between the instantons are two-body,} $i.\,e.$ involve only two instantons at a time:
\be
\E_{NA}^{\rm interaction}\
\approx\ \half\sum_{m\neq n} \E_{NA}^{\rm 2\,body}(m,n)\
\sim\ N_c\lambda M^3\,{a^4\over D^2}
\ee
while the multi-body interactions are $O(N_c\lambda M^3 a^6/D^4)$ or weaker.
Specifically,
\begin{align}
\E_{NA}^{\rm 2\,body}(m,n)\ &
=\ 2N_c\lambda M^3\times\left(
	a_ma_n\times{\eta^a_{4\nu}(X_m-X_n)_\nu\over|X_m-X_n|^2_{4D}}
	\times\half\tr\bigl(y_m^\dagger y_n^{}(-i\tau^a)\bigr)\right)^2\nonumber\\
&=\ {N_c\lambda M^3 a_m^2 a_n^2\over2|X_m-X_n|^2_{4D}}\times
	\tr^2\Bigl( y_m^\dagger y_n\,(-i\vec N_{mn}\cdot\vec\tau)_{3D}\Bigr)
\label{ENA2body}
\end{align}
where $\vec N_{mn}$ is the 3-vector part of the unit 4-vector
\be
N^\mu_{mn}\ \equiv\ \bigl(\vec N_{mn},N^4_{mn}\bigr)\
=\ {X_n^\mu-X_m^\mu\over|X_n-X_m|}\,.
\label{Ndef}
\ee

Now consider the Coulomb energy
\be
\label{ECintegral}
\E_C\ =\ {N_c\over4\lambda M}\int\!\!\!\!\int\!d^4x_1\,d^4x_2\,
{I(x_1)I(x_2)\over|x_1-x_2|^2}\,.
\ee
of the multi-instanton system.
Alas, we cannot obtain this Coulomb energy directly from the ADHM data via
integration by parts or some other clever trick but have to calculate it the hard way:
Derive the instanton density profile  from eq.~(\ref{Idensity}), decompose $I(x)$
into individual instantons plus interference terms, and finally perform the
integral~(\ref{ECintegral}) for some approximation to the integrand.

Our starting point is the $L_{mn}(x)$ matrix~(\ref{Ldef}) which governs the instanton
density profile.
For a system of small instantons separated by large distances, the diagonal elements of
this matrix are much larger than the off-diagonal elements.
Indeed, the diagonal matrix elements are given by
\be
L_{nn}(x)\
=\ \left(\left(\Gamma^\mu-x^\mu\right)^2\right)_{nn}\ +\ T_{nn}\
=\ \left(X_n^\mu-x^\mu\right)^2\ +\,\sum_{\ell\neq n}|\alpha^\mu_{\ell n}|^2\ +\ a_n^2
\label{Ldiag}
\ee
and their magnitudes are generally $O(D^2)$.
To be precise, in the inter-instanton space all of the $L_{nn}(x)$ are $O(D^2)$ or larger,
while in the neighborhood of an instanton $\#m$ the corresponding $L_{mm}(x)$ may
become as small as $O(a^2)$ but all the other $L_{nn}(x)$ remain $O(D^2)$.
On the other hand, the off-diagonal matrix elements are only $O(a^2)\ll D^2$,
\be
\label{Loffdiag}
\begin{split}
L_{m\neq n}\ &
=\ \left(\left(\Gamma^\mu-x^\mu\right)^2\right)_{mn}\ +\ T_{mn}\\[5pt]
&=\ (X_m+X_n-2x)^\mu \alpha_{mn}^\mu\
+\,\sum_{\ell\neq m,n}\alpha_{\ell m}^\mu\alpha_{\ell n}^\mu\
+\ a_ma_n\times\half\tr\bigl(y_m^\dagger y_n^{}\bigr)
\\[-3pt]
&=\ O(a^2).
\end{split}
\ee
This hierarchy between the diagonal and the off-diagonal matrix elements of $L$
allows us to evaluate $\log\det(L)$ as a power series in
in the ratios of the off-diagonal to diagonal elements:
\begin{align}
\begin{split}
\label{LDL}
\log\det(L)\ &
=\ \log\det\bigl(L_{\rm diag}\bigr)\
+\ \log\det\bigl(1\,+\,L_{\rm diag}^{-1}L_{\rm offdiag}\bigr)\\
&=\ \tr\log\bigl(L_{\rm diag}\bigr)\
+\ \tr\log\bigl(1\,+\,L_{\rm diag}^{-1}L_{\rm offdiag}\bigr)\\
&=\ \tr\log\bigl(L_{\rm diag}\bigr)\
+\ \tr\bigl(L_{\rm diag}^{-1}L_{\rm offdiag}\bigr)
-\ \half\tr\bigl(L_{\rm diag}^{-1}L_{\rm offdiag}L_{\rm diag}^{-1}L_{\rm offdiag}\bigr)
\qquad\\
&\qquad+\ \tfrac13\tr\bigl(L_{\rm diag}^{-1}L_{\rm offdiag}L_{\rm diag}^{-1}L_{\rm offdiag}
	L_{\rm diag}^{-1}L_{\rm offdiag}\bigr)\
+\ \cdots\\
&=\,\sum_n \log(L_{nn})\ +\ 0\
-\ \half\sum_{m\neq n}\frac{L_{mn}L_{nm}}{L_{mm}L_{nn}}\
+\ \tfrac13\sum_{\textstyle{{\rm different}\atop \ell,m,n}}
	\frac{L_{\ell m}L_{m n}L_{n\ell}}{L_{\ell\ell}L_{mm}L_{nn}}
+\ \cdots.
\end{split}
\end{align}
Consequently, the net instanton density profile of the system may written as a sum
of one-instanton, two-instanton, three-instanton, \etc, terms:
\begin{align}
I(x)\ &=\ {-1\over 16\pi^2}\,\square\square\,\log\det(L)\nonumber\\
&=\,\sum_n{\cal I}_n^{(1)}(x)\ +\ \half\sum_{m\neq n}{\cal I}_{mn}^{(2)}(x)\
+\ \tfrac16\sum_{\textstyle{{\rm different}\atop \ell,m,n}}{\cal I}_{\ell mn}^{(3)}(x)\
+\ \cdots\\
\intertext{where}
{\cal I}_n^{(1)}(x)\ &
=\ {-1\over 16\pi^2}\,\square\square\,\log\bigl(L_{nn}(x)\bigr),
\label{Ione}\\
{\cal I}_{mn}^{(2)}(x)\ &
=\ {+1\over 16\pi^2}\,\square\square\,\frac{L_{mn}L_{nm}}{L_{mm}L_{nn}}\,,
\label{Itwo}\\
{\cal I}_{\ell mn}^{(3)}(x)\ &
=\ {-2\over 16\pi^2}\,\square\square\,\frac{L_{\ell m}L_{mn}L_{n\ell}}%
	{L_{\ell\ell}L_{mm}L_{nn}}\,,
\label{Ithree}\\
\omit\hfil \it etc., etc.\nonumber
\end{align}
The one-instanton terms (\ref{Ione}) here have the form of stand-alone instanton profiles.
Indeed, rewriting eq.~(\ref{Ldiag}) as
\begin{align}
L_{nn}(x)\ &=\ r_n^2\ +\ \hat a_n^2\\
\text{where}\quad r_n^2\ &\equiv\ |x^\mu-X_n^\mu|^2\quad
\text{and}\quad\hat a_n^2\ =\ a_n^2\ +\,\sum_{m\neq n}|\alpha^\mu_{mn}|^2\
=\ a_n^2\ +\ O(a^4/D^2),\nonumber
\end{align}
we obtain
\be
{-1\over 16\pi^2}\,\square\square\,\log(L_{nn})\
=\ {-1\over 16\pi^2}\,\left(
	\frac{\partial^2}{\partial r_n^2}\,
	+\,\frac{3}{r_n}\,\frac{\partial}{\partial r_n}
	\right)^2 \log(r_n^2+\hat a_n^2)\
=\ \frac{(6/\pi^2)\,\hat a_n^4}{(r_n^2+\hat a_n^2)^4}\,,
\ee
which is precisely the density profile of a standalone instanton of radius $\hat a_n$.
Note that this radius is slightly larger than the original instanton radius $a_n$
--- this is a hidden effect of the interference from the other instantons $m\neq n$
--- while the visible effects of the interference come via the two-instanton terms
${\cal I}^{(2)}_{mn}(x)$, the three-instanton terms ${\cal I}^{(3)}_{\ell mn}(x)$, \etc{}
Deriving explicit formulae for all these terms is a straightforward (albeit rather tedious)
exercise, and the results are too cumbersome to present here.
Instead, let us simply estimate the magnitudes of the multi-instanton terms.

In the inter-instanton space --- $i.\,e.$ at $O(D)$ distances from {\sl all}
the instanton centers --- we have
\be
\frac{L_{mn}L_{nm}}{L_{mm}L_{nn}}\ \sim\ \frac{a^4}{D^4}\,,\quad
\frac{L_{\ell m}L_{mn}L_{n\ell}}{L_{\ell\ell}L_{mm}L_{nn}}\ \sim\ \frac{a^6}{D^6}\,,\quad
\it etc.
\ee
Moreover, all these expressions change rather slowly with $x$ --- on the scale $x\sim D$
--- so in the context of such expressions $\partial_\mu=O(1/D)$.
Consequently,
\be
\square\square\,\frac{L_{mn}L_{nm}}{L_{mm}L_{nn}}\ \sim\ \frac{a^4}{D^8}\,,\quad
\square\square\,\frac{L_{\ell m}L_{mn}L_{n\ell}}{L_{\ell\ell}L_{mm}L_{nn}}\
\sim\ \frac{a^6}{D^{10}}\,,\quad
\ldots,
\ee
and therefore
\be
\underline{\text{In the inter-instanton space}}\quad
{\cal I}^{(2)}_{mn}(x)\ \sim\ \frac{a^4}{D^8}\,,\quad
{\cal I}^{(3)}_{\ell mn}(x)\ \sim\ \frac{a^6}{D^{10}}\,,\quad
\ldots.
\label{Iestimates}
\ee
As promised in eq.~(\ref{Idecomp}), the interference between small well-separated
instantons is dominated by the two-instanton terms.
In fact, in the inter-instanton space the two-instanton terms have similar magnitudes to
the one-instanton terms,
\be
{\cal I}^{(2)}_{mn}(x)\ \sim\ \frac{a^4}{D^8}\ \sim\ {\cal I}^{(1)}_n(x).
\ee
However, the entire $I(x)$ in the inter-instanton space makes a negligible
contribution to the net Coulomb energy of the system,
\be
\E_C[{\rm IIS}]\
=\int\limits_{\rm IIS}\!\!d^4 x\,I(x)\times{N_c\over 2\lambda M}\sum_n{1\over|x^\mu-X_n^\mu|^2}\
\sim\ D^4\times\frac{a^4}{D^8}\times{N_c\over\lambda M D^2}\
\sim\ {N_c a^4\over\lambda M D^6}\,.
\ee
Therefore, the net Coulomb energy of the multi-instanton system may be summarized as
\be
\E_C^{\rm net}\ =\ {N_c\over4\lambda M}\times\left\{
	\sum_n{1\over\rho_n^2}\
	+\,\sum_{m\neq n}{1\over|X_m^\mu-X_n^\mu|^2}\
	+\ O(a^4/D^6)
	\right\},
\ee
where the first term inside the `$\{\cdots\}$' comprises the self-interactions
of individual instantons --- but with the effective charge radii $\rho_n$ that
include the effects of interference from the other instantons, ---
the second term is the direct Coulomb repulsion (\ref{Edirect})
between different instantons approximated as point charges
(or equivalently, as compact spherically-symmetric charges), and the third term
accounts for the $O(a^4/D^8)$ instanton density in the inter-instanton space.
Note that the self-interaction terms dominate the net Coulomb energy, so even
relatively small corrections to the effective charge radii $\rho_i$ can have an
effect comparable to the direct Coulomb repulsion between the instantons.

To work out the effects of interference on the charge radius $\rho_n$
of the $n^{\rm th}$ instanton we need to estimate (and then evaluate) the
interference terms ${\cal I}^{(2)}_{\ell m}$, ${\cal I}^{(3)}_{k\ell m}$, \etc,
for $x^\mu$ near the $n^{\rm th}$ instanton's center $X_n^\mu$.
In this neighborhood, the estimates (\ref{Iestimates}) do not work
--- or rather they work only for the interference terms between the {\it other instantons}
$\#k,\ell,m,\ldots\neq n$.

But the terms ${\cal I}^{(2)}_{mn}$, ${\cal I}^{(3)}_{\ell mn}$, \etc, that do involve
the $n^{\rm th}$ instanton itself turn out to be much larger than (\ref{Iestimates})
because of the $1/L_{nn}(x)$ factor that happens to be $O(1/a^2)$ instead of $O(1/D^2)$.
Moreover, this factor depends on $x$ much more rapidly than all the other
factors --- on the scale of $x\sim a$ instead of $x\sim D$ --- so taking
the space derivatives of this factor produces much larger results than taking the
same derivatives of the other factors.
Thus
\be
\begin{split}
\square\square\,\frac{L_{mn}L_{nm}}{L_{mm}L_{nn}}\ &
\approx\ \left(\square\square\,{1\over L_{nn}}\right)\times\frac{L_{mn}L_{nm}}{L_{mm}}\,,\\
\square\square\,\frac{L_{\ell m}L_{mn}L_{n\ell}}{L_{\ell\ell}L_{mm}L_{nn}}\ &
\approx\ \left(\square\square\,{1\over L_{nn}}\right)\times
	\frac{L_{\ell m}L_{mn}L_{n\ell}}{L_{\ell\ell}L_{mm}}\,,\\
\etc,\ &
\end{split}
\ee
and therefore
\begin{align}
{\cal I}_{mn}^{(2)}(x)\ &
=\ \left({1\over 16\pi^2}\,\square\square\,{1\over L_{nn}}\right)
	\times\left[\frac{L_{mn}L_{nm}}{L_{mm}}\right](@x=X_n)\
+\ \text{subleading,}
\label{TwoInear}\\
{\cal I}_{\ell mn}^{(3)}(x)\ &
=\ \left({1\over 16\pi^2}\,\square\square\,{1\over L_{nn}}\right)
	\times\left[-2\frac{L_{\ell m}L_{mn}L_{n\ell}}{L_{\ell\ell}L_{mm}}\right](@x=X_i)\
+\ \text{subleading,}
\label{ThreeInear}\\
\omit\hfil\it etc., etc.\nonumber
\end{align}
The first factor in all these interference terms evaluates to
\be
{1\over 16\pi^2}\,\square\square\left({1\over L_{nn}}\,=\,{1\over r_n^2+\hat a_n^2}\right)\
=\ {12\over\pi^2}\,{\hat a_n^2(\hat a_n^2+r_n^2)\over(\hat a_n^2+r_n^2)^5}\
\sim\ {1\over a^6}
\label{FirstFactor}
\ee
hence
\be
\underline{\text{In the vicinity of the $n^{\rm th}$ instanton}}\quad
{\cal I}_{mn}^{(2)}(x)\ \sim\ {1\over a^2 D^2}\,,\quad
{\cal I}_{\ell mn}^{(3)}(x)\ \sim\ {1\over D^4}\,,\quad
\ldots.
\ee
Since the standalone instanton's density is $I^{(1)}_n(x)\sim 1/a^4$, the interference
terms change the effective charge radius or rather
\be
{1\over\rho_n}\bigl[\text{due to}\ I^{(1)}_n\ \text{only}\bigr]\
=\mathop{\intop\!\!\!\!\intop}\limits_{{\rm near}\,X_n}\!
	d^4x\,d^4 x'\,{I^{(1)}_n(x)\times I^{(1)}_{n}(x')\over |x-x'|^2}\
=\ {4/5\over\hat a_n^2}
\ee
by
\begin{align}
\Delta_n\bigl[\text{due to}\ {\cal I}^2_{mn}\bigr]\ &
=\mathop{\intop\!\!\!\!\intop}\limits_{{\rm near}\,X_n}\!
	d^4x\,d^4 x'\,{2I^{(1)}_n(x)\times I^{(2)}_{mn}(x')\over |x-x'|^2}\
\sim\ {a^8\over a^2}\times{1\over a^4}\times{1\over a^2 D^2}\
\sim {1\over D^2}\,,\\
\Delta_n\bigl[\text{due to}\ {\cal I}^3_{\ell mn}\bigr]\ &
=\mathop{\intop\!\!\!\!\intop}\limits_{{\rm near}\,X_n}\!
	d^4x\,d^4 x'\,{2I^{(1)}_n(x)\times I^{(3)}_{\ell mn}(x')\over |x-x'|^2}\
\sim\ {a^8\over a^2}\times{1\over a^4}\times{1\over D^4}\
\sim {a^2\over D^4}\,,\\
\omit\hfil\it etc., etc.\nonumber
\end{align}
Therefore, {\it  the two-instanton interference terms ${\cal I}^2_{mn}$
affect the Coulomb energy by the amount comparable to the direct repulsion}
(between point-like instantons) {\it but the effect of interference terms
involving three or more instantons at once is suppressed by extra powers of $a^2/D^2$
and becomes negligible in the limit of small instantons separated by large distances.}
Since we have already verified similar behavior of the non-abelian energy of the
multi-instanton system, this completes the proof of our theorem.\quad
{\it Quod erat demonstrandum.}

To conclude this section, we would like to derive  explicit formulae
for the two-instanton interference terms (near one of the two instantons)
and hence the Coulomb interaction energy due to interference.
Starting from eq.~(\ref{TwoInear}) we have the first factor on the RHS
evaluated in eq.~(\ref{FirstFactor}), so let us evaluate the second factor.
For $x^\mu=X_m^\mu$, eq.~(\ref{Loffdiag}) gives us
\be
L_{mn}(x=X_I)\ =\ L_{nm}(x=X_I)\
=\ (X_m-X_n)^\mu\, \alpha_{mn}^\mu\
+\,\sum_{\ell\neq m,n}\alpha^\mu_{\ell m}\alpha^\mu_{\ell n}\
+\ a_m a_n\times\half\tr\bigl(y_m^\dagger y_n^{}\bigr)
\ee
where the first term on the RHS happens to vanish --- {\it cf.}\ eqs.~(\ref{ADHMsols}) ---
while the second term is $O(a^4/D^2)$,
which is much smaller than the $O(a^2)$ third term.
Thus,
\be
L_{mn}(x=X_n)\
=\ L_{nm}(x=X_n)\ \approx\ a_m a_n\times\half\tr\bigl(y_m^\dagger y_n^{}\bigr)
\ee
while
\be
L_{mm}(x=X_n)\ =\ |X_m-X_n|^2\ +\ \hat a_m^2\ \approx\ |X_m-X_n|^2,
\ee
hence
\be
\left[{L_{mn}L_{nm}\over L_{mm}}\right](x=X_n)\
\approx\ {\hat a_m^2\hat a_n^2\tr^2(y_m^\dagger y_n)\over 4|X_m-X_n|^2}
\ee
and therefore
\be
{\cal I}_{mn}^{(2)}(x\,{\rm near}\,X_n)\
\approx\ {3 \hat a_m^2\tr^2(y_m^\dagger n_j)\over \pi^2|X_m-X_n|^2}\times
	{\hat a_n^4(\hat a_n^2-r_n^2)\over(\hat a_n^2+r_n^2)^5}\
=\ {\hat a_m^2\tr^2(y_m^\dagger y_n)\over2|X_m-X_n|^2}\times
	{\hat a_n^2-r_n^2\over\hat a_n^2+r_n^2}\times{\cal I}^{(1)}_n(x).
\ee
It is easy to check that such two-instanton interference terms do not affect
the net charge of the  $n^{\rm th}$ instanton ---
\be
\int\limits_{{\rm near}\,X_n}\!\!d^4 x\,{\cal I}_{mn}^{(2)}(x)\ =\ 0
\ee
--- but they move the charge distribution close to the center $X_n$
since ${\cal I}^{(2)}_{mn}$ is positive for $r_n<\hat a_n$ but negative for $r_n>\hat a_n$.
Consequently, they decrease the effective charge radius $\rho_n$, which corresponds
to $\Delta_n>0$ and hence positive extra Coulomb energy.
Specifically,
\begin{align}
\!\Delta_i\bigl[\text{due to}\ {\cal I}^2_{mn}\bigr] &
=\mathop{\intop\!\!\!\!\intop}\limits_{{\rm instanton}\,\#n}\!
	d^4x\,d^4 x'\,{2I^{(1)}_n(x)\times I^{(2)}_{mn}(x')\over |x-x'|^2}
\nonumber\\
&=\, 144\,{\hat a_m^2\tr^2(y_m^\dagger y_n)\over|X_m-X_n|^2}\times\!\!
\mathop{\intop\!\!\!\!\intop}\limits_{\!0\,\,\!}^{\,\,\!\infty\!}\!\!dr\,r^3\,dr'r^{\prime3}\,
	{\hat a_n^4\over(\hat a_n^2+r^2)^4}\times
	{\hat a_n^4(\hat a_n^2-r^{\prime2})\over(\hat a_n^2+r^{\prime2})^5}
	\times{1\over\max(r^2,r^{\prime2})}\!
\nonumber\\
&=\, +{\hat a_m^2\over 5\hat a_n^2}\times{\tr^2(y_m^\dagger y_n)\over|X_n-X_j|^2}\,.
\label{FirstDelta}
\end{align}

Note however that this $\Delta_n$ is a correction of the instanton's charge
radius or (rather $1/\rho_n^2$) {\it starting from the standalone radius
$\hat a_n$ instead of the original $a_n$.}
Consequently, there is an additional correction
\be
\Delta_n^{\rm extra}\ =\ {4/5\over\hat a_n^2}\ -\ {4/5\over a_n^2}\
\approx\ -{4/5\over a_n^4}\times(\hat a_n^2-\hat a_n^2)\
=\ -{4/5\over a_n^4}\times\sum_{m\neq n}|\alpha_{mn}^\mu|^2.
\ee
The $\alpha^\mu_{mn}$ are spelled out in eqs.~(\ref{AlphaExpansion});
to the leading order
\be
\alpha^\mu_{mn}\ \approx\ \alpha^{(1)}_{\mu mn}\
=\ a_ma_n\times{\eta^a_{\mu\nu}(X_m-X_n)_\nu\over|X_m-X_n|^2}\times
\half\tr\bigl(y_m^\dagger y_n(-i\tau^a)\bigr)
\ee
Using $\eta^{(a}_{\mu\nu}\eta^{b)}_{\mu\lambda}=\delta^{ab}\delta_{\nu\lambda}$,
we obtain
\be
|\alpha^{(1)}_{\mu mn}|^2\
\approx\  {a_m^2 a_n^2\over|X_m-X_n|^2}\times\left(
	\sum_a\tfrac14\tr^2\bigl(y_m^\dagger y_n(-i\tau^a)\bigr)\,
	=\,1\,-\,\tfrac14\tr^2\bigl(y_m^\dagger y_n\bigr)
	\right)
\ee
and hence
\be
\Delta_n^{\rm extra}\
=\ -\frac15\sum_{m\neq n}{a_m^2\bigl(4-\tr^2(y_m^\dagger y_n^{})\bigr)
	\over a_n^2|X_m-X_n|^2}\
+\ O(a^2/D^4).
\label{SecondDelta}
\ee
Combining the corrections (\ref{FirstDelta}) and (\ref{SecondDelta}),
we obtain
\be
\Delta_n^{\rm net}\ =\ +\frac25\sum_{m\neq n}{a_m^2\over a_n^2}\times
{\tr^2(y_m^\dagger y_n^{})-2\over|X_m-X_n|^2}\
+\ O(a^2/D^4)
\ee
and therefore
\be
\E_C^{\rm net}\
=\ {N_c\over 4\lambda M}\left(\begin{aligned}
	\sum_n{4/5\over a_n^2}\ &
	+\,\sum_{m\neq n}{1\over|X_m-X_n|^2}\times\left(1\,
		+\,{1\over5}\left({a_m^2\over a_n^2}+{a_n^2\over a_m^2}\right)
		\times\bigl(\tr^2(y_m^\dagger y_n^{})-2\bigr)\right)\\
	&+\ O(a^2/D^4)
	\end{aligned}\right).
\label{ECnet}
\ee

Finally, let us combine the non-abelian and the Coulomb energies
of the system and re-organize the net energy into one-body, two-body, etc, terms:
\be\begin{split}
\E^{\rm total}\ &
=\,\sum_n\E^{\rm 1\,body}(n)\
+\ \half\sum_{m\neq n}\E^{\rm 2\,body}(m,n)\
+\ \tfrac16\!\!\sum_{\textstyle{{\rm different}\atop \ell,m,n}}\!\!\E^{\rm 3\,body}(\ell,m,n)\
+\ \cdots
\end{split}\ee
where
\begin{align}
\E^{\rm 1\,body}(n)\ &
=\ N_c M\left(
	\lambda M^2\times\bigl(X_n^4\bigr)^2\,
	+\,{\lambda M^2\over2}\times a_n^2\,
	+\,{1\over 5\lambda M^2}\times{1\over a_n^2}
	\right),\label{Eonebody}\\[5pt]
\E^{\rm 2\,body}(m,n)\ &
=\ {N_c\over2\lambda M}\times{1\over|X_m-X_n|^2_{4D}}
\left(\begin{aligned}
	\lambda^2 M^4 &
	\times a_m^2 a_n^2\times\tr^2\bigl(y_m^\dagger y_n^{}(-i\vec N_{mn}\cdot\vec\tau)\bigr)\\
	{}+\,1\, &
	+\,{1\over5}\left({a_m^2\over a_n^2}+{a_n^2\over a_m^2}\right)\times
	\bigl(\tr^2(y_m^\dagger y_n^{})-2\bigr)\\\
	&+\, O\left({a^2\over D^2}\sim{1\over \lambda M^2 D^2}\right)
	\end{aligned}\right),\label{Etwobody}\\
\E^{\rm 3\,body}(\ell,m,n)\ &
=\ O\left({N_c a^2\over\lambda M^2 D^4}\sim{N_c\over\lambda^2 M^3 D^4}\right),
\quad\text{\etc, \etc}
\end{align}
When comparing magnitudes of the non-abelian and the Coulomb terms here we have used
$a\sim1/(\sqrt\lambda M)$.
In fact, we may be more precise by noticing that for small instantons distant from
each other, the one-body potential energies for the instanton radii $a_i$
are much larger than the two-body, \etc, interactions between different instantons.
Consequently, in the minimal-energy or near-minimal-energy configuration of the
multi-instanton system, the instanton radii will be close to the equilibrium radius
of a stand-alone instanton,
\be
a_n\ =\ a_0\ +\ O(a^3/D^2)
\ee
where
\be
a_0\ =\ {\root 4\of{2/5}\over\sqrt\lambda M}
\label{InstantonRadius}
\ee
minimizes the one-body potential energy~(\ref{Eonebody}).
Plugging this equilibrium radius into eq.~(\ref{Etwobody}), we obtain
a simpler formula for the two-instanton interaction energy:
\be
\E^{\rm 2\,body}(m,n)\
=\ {2N_c\over 5\lambda M}\times{1\over|X_m-X_n|^2_{4D}}\times\left[
	{1\over2}\
	+\ \tr^2\Bigl(y_m^\dagger y_n^{}\Bigr)\
	+\ \tr^2\Bigl(y_m^\dagger y_n^{}\,(-i\vec N_{mn}\cdot\vec\tau)\Bigr)\
	\right].
\label{KZ}
\ee
Note that the expression inside `$[\cdots]$' is always positive, so the two-body forces
between the instantons are always repulsive,
regardless of the  instantons' $SU(2)$ orientations.
However, the orientations do affect the strength of the repulsion: two instantons
with similar orientations repel each other 9 times stronger then the instantons at the
same distance from each other but whose orientations differ by a $180^\circ$ rotation
(in $SO(3)$ terms) around a suitable axis.
{\it This fact will be at the core of our analysis of instanton crystals in
subsequent sections.}

%
\section{Linear Chains of Instantons}
Consider an infinite set of instantons arranged in a crystalline lattice.
In light of eq.~\eqref{KZ}, the orientations of instantons at different lattice sites affects
the system's energy just as much as the lattice geometry.
In this section we shall see how this works for a simple one-dimensional
low-density lattice
--- $i.\,e.$ an infinite chain of small instantons located along a straight line at
\be
X_n^\mu\ =\ (nD,0,0,0),\quad n\in\Z
\label{chain1D}
\ee
where the lattice spacing $D$ is much larger than the instanton radius $a$.

Before we go any further, we need a reason why the instantons would form a 1D chain
rather then spread out into all 3 flat space dimensions like holographic baryons
(or real-life baryons) in the compressed nuclear matter.
In principle, we may sidestep this question by simply freezing the $X_n^\mu$
moduli of the instantons and allow only their orientation moduli $y_n$ to vary,
but then it would be hard to place such a system in any kind of a physical context.
Alternatively, we can make 5D gauge coupling depend on the $x_2$ and $x_3$ coordinates
as well as the holographic dimension $x_4$
\be
{8\pi ^2\over g^2_5(x_2,x_3,x_4)}\ =\ N_c\lambda M\Bigl(
	1\,+\,M_4^2 x_4^2\,+\,M_3^2x_3^2\,+\,M_2^2x_2^2\,+\,O(M^4x^4)\Bigr),\qquad
M_4\equiv M,
\label{G5Curved}
\ee
which creates an effective potential for the instanton centers $X_n$,
\be
\E^{(1)}(n)\ =\ N_c\lambda MM_4^2\times(X_n^4)^2\
+\ N_c\lambda MM_3^2\times(X_n^3)^2\ +\ N_c\lambda MM_2^2\times(X_n^2)^2\
+\ O(N_c\lambda M^5 X^4).
\label{ChainPotential}
\ee
At low instanton densities, this potential makes the instantons line up
along the $x^1$ axis, while
at higher densities, the repulsion between the instantons becomes stronger than this potential
and they spread out into the other dimensions.
We shall address such spreading out in the following section \S5, but for the moment
let us focus on the low-density 1D chains~(\ref{chain1D}).

In a bigger holographic picture, a $g_5$ that depends on three coordinates $x_{2,3,4}$
indicates that the flavor branes are curved in all 3 of these directions,
\be
U[\text{the energy scale}]\ \approx\ U_0\ +\ {X_4^2\over 2R_4}\
+\ {X_3^2\over 2R_3}\ +\ {X_2^2\over 2R_2}\ +\ O(X^4/R^3),
\label{BigGeometry}
\ee
which is quite different from the usual holographic setup where the energy-scale
coordinate $U$ depends only on the $x^4$.
But in this article we are not going to explore the geometric aspects of such curvature.
Instead, we shall simply use the gauge coupling~(\ref{G5Curved}) in an
approximately flat space (for distances${}\ll1/M$) as a tool to put instantons into
a 1D lattice.
In our next article \cite{OurNextPaper},
we shall  make $g_2$ depend only on $x_3$ and $x_4$
to put instantons into a 2D lattice in the $(x_1,x_2)$ plane, and eventually go back to
$g_5$ depending only on the $x_4$ to make 3D instanton lattices.

For simplicity, let us assume that the $g_5$ is much less sensitive to the $x_2$ and
$x_3$ coordinates than the $x_4$, thus $M_2^2,M_3^3\ll M_4^2$.
In this limit, the  $M_2^2$ and $M_3^3$ parameters give rise to the $x_2^2$ and $x_3^3$
terms in the one-body instanton potential~(\ref{ChainPotential}), but their effect on
the instanton radius $a$ or the two-body forces between the instantons may be neglected
(compared to the effect of the $M^2_4x_4^2$ term).
Therefore, the net two-body forces between the instantons remain approximately as
in eq.~(\ref{KZ}).

Moreover, for the 1D lattice geometry~(\ref{chain1D}) of the instanton centers,
we have $|X_m-X_n|^2=D^2\times(m-n)^2$ while $\vec N_{mn}=(\pm1,0,0)$.
Consequently, the net energy of the instanton chain as a function of the
instantons' orientations $y_n$ may be summarized as
\be
\E^{\rm int}\ =\ {N_c\over5\lambda M D^2}\times\sum_{m\neq n}
\frac1{(m-n)^2}\times\Bigl[ \half\,
	+\,\tr^2\bigl(y_m^\dagger y_n^{}\bigr)\,
	+\,\tr^2\bigl(y_m^\dagger y_n^{}(-i\tau_1)\bigr)
	\Bigr]\,.
\label{Echain}
\ee
To minimize this energy, each pair of instantons $m$ and $n$ wants to have
$y_m^\dagger y_n^{}$ to be a linear combination of $i\tau_2$ and $i\tau_3$;
in $SO(3)$ terms, this corresponds to a relative rotation (between the 2 instantons)
through a $180^\circ$ angle around some axis${}\perp x_1$.
Alas, this cannot be achieved for all instanton pairs at once;
indeed, if we minimize the energies of the $(n,m)$ and the $(n,\ell)$ pairs,
then the energy of the $(m,\ell)$ pair would be maximal instead of minimal:
\be
\begin{split}
\bf If\quad &
y_n^\dagger y_m^{}\ =\ i\vec n_1\cdot\vec\tau\quad\text{\bf\&}\quad
y_n^\dagger y_\ell^{}\ =\ i\vec n_2\cdot\vec\tau\quad
\text{for}\ \vec n_1,\vec n_2\perp x_1\\
\bf Then\quad &
y_m^\dagger y_\ell^{}\ =\ (\vec n_1\cdot\vec\tau)(\vec n_2\cdot\vec\tau)\
=\ (\vec n_1\cdot\vec n_2)\ +\ i(\vec n_1\times\vec n_2)\cdot\vec\tau\quad
\text{for}\ \vec n_1\times\vec n_2\,\parallel\,x_1\\
\Longrightarrow\quad&
\tr^2\bigl(y_m^\dagger y_\ell^{}\bigr)\,+\,\tr^2\bigl(y_m^\dagger y_\ell^{}(-i\tau_1)\bigr)\
=\ 4(\vec n_1\cdot\vec n_2)^2\,+\,4(\vec n_1\times\vec n_2)_1^2\ =\ 4\
\text{(maximum).}
\end{split}
\ee
So our best bet is to first minimize the energies of the most expensive pairs --- the
nearest neighbors $(m,m+1)$ --- and then worry about the less expensive pairs of instantons.
Thus, we want
\be
\label{Condition}
\forall m:\quad
y_m^\dagger y_{m+1}^{}\ =\ \cos\psi_m\times(i\tau_3)\ +\ \sin\psi_m\times(i\tau_2)\quad
\text{for some angle}\ \psi_m\,,
\ee
and the most general solution for this set of equations (modulo a common $SU(2)$
symmetry of all the $y_m$) is
\be
\label{BigFamily}
y_n\ =\ \exp\bigl(i\phi_n\tau_1\bigr)\times(i\tau_3)^n\
=\begin{cases}
	\pm[\cos\phi_n\times1\,+\,\sin\phi_n\times(i\tau_1)] &
	\text{for even}\ n,\\
	\pm[\cos\phi_n\times(i\tau_3)\,+\,\sin\phi_n\times(i\tau_2)] &
	\text{for odd}\ n,\\
	\end{cases}
\ee
for some  angles $\phi_n$.
Indeed, for any set of the angles $\phi_n$ we have
\be
\begin{split}
y_m^\dagger y_{m+1}^{}\ &
=\ (-i\tau_3)^m\times\exp\bigl(i(\phi_{m+1}-\phi_m)\tau_1\bigr)\times(i\tau_3)^{m+1}\\
&=\ \exp\bigl( (-1)^m i(\phi_{m+1}-\phi_m)\tau_1\bigr)\times(i\tau_3)\\
&=\ \cos(\phi_{m+1}-\phi_m)\times(i\tau_3)\ +\ (-1)^m\sin(\phi_{m+1}-\phi_m)\times(i\tau_2),
\end{split}
\ee
which agrees with eqs.~(\ref{Condition}) for $\psi_m=(-1)^m(\phi_{m+1}-\phi_m)$.
Clearly, any set of $\psi_m$ can be obtained for suitable $\phi_n$, so eqs.~(\ref{BigFamily})
indeed describe all the solutions to the eqs.~(\ref{Condition}).

Eqs.~(\ref{BigFamily}) for various angles $\phi_n$ define a big family of
instanton configurations.
Surprisingly, all these configurations have exactly the same net energies!
Indeed, for all sets of $\phi_n$, all instanton pairs $(m,n)$ with odd $m-n$ have minimal
energies while pairs with even $m-n$ have maximal energies:
\be
\label{BigFamilyTwists}
\begin{split}
\text{for odd}\ m-n,\quad y_m^\dagger y_n^{}\ &
=\ \cos(\phi_n-\phi_m)\times(\pm i\tau_3)\ +\ \sin(\phi_n-\phi_m)\times(\pm i\tau_2)\\
\Longrightarrow\ &
\half\,+\,\tr^2\bigl(y_m^\dagger y_n^{}\bigr)\,
	+\,\tr^2\bigl(y_m^\dagger y_n^{}(-i\tau_1)\bigr)\ ={}\\
&=\ \half\,+\,0\,+\,0\
=\ \half\,,\\
\text{for even}\ m-n,\quad y_m^\dagger y_n^{}\ &
=\ \cos(\phi_n-\phi_m)\times(\pm1)\ +\ \sin(\phi_n-\phi_m)\times(\pm i\tau_1)\\
\Longrightarrow\ &
\half\,+\,\tr^2\bigl(y_m^\dagger y_n^{}\bigr)\,
	+\,\tr^2\bigl(y_m^\dagger y_n^{}(-i\tau_1)\bigr)\ ={}\\
&=\ \half\,+\,4\cos^2(\phi_n-\phi_m)\,+\,4\sin^2(\phi_n-\phi_m)\
=\ \tfrac92\,.
\end{split}
\ee
Consequently, regardless of the angles $\phi_m$,
the net energy  per instanton of the 1D lattice is
\be
\begin{split}
\E^{\rm interaction}_{\rm per\,instanton}\ &
=\ {N_c\over 5\lambda MD^2}\times\sum_{m\neq n}^{{\rm fixed}\,n}
	\frac1{(m-n)^2}\times\Bigl[ \half\,
		+\,\tr^2\bigl(y_m^\dagger y_n^{}\bigr)\,
		+\,\tr^2\bigl(y_m^\dagger y_n^{}(-i\tau_1)\bigr)
		\Bigr]\\
&=\ {N_c\over 5\lambda MD^2}\times\sum_{\ell=m-n\neq0}\frac1{\ell^2}\times
	\begin{cases}
		\frac12 & \text{for odd}\ \ell,\\
		\frac92 & \text{for even}\ \ell,
		\end{cases}\\
&=\ {N_c\over 5\lambda MD^2}\times\left(
	\frac12\sum_{{\rm odd}\,\ell}\frac1{\ell^2}\,
	+\,\frac92\sum_{{\rm even}\,\ell\neq0}\frac1{\ell^2}
	\right)\\
&=\ {N_c\over 5\lambda MD^2}\times\left(
	\frac12\times\frac{\pi^2}{4}\ +\ \frac92\times\frac{\pi^2}{12}\
	=\ \frac{\pi^2}{2}\right).
\end{split}
\ee

Now let's consider the instantons' orientations $y_n$
from the group-theoretical point of view.
Each $y_n$ is an $SU(2)$ matrix, but its overall sign is  irrelevant, so
we only care about the $y_n$ (modulo sign), which belongs to the $SU(2)/\Z_2\cong SO(3)$.
In a generic lowest-energy configuration (\ref{BigFamily}) of the 1D chain,
the instanton orientations span an $SO(2)\times\Z_2$ subgroup of
the $SO(3)$ corresponding to the rotational symmetries of a cylinder ---
rotations through arbitrary angles around the $x_1$ axis, and $180^\circ$
rotations around axes${}\perp x_1$.
The $y_n$ alternate between the two types of rotations, but apart from that
they generically do not follow any regular patterns.

However, the family (\ref{BigFamily}) also contains some regular patterns
in which the orientations $y_n$ (modulo sign) follow a repeating cycle of finite length $p$;
moreover, the values of $y_n$  span a discrete subgroup of the cylindrical symmetry
$SO(2)\times\Z_2$.
Here are some examples:
\begin{itemize}
\item
The {\it anti-ferromagnetic chain}, with 2 alternating instanton orientations:
\be
y_{{\rm even}\,n}\ =\ \pm1,\quad y_{{\rm odd}\,n}\ =\ \pm i\tau_3\,.
\label{AntiFerro}
\ee
In this configuration --- which obtains for $\phi_n\equiv0$ ---
the $y_n$ (modulo sign) span a $\Z_2$ subgroup of the $SO(2)\times\Z_2$.

\item
Period${}=4$ configuration spanning the {\it Klein group}
of $180^\circ$ rotations around the 3 Cartesian axes\footnote{%
	As a subgroup of $SO(3)$, the Klein group is abelian and isomorphic to $\Z_2\times \Z_2$.
	But its covering group in $SU(2)$ is  non-abelian group and isomorphic
	to the group of unit quaternions $\pm 1,\pm i,\pm j,\pm k$.
	}:
\be
y_{n\equiv0\!\!\!\pmod4}\ =\ \pm1,\quad
y_{n\equiv1\!\!\!\pmod4}\ =\ \pm\tau_3,\quad
y_{n\equiv2\!\!\!\pmod4}\ =\ \pm\tau_1,\quad
y_{n\equiv3\!\!\!\pmod4}\ =\ \pm\tau_2.
\label{Klein}
\ee

\item
Period${}=2k=6,8,10,\ldots$ configurations spanning {\it prismatic groups}
$\Z_k\times\Z_2$:
\be
\label{Prismatic}
\begin{split}
y_{{\rm even}\,n}\ &
=\ \cos\frac{\pi n}{2k}\times1\ +\ \sin\frac{\pi n}{2k}\times(i\tau_1),\\
y_{{\rm odd}\,n}\ &
=\ \cos\frac{\pi(n-1)}{2k}\times(i\tau_3)\ +\ \sin\frac{\pi(n-1)}{2k}\times(i\tau_2).
\end{split}
\ee

\item
Period${}=2k=6,8,10,\ldots$ configurations spanning {\it dihedral groups} $D_{2k}$,
which obtain for $\phi_n=n\times(\pi/2k)$, $i.\,e.$
\be
\label{Dihedral}
\begin{split}
y_{{\rm even}\,n}\ &
=\ \cos\frac{\pi n}{2k}\times1\ +\ \sin\frac{\pi n}{2k}\times(i\tau_1),\\
y_{{\rm odd}\,n}\ &
=\ \cos\frac{\pi n}{2k}\times(i\tau_3)\ +\ \sin\frac{\pi n}{2k}\times(i\tau_2).
\end{split}
\ee

\end{itemize}\par

There is a wider class of regular configurations --- we shall call them link-periodic
--- in which the $y_n$ themselves are not periodic,
but the relative rotations $y_n^\dagger y_{n+1}^{}$ between nearest neighbors
follow a periodic pattern.
In terms of the $\phi_n$ angles, this corresponds to periodic differences $\phi_{n+1}-\phi_n$,
for example
\be
\phi_n\ =\ n\varphi\ -\ \half(-1)^n\theta\quad\Longrightarrow\quad
\phi_{n+1}\,-\,\phi_n\ =\begin{cases}
	\varphi+\theta & \text{for even}\ n,\\
	\varphi-\theta & \text{for even}\ n.
	\end{cases}
\label{LinkPeriodic}
\ee
For rational $\varphi/\pi$ this pattern produces a periodic array of
instantons' orientations $y_n$
--- for example, the dihedral cycle for $\varphi=\pi/2k$ and $\theta=0$,
or the prismatic cycle for $\varphi=\pi/2k$ and $\theta=-\varphi/2$.
But for irrational $\varphi$'s
the orientations $y_n$ themselves do not have a finite period;
instead, they wind irrationally around the cylinder group $SU(2)\times\Z_2$.

\medskip
\centerline{$\star\quad\star\quad\star$}
\medskip

All in all, the 1D lattice of instantons has a huge degenerate family~(\ref{BigFamily})
of lowest-energy configurations.
However, this degeneracy obtains only in the $M_2,M_3\ll M_4$ limit where the
5D gauge coupling (\ref{G5Curved}) depends mostly on the holographic coordinate $x_4$
and only a little bit on the $x_2$ and $x_3$.
For the larger $M_2,M_3\sim M_4$, the degeneracy is lifted and there is a unique
lowest-energy configuration of the 1D lattice, namely the link-periodic
array~(\ref{LinkPeriodic}) whose $\varphi$ and $\theta$ parameters depends
on the $M_2/M_3$ ratio.

To see how this works we note that for $M_2\sim M_3\sim  M_4$, all 3 coordinates
$x_{2,3,4}$ transverse to the 1D lattice play similar roles, and the
3 terms $M_4^2x_4^2+M_3^2x_3^2+M_2^2x_2^4$ inside `$(\cdots)$' in eq.~(\ref{G5Curved})
give rise to similar contributions to the non-abelian energy of a multi-instanton system:
\be
\E_{NA}^{\rm net}\
=\ N_c\lambda M\times\!\int\!\!d^4x\,I(x)\times\sum_{\mu=2,3,4}M_\mu^2(x_\mu)^2
\ee
(where $M_4^2$ is the same as $M^2$).
Consequently, the two-body interaction energy between the instantons becomes more
complicated then in eq.~\eqref{ENA2body}.
Instead, generalizing eq.~\eqref{ENAprelim} we obtain
\be
\E_{NA}^{\rm net}\
=\ N_c\lambda M\sum_{\mu=2,3,4}M_\mu^2\left\{
	\sum_n\left( \left(X_n^\mu\right)^2+\half a_n^2\right)\
	+\,\sum_{n\neq m}\left(\alpha^\mu_{nm}\right)^2
	\right\}
\ee
and hence to the leading order in expansion~\eqref{AlphaExpansion},
\begin{align}
\E_{NA}^{\rm net}\ &
\approx\ \sum_n\E_{NA}^{(1)}(n)\ +\ \half\sum_{m\neq n}\E_{NA}^{(2)}(m,n)\\
\intertext{where}
\E_{NA}^{(1)}(n)\ &
=\ \sum_{\mu=2,3,4}N_c\lambda M M_\mu^2(X_n^\mu)^2\
+\ \half N_c\lambda M(M_4^2+M_3^2+M_2^2)\times a_n^2
\label{ENA1bodyM}\\
\intertext{and}
\E_{NA}^{(2)}(m,n)\ &
=\ {a_n^2 a_m^2\over 2|X_m-X_n|^2}\times\sum_{\mu=2,3,4}N_c\lambda M M_\mu^2\times
	\Bigl(\eta^a_{\mu\nu} N^\nu_{mn}\,\tr\bigl( y_m^\dagger y_n^{}(-i\tau^a)\bigr)\Bigr)^2.
\label{ENA2bodyM}
\end{align}
Note that the coefficient of each $a^2_n$ in the one-body potential (\ref{ENA1bodyM}) depends
on the $M_3^2$ and the $M_2^2$ as well as the $M_4^2\equiv M^2$.
Consequently, when we combine this potential with the Coulomb one-body potential for the
instanton radius $a$, the equilibrium radius ends up smaller
than in eq.~(\ref{InstantonRadius}), namely
\be
a_0^4\ =\ {2/5\over\lambda^2 M^2(M_4^2+M_3^2+M_2^2)}\,.
\label{InstantonRadiusM}
\ee
Plugging this radius into eq.~(\ref{ENA2bodyM}) and combining with the Coulomb
two-body interaction term from eq.~(\ref{ECnet}), we obtain the net two-body
force in the background with $M_2\sim M_3\sim M_4\equiv M$:
\be
\E_{net}^{(2)}(m,n)\
=\ {2N_c\over 5\lambda M}\times{1\over|X_m-X_n|^2}\times
\left(\begin{aligned}
	\half\,&
	+\,\tr^2\bigl(y_m^\dagger y_n^{}\bigr)\\
	&+\sum_{\mu=2,3,4}C_\mu\times
		\Bigl(\eta^a_{\mu\nu} N^\nu_{mn}\,\tr\bigl( y_m^\dagger y_n^{}(-i\tau^a)\bigr)\Bigr)^2
	\end{aligned}\right)
\label{E2bodyM}
\ee
where
\be
C_\mu\ \buildrel{\rm def}\over=\ {M_\mu^2\over M_4^2+M_3^2+M_2^2}\,,\quad
C_4\,+\,C_3\,+\,C_2\ =\ 1.
\label{Cdef}
\ee

For instantons located along a straight line, this formula can be simplified a bit using
$N^\nu_{mn}\equiv(\pm1,0,0,0)$ and hence
\be
\eta^a_{\mu\nu} N^\nu_{mn}\,\tr\bigl( y_m^\dagger y_n^{}(-i\tau^a)\bigr)\
=\begin{cases}
	\pm\tr\bigl( y_m^\dagger y_n^{}(-i\tau^1)\bigr) & {\rm for}\ \mu=4,\\
	\pm\tr\bigl( y_m^\dagger y_n^{}(-i\tau^2)\bigr) & {\rm for}\ \mu=3,\\
	\pm\tr\bigl( y_m^\dagger y_n^{}(-i\tau^3)\bigr) & {\rm for}\ \mu=2.
	\end{cases}
\ee
Consequently, the net interaction energy of the 1D lattice is
\begin{align}
\label{EchainM}
\E^{\rm int}_{\rm net}\
\equiv\ \frac12\sum_{n\neq m}\E^{(2)}(n,m)\ &
=\ {N_c\over 5\lambda M D^2}\times\sum_{m\neq n}\frac{Q(m,n)}{(m-n)^2}\\
\intertext{where}
\label{Qdef}
Q(m,n)\
\buildrel {\rm def}\over=\
\half\,+\,\tr^2\bigl(y_m^\dagger y_n^{}\bigr)\,&
+\,C_4\,\tr^2\bigl(y_m^\dagger y_n^{}(-i\tau^1)\bigr)\\
&+\,C_3\,\tr^2\bigl(y_m^\dagger y_n^{}(-i\tau^2)\bigr)\
+\,C_2\,\tr^2\bigl(y_m^\dagger y_n^{}(-i\tau^3)\bigr).
\qquad\nonumber
\end{align}

Without loss of generality we assume $M_4\ge M_3\ge M_2$ and hence $C_4\ge C_3\ge C_2$
--- otherwise we simply re-label the coordinates $x_{2,3,4}$.
Thus, the lowest-energy relative orientation of an instanton pair $(m,n)$ is
$y_m^\dagger y_m^{}=\pm i\tau^3$.
If this orientation is unachievable, the next best bet is a linear combination of
$i\tau^3$ and $i\tau^2$.
If that is also unachievable, the third-best choice would be a linear combination
of all three $i\tau^{1,2,3}$; in $SO(3)$ terms, this corresponds to the $180^\circ$
rotation around a generic axis.
Finally, rotations through other angles would be the least attractive option which the
instantons $\#m$ and $\#n$ would rather avoid --- unless they are
forced to it by interactions with the other instantons between $m$ and $n$.

In this setting, it is not intuitively obvious how to balance the energy cost of
nearest-neighbor interactions versus next-to-nearest neighbors and more distant
instanton pairs.
Instead of intuition, we have performed a computer experiment using a lattice of
200 $SU(2)$ matrices $y_n$.
In each run, we start with the $y_n$ being independent random elements
of the $SU(2)$ group, and then let them evolve towards a
minimum of the energy function (\ref{EchainM}) via the
relaxation method.
That is, we let the $y_n$ evolve with time according to
\be
\frac{d y_n(t)}{dt}\ =\ -K\times\frac{\delta\E^{\rm int}_{\rm net}}{\delta y_n}
\label{Relax}
\ee
where $K$ is a constant mobility factor and the derivative with respect to an $SU(2)$
matrix $y_n$ is defined as
\be
\frac{\delta\E}{\delta y_n}\ \buildrel{\rm def}\over=\
y_n\,\Bigl((-i\vec\tau)\cdot\left.
	\nabla_s\E\bigl(y_n\to y_n(1+i\vec s\cdot\vec\tau)\bigr)
	\right|_{\vec s=0}\Bigr).
\ee
Each run ends when all the $y_n$ seem to converge to
an equilibrium configuration and their derivatives~\eqref{Relax} become very small.
We  made many runs for different $C_3$ and $C_4$ parameters, and here
is what we have found:
\par\bigskip\goodbreak
\begin{itemize}
\item
In all equilibrium configurations for
backgrounds with $C_3<C_4$, all twists $y_n^\dagger y_{n+1}^{}$ between
nearest neighbors have form
\be
y_n^\dagger y_{n+1}^{}\
=\ \cos\psi_n\times i\tau^3\ +\ \sin\psi_n\times i\tau^2\
+\ ({\rm tiny})\times i\tau^1\ +\ ({\rm tiny})\times 1,
\label{ChainMEquilibrium}
\ee
where the tiny coefficients of $1$ and $i\tau^1$ are artefacts of imperfect convergence
to equilibrium (they get smaller when we allow more time for convergence).
\item[$\circ$]
The backgrounds with $M_3=M_4\ \Longrightarrow\ C_3=C_4$
are trickier due to symmetry between $i\tau^1$ and
$i\tau^2$ components of the $SU(2)$ matrices.
(Or all three $i\tau^{1,2,3}$ matrices for $M_2=M_3=M_4\ \Longrightarrow\ C_2=C_3=C_4$.)
In equilibrium configurations for such symmetric backgrounds, the nearest-neighbor
twists $y_n^\dagger y_{n+1}^{}$ generally do have $i\tau^1$ components.
However, all such components can be eliminated by a suitable global symmetry of the system,
$y_n\to y_n\times G$, same $G\in SU(2)$ for all $y_n$, and then all
the nearest-neighbor twists $y_n^\dagger y_{n+1}^{}$ take the form~(\ref{ChainMEquilibrium}).
	
\item
For most nearest-neighbor pairs, the angles $\psi_n$ in eqs.~(\ref{ChainMEquilibrium})
take values $\pm\varphi\pmod\pi$ where $\varphi$ depends on the $C_{2,3,4}$ parameters.
Moreover, the $\pm$ sign here tends to alternate between odd and even $n$.
However, some pairs do not follow these rules.
\item[{}]
In general, our runs end up with patterns of $(-1)^n\psi_n$ that look  like this:
\be
\psset{unit=9.5mm}
\begin{pspicture}(-0.5,-1.5)(15,+2)
\psline[linewidth=0.5pt]{->}(0,0)(14.5,0)
\uput[0](14.5,0){$n$}
\psline[linewidth=0.5pt]{>->}(0,-1.5)(0,+1.5)
\uput[90](0,1.5){$(-1)^n\psi_n$}
\psline[linestyle=dotted,linewidth=1pt](0,+1)(14,+1)
\psline[linestyle=dotted,linewidth=1pt](0,-1)(14,-1)
\uput[180](0,+1){$+\varphi$}
\uput[180](0,-1){$-\varphi$}
\pscircle*(0.5,+1){0.1}
\pscircle*(1.0,+1){0.1}
\pscircle*(1.5,+1){0.1}
\pscircle*(2.0,+1){0.1}
\pscircle*(2.5,+1){0.1}
\pscircle*(3.0,+1){0.1}
\pscircle*(3.5,+1){0.1}
\pscircle*(4.0,+0.9){0.1}
\pscircle*(4.5,+0.3){0.1}
\pscircle*(5.0,-0.3){0.1}
\pscircle*(5.5,-0.9){0.1}
\pscircle*(6.0,-1){0.1}
\pscircle*(6.5,-1){0.1}
\pscircle*(7.0,-1){0.1}
\pscircle*(7.5,-1){0.1}
\pscircle*(8.0,-1){0.1}
\pscircle*(8.5,-1){0.1}
\pscircle*(9.0,-1){0.1}
\pscircle*(9.5,-0.8){0.1}
\pscircle*(10.0,0){0.1}
\pscircle*(10.5,+0.8){0.1}
\pscircle*(11.0,+1){0.1}
\pscircle*(11.5,+1){0.1}
\pscircle*(12.0,+1){0.1}
\pscircle*(12.5,+1){0.1}
\pscircle*(13.0,+1){0.1}
\pscircle*(13.5,+1){0.1}
\pscircle*(14.0,+1){0.1}
\end{pspicture}
\ee
Physically, this corresponds to the 1D lattice having two degenerate ground states,
one with $\psi_n=+(-1)^n\varphi$ and the other with $\psi_n=-(-1)^n\varphi$.
In our numerical runs, some parts of the lattice converge to one of these states
while other parts converge to the other state;
altogether, we end up with several domains separated by walls.
In a perfect simulation, the walls would eventually move towards each other and annihilate,
so the whole lattice would end up in the same ground state.
But this process is so slow that our numeric simulation stops before it barely begins,
thus we always end up with multiple domains instead of a single ground state for
the whole lattice.

\item[$\star$]
The bottom line of our numerical simulations is that the ground state of the 1D
instanton lattice always has a link-periodic instanton orientations~(\ref{LinkPeriodic})
(up to a global symmetry, if any) for a
periodicity angle $\varphi$ that depends on ratios of parameters $M_2:M_3:M_4$.
Curiously,  it actually depends only on the $M_2:M_3$ ratio regardless of the $M_4$
(as long as $M_4\ge M_3\ge M_2$).
\end{itemize}

Once we know the ground state of the instanton orientations in a 1D lattice is
link-periodic, we may calculate the periodicity angle $\varphi$ analytically.
Indeed, applying eqs.~(\ref{BigFamilyTwists}) to the link-periodic array~(\ref{LinkPeriodic})
of instanton orientations, we have:
\begin{align}
\text{For even}\ n-m,\quad y_m^\dagger y_n^{}\ &
=\ \cos((n-m)\varphi)\times i^{n-m}\
	+\ \sin((n-m)\varphi)\times i^{n+m+1}\tau^1
	\nonumber\\
{}\Longrightarrow\ Q(m,n)\ &
=\ \half\ +\ 4\cos^2((n-m)\varphi)\ +\ 4C_4\,\sin^2((n-m)\varphi)\\
&=\ \tfrac52\ +\ 2C_4\ +\ 2(1-C_4\,=\,C_3+C_2)\times\cos(2(n-m)\varphi),
\nonumber\\[5pt]
\text{For odd}\ \ n-m,\quad y_m^\dagger y_n^{}\ &
=\ \cos((n-m)\varphi\pm\theta)\times i^{n-m}\tau^3\
	+\ \sin((n-m)\varphi\pm\theta)\times i^{n+m}\tau^2
	\quad\nonumber\\
{}\Longrightarrow\ Q(m,n)\ &
=\ \half\ +\ 4C_2\,\cos^2((n-m)\varphi\pm\theta)\ +\ 4C_3\,\sin^2((n-m)\varphi\pm\theta)
\nonumber\\
&=\ \half\ +\ 2(C_3+C_2)\ -\ 2(C_3-C_2)\times\cos(2(n-m)\varphi\pm2\theta),
\end{align}
where the $\pm$ sign is $(-1)^n$.
To calculate average interaction energy per instanton in the lattice, we should average
$Q(n,m)$ between odd and even $n$ for fixed $\ell=n-m$; indeed
\be
\E^{\rm interaction}_{\rm per\,instanton}\
=\ {N_c\over 5\lambda MD^2}\times \left\langle
	\,\smash{\sum_{m\neq n}^{{\rm fixed}\, n}}\vphantom{\sum}
	\frac{Q(m,n)}{(n-m)^2}
	\right\rangle^{\rm average}_{{\rm over}\,n}\
=\ {N_c\over 5\lambda MD^2}\times\sum_{\ell\neq 0}
\frac{\overline Q(\ell)}{\ell^2}\,.
\ee
Thus,
\be
\begin{split}
\text{for even}\ \ell,\quad\overline Q(\ell)\ &
=\ \tfrac52\ +\ 2C_4\ +\ 2(C_3+C_2)\cos(2\ell\varphi),\\
\text{for odd}\ \ \ell,\quad\overline Q(\ell)\ &
=\ \tfrac12\ +\ 2(C_3+C_2)\ -\ 2(C_3-C_2)\times\cos(2\theta)\times\cos(2\ell\varphi),
\end{split}
\ee
and therefore
\begin{align}
\E^{\rm interaction}_{\rm per\,instanton}\ &
=\ {N_c\over 5\lambda MD^2}\times\left(\begin{aligned}
	&(\tfrac52+2C_4)\times
	\left(\sum_{{\rm even}\,\ell\neq0}{1\over\ell^2}\,
		=\,\frac{\pi^2}{12}\right)\\
	&+\ 2(C_3+C_2)\times
	\left(\sum_{{\rm even}\,\ell\neq0}{\cos(2\ell\varphi)\over\ell^2}\,
		=\,\frac{\pi^2}{12}\,-\,\pi|\varphi|\,+\,2\varphi^2\right)\\
	&+\ (\half+2C_3+2C_2)\times
	\left(\sum_{{\rm odd}\,\ell}{1\over\ell^2}\,
		=\,\frac{\pi^2}{4}\right)\\
	&-\ 2(C_3-C_2)\cos(2\theta)\times
	\left(\sum_{{\rm odd}\,\ell}{\cos(2\ell\varphi)\over\ell^2}\,
		=\,\frac{\pi^2}{4}\,-\,\pi|\varphi|\right)\\
	&\quad\langle\!\langle\,\text{where the sums are evaluated assuming}\
	|\varphi|<\tfrac\pi2\,\rangle\!\rangle
	\end{aligned}\right)\nonumber\displaybreak[0]\\[15pt]
&=\ {N_c\over 5\lambda MD^2}\times\left(\begin{aligned}
	\frac{\pi^2}{2}\times &
	\Bigl(1\,+\,C_3\,+\,C_2\,-\,(C_3-C_2)\cos(2\theta)\Bigr)\\
	-\ 2\pi|\varphi|\times &
	\Bigl( C_3\,+\,C_2\,-\,(C_3-C_2)\cos(2\theta)\Bigr)\\
	+\ 4\varphi^2\times &
	\Bigl(C_3\,+\,C_2)\\
	\end{aligned}\right).
\end{align}
Minimizing this expression WRT the $\varphi$ and $\theta$ produces
4 degenerate minima, namely
\be
\vcenter{\openup 1\jot \ialign{
	#\quad\hfil & $\displaystyle{\varphi\ =\ #\,}$,\quad\hfil & $\theta\ =\ #$,\hfil\cr
	(1) & +\frac\pi2\times\frac{C_2}{C_2+C_3} & 0\cr
	(2) & -\frac\pi2\times\frac{C_2}{C_2+C_3} & 0\cr
	(3) & -\frac\pi2\times\frac{C_3}{C_2+C_3} & \tfrac\pi2\cr
	(4) & +\frac\pi2\times\frac{C_3}{C_2+C_3} & \tfrac\pi2\cr
	}}
\ee
however the last two minima are physically equivalent to the first two.
Thus, in agreement with our computer `experiments', the 1D instanton lattice
has 2 degenerate ground states related by the $\varphi\to-\varphi$ symmetry.
The value of $|\varphi|$ also agrees with our `experimental data':
\be
\theta\ =\ 0,\quad
\varphi\ =\ \pm\frac\pi2\times\frac{M_2^2}{M_2^2+M_3^2}\quad
\text{regardless of the}\ M_4^2
\label{BestPhi}
\ee
(as long as $M_4\ge M_3\ge M_2$).
In particular, for rational ratios of $M_2^2/M_3^2$, the angle $\varphi$ is rational
(in units of $\pi$) and the instanton orientations $y_n$ become periodic.
Specifically, for
\be
\frac{M_2^2}{M_3^2}\ =\ \frac{p}{q}\,,\quad p<q,\quad \mathop{\rm gcd}(p,q)=1,
\ee
the $y_n$ (modulo signs) repeat with period $2(q+p)$ while spanning the $D_{2(q+p)}$
dihedral subgroup of the $SO(3)\cong SU(2)/\Z_2$, for example see
eq.~(\ref{Dihedral}) for $p=1$ and $q=k-1$.

Two particularly interesting $M_2^2:M_3^2$ ratios need special handling,
$M_2^2=M_3^2$ and $M_2^2\ll M_3^2$.
For $M_2=M_3$ the background has a rotational symmetry in the $x^{2,3}$ plane; for the
1D instanton lattice, this translates into the $U(1)$ symmetry between the $i\tau^2$
and $i\tau^3$ directions in the $SU(2)$.
Consequently, instead of two discrete ground states (\ref{BestPhi}) we have a continuous family:
\be
\varphi\ =\ \frac\pi2\,,\quad \theta\ =\ \rm anything.
\ee
For all configurations in this family, the instanton orientations $y_n$ (modulo signs)
repeat with period~4 while spanning the Klein groups $\Z_2\times\Z_2$ of $180^\circ$
rotations around 3 mutually $\perp$ axes.
For $\theta=-\frac\pi4$ the $y_n$ are spelled out in eq.~(\ref{Klein});
for other values of $\theta$ we have similar cycles related by the $U(1)\subset SU(2)$
symmetry.

Finally, for the very asymmetric background with $M_2^2\ll M_3^2$, the two ground states
(\ref{BestPhi}) become indistinguishable as $\varphi\to\pm0$.
In this limit, the instanton orientations form the anti-ferromagnetic order~(\ref{AntiFerro}).

\section{Instanton Zigzags}
In this section we consider the most likely
first step in the transition between 1D and 2D
instanton lattices --- a lattice with one infinite dimension $x^1$
while the other dimension $x^2$ has just two layers.
Since the instantons repel each other, the two layers should be staggered in the
$x^1$ direction, so the whole lattice looks like a single zigzag-shaped chain:
\be
\begin{pspicture}(-7,-1.2)(+7.4,+1.6)
\psline[linewidth=0.5pt]{>->}(-7,0)(+7,0)
\uput[r](7,0){$x^1$}
\psline[linewidth=0.5pt]{>->}(0,-1.2)(0,+1.2)
\uput[u](0,1.2){$x^2$}
\psline[linestyle=dotted](-6.5,+0.4)(+6.5,+0.4)
\psline[linestyle=dotted](-6.5,-0.4)(+6.5,-0.4)
\multido{\n=-6+2}{7}{\pscircle*[linecolor=blue](\n,+0.4){0.1}}
\multido{\n=-5+2}{6}{\rput(\n,0){%
	\psline[linecolor=magenta,linewidth=0.5pt](-1,+0.4)(0,-0.4)(+1,+0.4)
	\pscircle*[linecolor=blue](0,-0.4){0.1}
	}}
\end{pspicture}
\label{ZigzagPicture}
\ee
In such a zigzag, the instanton $\#n$ has its center at
\be
X_n^1\ =\ n\times D,\quad
X_n^2\ =\ (-1)^n\times\epsilon,\quad
X_n^3\ =\ X_n^4\ =\ 0,
\label{Zigzag}
\ee
for some parameters $D$ and $\epsilon$; we shall refer to $D$ as the {\it lattice spacing}
and to $\epsilon$ as the {\it zigzag amplitude.}

To make the instantons form a zigzag lattice, we use the same background
$$
{8\pi ^2\over g^2_5(x_2,x_3,x_4)}\ =\ N_c\lambda M\Bigl(
	1\,+\,M^2 x_4^2\,+\,M_3^2x_3^2\,+\,M_2^2x_2^2\,+\,O(M^4x^4)\Bigr),
\eqno(\ref{G5Curved})
$$
we have used in the last section to make a straight chain but increase the instanton
density $\rho=1/D$ until the repulsion between the instantons pushes them away from
their neighbors in some transverse direction.
For $M_2<M_3\le M_4$, the initial breakout from the straight line
is going to be in the $\pm x^2$ directions, with opposite signs for the nearest neighbors
(to increase the distance between them), hence the zigzag geometry~(\ref{Zigzag}).
For higher densities --- and hence stronger repulsion between the instantons ---
we will get more complicated lattices: multiple layers in the $x^2$ direction, and
eventually breaking out into the $x^3$ and $x^4$ dimensions.
However, the first geometry right after the straight 1D lattice ought to be the zigzag.

Clearly, the above heuristic argument is not a proof.
The proof --- or disproof --- will come from a computer simulation of multi-instanton systems
where the instantons are allowed to move independently in all directions as well as to change
their orientations.
We shall describe such a simulation --- and its results --- in our next paper \cite{OurNextPaper}
in this project.
Meanwhile, in the present paper we shall simply assume that the instanton centers from a
zigzag-shaped chain \eqref{ZigzagPicture} and focus on the instantons' orientations.

Since our formulae for the instanton interactions presume distances $|X_m-X_n|$ much larger
than the instanton radius $a$, we need the transition from a straight 1D chain to a zigzag
to happen at a critical lattice spacing $D_c\gg a$,
which calls for
\be
M_2\ \ll\ M_4\ \equiv\ M.
\label{ZigHierarchy}
\ee
Indeed, consider an instanton that have moved a little bit away from the $x^1$ axis
in the $x^2$ direction.
The restoring force on this instanton due to $x^2$ dependence of the gauge coupling is
\be
F_{\rm rest}\ = -X_2\times 2N_c\lambda MM_2^2
\ee
while the  repulsion from the neighboring instantons pushes it further away with
the net force
\be
F_{\rm rep}\
=\ -\frac{\partial}{\partial X_2} O\left({N_c\over\lambda M}\times{1\over O(D^2)+X_2^2}\right)\
=\ +{X_2\over D^4}\times O\left({N_c\over\lambda M}\right).
\ee
At the critical lattice spacing, the two forces balance each other, hence
\be
D_c^4\ =\ {O(1)\over\lambda^2 M^2 M_2^2}\,.
\ee
At the same time,
\be
a^4\ =\ {2/5\over\lambda^2 M^2(M^2_4+M_3^2+M_2^2)}\ =\ {O(1)\over\lambda^2 M^2 M_4^2}\,,
\ee
so to assure $D_c\gg a$ we need $M_2\ll M_4\equiv M$.

Note however that while we need $M_2\ll M_4$, there are no restrictions on the $M_3$
parameter except $M_2\le M_3\le M_4$.
Consequently, we may consider zigzags in a variety of backgrounds with
different $M_3/M_4$ ratios --- ranging from $M_2=M_3\ll M_4$ to $M_2\ll M_3=M_4$ ---
and hence different ways in which the  forces between two instantons depend on their
relative orientations.
For small $M_3\ll M_4$, the two-instanton interactions are spelled out in eq.~(\ref{KZ})
while for finite $M_3/M_4$ ratios we need to use a more complicated formula~(\ref{E2bodyM})
(with $C_2\approx0$ but finite $C_3$ and $C_4$).
For a zigzag where all the instantons lie in the $x^{1,2}$ plane and hence
all the $N_{mn}^\mu$ have form $N^\mu_{mn}=(*,*,0,0)$, we may simplify this formula
using the identity
\be
\sum_{\mu=2,3}\Bigl(\eta^a_{\mu\nu} N^\nu_{mn}
	\tr\bigl(y_m^\dagger y_n^{}(-i\tau^a)\bigr)\Bigr)^2\
=\,\sum_{a=1,2} \tr^2\bigl(y_m^\dagger y_n^{}(-i\tau^a)\bigr)\quad
({\rm for} N^\nu_{mn}=(*,*,0,0)).
\ee
Consequently, eq.~(\ref{E2bodyM}) can be rearranged as
\begin{subequations}
\label{E2D}
\begin{align}
\E^{\rm int}_{\rm net}[{\rm zigzag}]\ &
=\ {N_c\over5\lambda M}\sum_{m\neq n}\frac{Q_z(m,n)}{|X_m-X_n|^2}\\
\text{where}\quad Q_z(m,n)\ &
=\ \half\ +\ \tr^2\bigl(y_m^\dagger y_n^{}\bigr)\
+\ C_3\sum_{a=1,2} \tr^2\bigl(y_m^\dagger y_n^{}(-i\tau^a)\bigr)\
\nonumber\\
&\hskip2.6em+\ (1-2C_3)\times\tr^2\bigl(y_m^\dagger y_n^{}
	(-i\vec\tau\cdot\vec N_{mn})\bigr).
\end{align}
\end{subequations}
Note that in general, the force between two instantons is not central --- it
depend not only on the distance between the instantons and
their relative orientation $y_m^\dagger y_n^{}$ of their isospins,
but  also on the direction $\vec N_{mn}$ of their separation $\vec X_n-\vec X_n$
in space.
However, in the background with $M_3=M_4$ --- which is the background
we have used in our previous paper \cite{Kaplunovsky:2012gb} (albeit in different notations) ---
$C_3=\half$, the last term in eq.~(\ref{E2D}b) goes away,
and the force becomes central (but isospin-dependent).

The forces between instantons will be important in our next article
\cite{OurNextPaper} where we shall
allow them to move in the $x^{1,2}$ plane seeking the lowest-energy lattice.
For the moment, we simply assume that the instantons somehow form a zigzag of some particular
lattice spacing $D$ and amplitude $\epsilon$ and focus on their orientation moduli $y_n$.
Our immediate goal is to find the lowest-energy configuration of the orientations $y_n$
as a function of the $\epsilon/D$ and $M_3/M_4$ ratios.

\begin{figure}[ht]
\def\colordot[#1](#2){\pscircle*[linecolor=#1](#2){2pt}}
\psset{unit=10cm,linewidth=1pt}
\centering
\begin{pspicture}(-0.1,-0.1)(1.1,1.1)
\psframe(0,0)(1.0,1.0)
\input{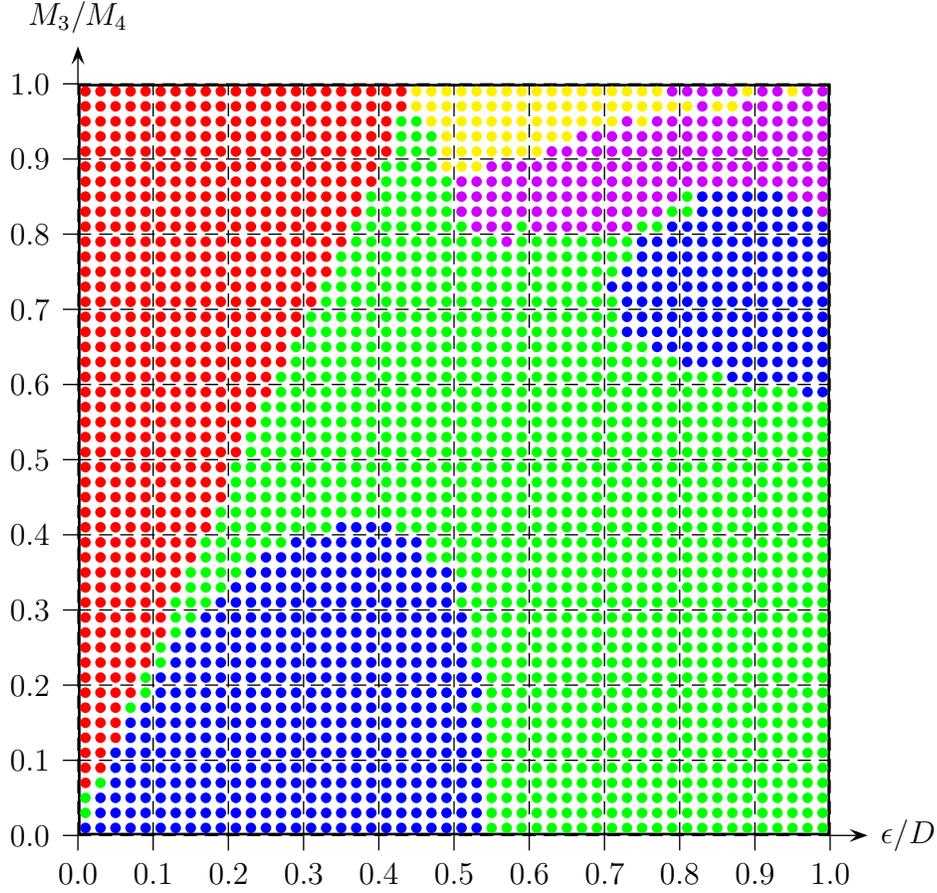}
\psset{unit=1cm,linewidth=0.5pt,arrowscale=2}
\multido{\n=0+1}{11}{%
	\psline(-0.2,\n)(0,\n)
	\psline[linestyle=dashed](0,\n)(10,\n)
	\uput[l](-0.2,\n){0.\n}
	\psline(\n,-0.2)(\n,0)
	\psline[linestyle=dashed](\n,0)(\n,10)
	\uput[d](\n,-0.2){0.\n}
	}
\uput*[l](-0.2,10){\enspace 1.0}
\uput*[d](10,-0.2){1.0}
\psline{->}(10,0)(10.5,0)
\uput[r](10.5,0){$\epsilon/D$}
\psline{->}(0,10)(0,10.5)
\uput[u](0,10.5){$M_3/M_4$}
\end{pspicture}
\caption{%
    Rough distribution of instanton orientation patterns in the zigzag parameter space.
    This diagram is obtained from a numerical relaxation method.
    }
\label{ZigzagRoughDiagram}
\end{figure}

{\it A priori,} we do not know if the orientations --- or rather the relative
orientations $y_n^\dagger y_{n+1}$ of the neighbor instantons --- follow any
particular pattern or patterns.
To find such pattern(s) we used numerical simulations:
starting with completely random $y_n$,
we let them evolve towards a minimum of the energy
function~(\ref{E2D}) according to eqs.~(\ref{Relax});
when the evolution seems to stop, we look at the relative orientations
of the nearest neighbors $y_n^\dagger y_{n+1}$ to see if they form any recognizable
pattern.
Repeating this procedure for different combination of the  $\epsilon/D$ and $M_3/M_4$
parameters, we saw five distinct patterns of orientations (or rather four distinct
patterns and one confused mess).
Figure~\ref{ZigzagRoughDiagram} 
diagrams the distribution
of these patterns in the parameter space.

\smallskip
Here is the color key to this figure:
\begin{itemize}
\item[$\red\bullet$]
The red dots on  figure~\ref{ZigzagRoughDiagram} denote the anti-ferromagnetic pattern ({\red AF})
of instanton orientations in which
the nearest neighbors  always differ by a $180^\circ$ rotation around
the third axis,
\be
\label{RedPattern}
y_n\ =\left.\begin{cases}
	\pm1 & {\rm for\ even}\ n,\\
	\pm i\tau_3 & {\rm for\ odd}\ n,
	\end{cases}\right\},\qquad
{\rm same}\ y_n^\dagger y_{n+1}^{}\ =\ i\tau_3\ {\rm for\ all}\ n.
\ee

\goodbreak
\item[$\yellow\bullet$]
The yellow dots denote another abelian pattern ({\yellow AB})
in which all nearest neighbors
differ by the same $U(1)\subset SU(2)$ rotation,
but now the rotation angle $\rm is<180^\circ$,
\be
\label{YellowPattern}
{\rm same}\ y_n^\dagger y_{n+1}^{}\
=\ \exp\bigl(\tfrac{i}{2}\phi\tau_3\bigr)\ {\rm for\ all}\ n,\quad
0<\phi<\pi.
\ee

\item[$\blue\bullet$]
The blue dots denote a non-abelian link-periodic pattern ({\blue NA1})
in which the relative rotation between nearest neighbors is always
through a $180^\circ$ angle, but the direction of rotation alternates between
two different axes in the (12) plane, one axis for the odd-numbered instantons
and the other for the even-numbered.
In $SU(2)$ terms,
\be
\label{BluePattern}
\begin{split}
y_{2k}^\dagger y_{2k+1}^{}\ &
=\ \exp\bigl(\tfrac{i\pi}{2}\,\vec n_e\cdot\vec\tau\bigr)\
=\ i\vec n_e\cdot\vec\tau\
=\ +iA\tau_1\ +\ iB\tau_2\,,\\
y_{2k+1}^\dagger y_{2k+2}^{}\ &
=\ \exp\bigl(\tfrac{i\pi}{2}\,\vec n_o\cdot\vec\tau\bigr)\
=\ i\vec n_o\cdot\vec\tau\
=\ +iA\tau_i\ -\ iB\tau_2\,,
\end{split}
\ee
for some $A,B\neq0$ ($A^2+B^2=1$).

\item[$\green\bullet$]
The green dots denote another non-abelian link-periodic pattern ({\green NA2}).
Again, the relative rotation between nearest neighbors is always
through a $180^\circ$ angle, but the direction of rotation alternates between
two different axes.
However, this time the two axes no longer lie within the (12) plane, thus
\be
\label{GreenPattern}
\begin{split}
y_{2k}^\dagger y_{2k+1}^{}\ &
=\ iA\tau_1\ +\ iB\tau_2\ +\ iC\tau_3\,,\\
y_{2k+1}^\dagger y_{2k+2}^{}\ &
=\ iA\tau_1\ -\ iB\tau_2\ -\ iC\tau_3\,,\\
\end{split}
\ee
where $A,B,C\ \rm all\neq0$ ($A^2+B^2+C^2=1$).

\item[$\purple\bullet$]
Finally, the purple dots denote a confused non-abelian pattern {\purple NAX}
in which the nearest neighbors
seem to differ by rotations through angles $\phi<180^\circ$ around 2 alternating axes.
In $SU(2)$ terms, we see something like
\be
\label{PurplePattern}
\begin{split}
y_{2k}^\dagger y_{2k+1}^{}\ &
=\ +iA\tau_1\ +\ iB\tau_2\ +\ iC\tau_3\ +\ D,\\
y_{2k+1}^\dagger y_{2k+2}^{}\ &
=\ -iA\tau_1\ +\ iB\tau_2\ +\ iC\tau_3\ +\ D\,,\\
\end{split}
\ee
where $A,B,C,D\ \rm all\neq0$ ($A^2+B^2+C^2+D^2=1$),
but this picture is rather noisy:
the coefficients $A,B,C,D$  fluctuate from $n$ to $n$ and
from run to run much stronger than their analogues in the other 4 patterns
(\ref{RedPattern})--(\ref{GreenPattern}),
and even the relative signs between $A,B,C,D$ flip every few lattice sites
--- the long-range correlations are quite poor.
\end{itemize}

Physically, the disorder in the purple part of the phase diagram~\ref{ZigzagRoughDiagram}
indicates a nearby first-order phase transition between
incompatible orientation patterns.
Near such a transition, our relaxation method fails to converge to a homogeneous phase
that happens to have lowest energy for the $(\epsilon/D,M_3/M_4)$ parameters in question.
Instead, it leads to a mishmash of short domains of two nearly-degenerate phases;
in fact, the domains are only a few lattice sites long, so they are
significantly distorted by the boundary effects.
The resulting mess is not even a local minimum of the zigzag's energy function,
but once our simulation reaches this point,
further evolution of the instanton orientations $y_n$ becomes very slow.
Given very long time (and a better numeric algorithm) we would eventually see
the domains of each phase growing longer and eventually merging into a single phase
--- but in our calculations we have simply run out of time and patience before this
can happen.

The bottom line of our numeric analysis is that the instanton zigzag has 4 or 5
distinct orientation phases --- we can not tell if  {\purple NAX} is a real phase
or an artefact of mixed domains of other phases that we could not resolve.
To answer that question --- and also to get a better map of phase boundaries ---
we need to go back to analytic work.

Our starting point is the {\sl qualitative} result of our numerical analysis:
all  5 patterns (\ref{RedPattern})--(\ref{PurplePattern}) we have seen
--- or might have seen ---
can be fit into a single anzatz:
\begin{align}
y_n^{}\ &
=\ \exp\bigl(in\,\tfrac\phi2\,\tau_2\bigr)
\times\exp\Bigl(i(\tfrac\alpha2+(-1)^n\tfrac\beta2)\tau_1\Bigr),
\label{ZigYAnsatz}\\
y_n^\dagger y_{n+1}^{}\ &
=\ \cos\tfrac\phi2\,\cos\beta\times1\
-\ (-1)^n\cos\tfrac\phi2\,\sin\beta\times i\tau_1
\nonumber\\
&\qquad+\ \sin\tfrac\phi2\,\cos\alpha\times i\tau_2\
+\ \sin\tfrac\phi2\,\sin\alpha\times i\tau_3
\label{ZigLinkAnzatz}
\end{align}
for some angles $\phi,\alpha,\beta$.
Indeed, the {\red AF} phase obtains for \cline\red{\phi=\pi,\ \alpha=\tfrac\pi2};
the  {\yellow AB} phase obtains for
\cline\yellow{0<\phi<\pi,\ \alpha=\tfrac\pi2,\ \beta=0};
the  {\blue NA1} phase obtains for
\cline\blue{0<\phi<\pi,\ \alpha=0,\ \beta=\tfrac\pi2};
the  {\green NA2} phase obtains for
\cline\green{0<\phi<\pi,\ 0<\alpha<\tfrac\pi2,\ \beta=\tfrac\pi2};
and the {\purple NAX} phase --- if it exists at all --- obtains for generic values of
all three angles, \cline\purple{0<\phi<\pi,\ 0<\alpha<\tfrac\pi2,\ 0<\beta<\tfrac\pi2}.

{\it Given} the ansatz \eqref{ZigYAnsatz}, we may calculate the zigzag's net energy
as an analytic function of the angles $\phi,\alpha,\beta$ and parameters $(\epsilon/D,M_3/M_4)$.
The calculation is presented in Appendix~A, and the result is
\begin{align}
\label{ZigNetEnergy}
\E^{\rm zigzag}_{\rm per\,instanton}\ &
=\ {\pi^2N_c\over 20\lambda MD^2}\times{\cal F}(\phi,\alpha,\beta;C_3,\epsilon/D)
+\ {N_c\lambda MM_2^2\times\epsilon^2},\\[5pt]
\noalign{\penalty 0}
\begin{split}
\label{Fdef}
{\rm for}\quad{\cal F}\ =\ \frac32\ &
+\ \Bigl(-1\,+\,C_3(\cos^2\!\alpha\cos^2\!\beta+\sin^2\!\alpha\sin^2\!\beta)\Bigr)
	\times\Sigma_0(\phi)\\
&+\ \Bigl(\frac52+\cos^2\!\alpha-\sin^2\!\beta\Bigr)\times\Sigma_1(\epsilon/D)\
+\ \Bigl(2-\cos^2\!\alpha-\sin^2\!\beta\Bigr)\times\Sigma_2(\phi,\epsilon/D)\\
&+\ (1-2C_3)(\sin^2\!\beta-\cos^2\!\alpha)\times\Sigma_3(\epsilon/D)\\
&+\ (1-2C_3)(\sin^2\!\beta+\cos^2\!\alpha)\times\Sigma_4(\phi,\epsilon/D)\\
&-\,2(1-2C_3)\cos\alpha\sin\beta\times\Sigma_5(\phi,\epsilon/D),
\end{split}
\end{align}
\par\goodbreak\noindent
where
\begin{subequations}
\label{Sigmas}
\begin{align}
\Sigma_0\ &=\ {4\phi(\pi-\phi)\over\pi^2}\,,\\
\Sigma_1\ &=\ {\tanh(\pi\epsilon/D)\over (\pi\epsilon/D)}\,,\\
\Sigma_2\ &=\ {\sinh((\pi-2\phi)\epsilon/D)\over (\pi\epsilon/D)\cosh(\pi\epsilon/D)}\,,\\
\Sigma_3\ &=\ {1\over\cosh^2(\pi\epsilon/D)}\,,\\
\Sigma_4\ &=\ {\cosh(2\phi\epsilon/D)\over\cosh^2(\pi\epsilon/D)}\
-\ {2\phi\over\pi}\times{\cosh((\pi-2\phi)\epsilon/D)\over\cosh(\pi\epsilon/D)}\,,\\
\Sigma_5\ &=\ {\sinh(2\phi\epsilon/D)\over\cosh^2(\pi\epsilon/D)}\
+\ {2\phi\over\pi}\times{\sinh((\pi-2\phi)\epsilon/D)\over\cosh(\pi\epsilon/D)}\,,
\end{align}
\end{subequations}

In  Appendix A we also minimize $\cal F$ --- and hence the net energy per instanton ---
with respect to the orientation parameters $\alpha$, $\beta$, and $\phi$ for fixed $C_3$
and $\epsilon/D$.
In particular, we show that the minimum never lies at generic values of the angles $\phi,\alpha,\beta$
--- which would correspond to the {\purple NAX} phase.
This means that {\it the {\purple NAX} phase does not really exist} --- its appearance
in the purple region on the figure~\ref{ZigzagRoughDiagram}
is an artefact of poor convergence near a first-order phase transition (see below).

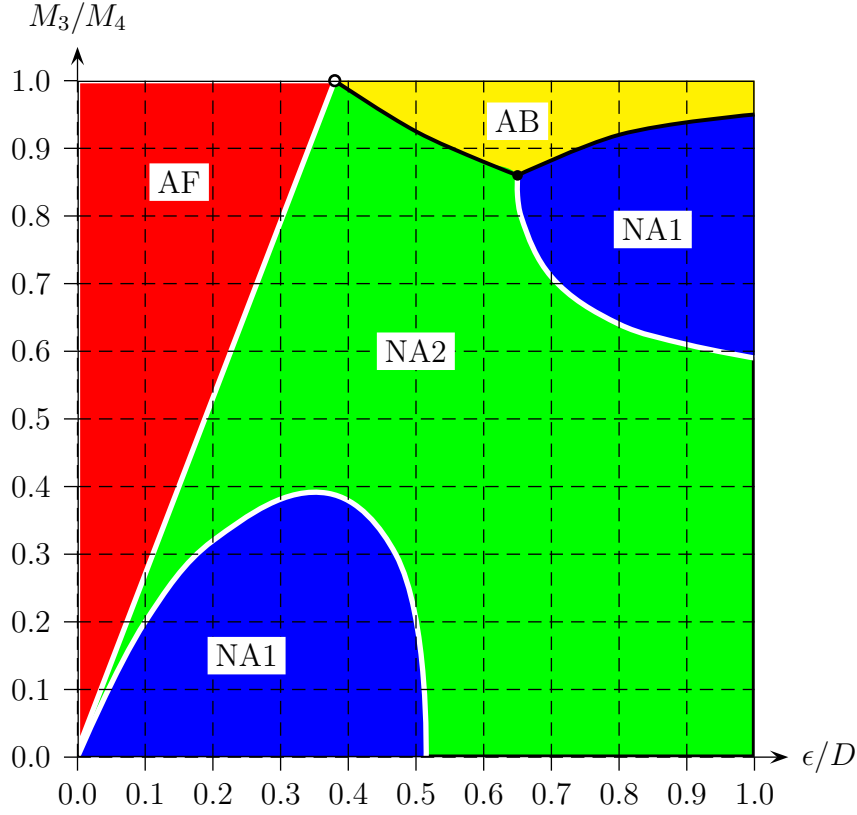
\begin{figure}[t]
\centering
\psset{unit=9mm,linewidth=1pt}
\begin{pspicture}(-1,-1)(11,11)
\psframe[fillstyle=solid,fillcolor=green](0,0)(10,10)
\pspolygon[fillstyle=solid,fillcolor=red,linecolor=white,linewidth=2pt]%
	(0,0.1)(0,10)(3.8,10)
\pscurve[fillstyle=solid,fillcolor=blue,linecolor=white,linewidth=2pt]%
	(0,0)(1,2)(2,3.2)(3.7,3.9)(4.7,3)(5.15,0)
\pscustom[fillstyle=solid,fillcolor=blue,linestyle=none]{%
	\pscurve(10,9.5)(8,9.2)(6.5,8.6)
	\pscurve(6.5,8.6)(6.5,8.3)(6.55,8)(7,7.1)(8,6.4)(9,6.1)(10,5.9)
	\psline(10,5.9)(10,9.5)
	}
\pscustom[fillstyle=solid,fillcolor=yellow,linestyle=none]{%
	\pscurve(10,9.5)(8,9.2)(6.5,8.6)
	\pscurve(6.5,8.6)(6,8.8)(5,9.25)(3.8,10)
	\psline(3.8,10)(10,10)(10,9.5)
	}
\pscurve[linecolor=white,linewidth=2pt](6.5,8.6)(6.5,8.3)(6.55,8)(7,7.1)(8,6.4)(9,6.1)(10,5.9)
\pscurve[linecolor=black,linewidth=1.5pt](10,9.5)(8,9.2)(6.5,8.6)
\pscurve[linecolor=black,linewidth=1.5pt](6.5,8.6)(6,8.8)(5,9.25)(3.8,10)
\pscircle[fillcolor=white,fillstyle=solid](3.8,10){2.5pt}
\pscircle*(6.5,8.6){2pt}
\psset{linewidth=0.5pt,arrowscale=2}
\psframe(0,0)(10,10)
\multido{\n=0+1}{10}{%
	\psline(-0.2,\n)(0,\n)
	\psline[linestyle=dashed](0,\n)(10,\n)
	\uput[l](-0.2,\n){0.\n}
	\psline(\n,-0.2)(\n,0)
	\psline[linestyle=dashed](\n,0)(\n,10)
	\uput[d](\n,-0.2){0.\n}
	}
\psline(-0.2,10)(0,10)
\uput[l](-0.2,10){1.0}
\psline(10,0)(10,-0.2)
\uput[d](10,-0.2){1.0}
\psline{->}(10,0)(10.5,0)
\uput[r](10.5,0){$\epsilon/D$}
\psline{->}(0,10)(0,10.5)
\uput[u](0,10.5){$M_3/M_4$}
\rput*(1.5,8.5){AF}
\rput*(2.5,1.5){NA1}
\rput*(8.5,7.8){NA1}
\rput*(5,6){NA2}
\rput*(6.5,9.4){AB}
\end{pspicture}
\caption{%
	Phase diagram of the instanton orientation patterns in the zigzag parameter space.
	This diagram obtains from the  mostly-analytical calculation in Appendix~A.
	The lines separating different phases indicate the order of the phase transition:
	a white line for a second-order transition and a black line for a first-order transition.
	}
\label{ZigzagDiagram}
\end{figure}

On the other hand, for appropriate values of the $\epsilon/D$ and $M_3/M_4$ parameters,
the zigzag energy (\ref{ZigNetEnergy}) has global minima corresponding to the
other 4 phases (\ref{RedPattern})--(\ref{GreenPattern}) of instanton orientations.
Figure~\ref{ZigzagDiagram} on the next page
shows their distributions in the parameter space.
Curiously, one of the non-abelian phases --- the {\blue NA1} --- appears in two different regions
of the parameter space --- one at low $M_3/M_4$ and small zigzag amplitudes, and
the other at higher $M_3/M_4$ and higher amplitudes ---
while the intermediate region is occupied by the different non-abelian phase {\green NA2}.
Both phases transitions between the two non-abelian phases are  second-order.
Likewise, the transition between the non-abelian {\green NA2} and the abelian anti-ferromagnetic
phase {\red AF} is also  second-order.

On the other hand, the transitions between the other abelian {\yellow AB} phase and the
non-abelian phases {\blue NA1} and {\green NA2} are first-order.
In fact, these are precisely the first-order transitions that confused our numeric analysis
in the purple region of the figure~\ref{ZigzagRoughDiagram} into appearance of the fifth
orientation phase {\purple NAX}.

Finally, the triple point at $(M_3/M_4)=1$, $(\epsilon/D)\approx0.38$ between the
{\red AF}, {\green NA2}, and {\yellow AB} phases is {\it critical}.
At that point, the transitions between all 3 phases are second-order.
Indeed, in our previous paper~\cite{Kaplunovsky:2012gb} --- where the analysis was limited
to zigzags with $M_3=M_4$ (in present notations) --- we saw a second-order transition
between the two abelian phases {\red AF} and {\yellow AB}.

\bigskip\goodbreak\centerline{$\star\quad\star\quad\star$}\bigskip\nobreak

Thus far, we have focused on the instanton orientations $y_n$ for a given zigzag geometry.
That is, we have not only assumed that the instanton centers form a zigzag, but we have
also treated the zigzag amplitude $\epsilon$ as an independent input parameter, just like
the lattice spacing $D$ or the $M_2,M_3,M_4$ parameters of the 5D gauge theory.
Indeed, figure~\ref{ZigzagDiagram} shows the transitions between different orientation
phases as function of independent, freely-adjustable parameters $\epsilon/D$ and $M_3/M_4$.
But physically, the zigzag amplitude $\epsilon$ is a dynamical modulus whose value follows
from minimizing the net energy of the multi-instanton system.

In fact, all instanton center coordinates $X_n^\mu$ are dynamical moduli, which raises
two problems.
First, in some situations the lowest-energy configuration
of the instanton lattice may be more complicated than a zigzag or a straight chain;
we shall address this issue in our next paper~\cite{OurNextPaper}.
Second, for some lattice spacings $D$, a uniform lattice of any kind --- a straight chain,
a zigzag, or anything else --- may be unstable against breaking into domains of different phases
with different densities.

By way of analogy, consider a fluid governed by an equation of state such as Van-der-Waals.
Formally, this equation allows a uniform fluid to have any density (up to some maximum).
But in reality, at sub-critical temperatures one may have a low-density gas or a high-density
liquid, but there are no uniform fluids with intermediate densities.
If we constrain the overall volume $V$ of some amount of fluid such that its {\it average density}
would fall into the intermediate range, we would not get a uniform fluid; instead, part of
the volume would be filled by the higher-density liquid while the other part by the lower-density gas.
In the same way, if we fix the overall length $L$ of the $x^1$ axis occupied by some large number $N$
of instantons, their lowest-energy configuration is not necessarily a uniform lattice of some kind.
Instead, for some {\it average densities} $\rho=N/L$ we we would have $L$ split into domains
of two different lattices of different densities, one denser than $N/L$ and the other less dense.

To keep the fluid uniform, one should control its pressure $P$ rather than the volume $V$;
consequently, the preferred phase follows from minimizing the free enthalpy $G=E-ST+PV$
rather than the free energy $F=E-ST$.
Likewise, for the 1D lattice of instantons, we should control the net compression force $\bf F$
along the $x^1$ axis rather than the net length $L$ or the lattice spacing $D$.
Also, we should minimize the free enthalpy of the lattice
$G=E-ST+L\bf F$, but since we work at zero temperature all we need is the ordinary enthalpy $H=E+L\bf F$.
Equivalently, we may minimize the {\it non-relativistic chemical potential}%
\be
\hat\mu\ =\ \mu_{\rm rel}\ -\ M_{\rm baryon}\ =\ {G_{\rm non-rel}\over N}\
\mathop{\pspicture[shift=0.11](0,0)(1,0)\psline{->}(0,0)(1,0)\endpspicture}\limits_{T=0}\
{E+L{\bf F}\over N}\ =\ \E\ +\ {{\bf F}\over\rho}\,.
\label{ChemPot}
\ee
(We focus on the non-relativistic chemical potential $\hat\mu=\mu-M_{\rm baryon}$
because the relevant scale of $\hat\mu$ would be {\it much} smaller than the baryon mass.
Indeed, according to eq.~\eqref{CPE} below, $\hat\mu\sim N_cM_2\ll N_cM\ll N_cM\lambda\sim M_b$.)

Thus, through the remainder of this section, we are going to impose a compression force $\bf F$
on the multi-instanton system, {\it assume} that the instantons
form a uniform zigzag of some lattice spacing $D$ and amplitude $\epsilon$
--- or a uniform straight chain for $\epsilon=0$ --- and vary $D$, $\epsilon$
and the orientation moduli $(\phi,\alpha,\beta)$ of the zigzag to
seek the minimum of the NR chemical potential \eqref{ChemPot} for any given
combination of $\bf F$ and $M_3/M_4$.


We begin by changing variables from $\epsilon$ to $\xi=\epsilon/D$, combining eqs.~\eqref{ChemPot}
and~\eqref{ZigNetEnergy}, and rewriting the result as
\be
\hat\mu\
=\ {\bf F}\times D\ +\ N_c\lambda MM_2^2\times D^2\xi^2\
+\ {\pi^2 N_c\over 20\lambda M}\times{{\cal F}_m(\xi;M_3/M_4)\over D^2}
\label{CPM}
\ee
where
\be
{\cal F}_m(\xi;M_3/M_4)\ =\ \min_{\alpha,\beta,\phi}{\cal F}(\alpha,\beta,\phi;(\xi=\epsilon/D),M_3/M_4)
\ee
for $\cal F$ from eq.~\eqref{Fdef}.
Note that although we know $\cal F$ as an  analytic function of all its arguments,
and we know how to minimize it analytically with respect to the $\alpha$ and $\beta$ angles
({\it cf.}\ Appendix~A), the minimization with respect to the $\phi$ angle has to be done numerically,
so we do not have an explicit formula for the ${\cal F}_m(\xi;M_3/M_4)$.
Nevertheless, it is fairly easy to calculate numerically both the ${\cal F}_m$ function itself
and its derivative $\partial{\cal F}_m/\partial\xi$.

Before we seek the global minimum of the chemical potential~\eqref{CPM}, let's find all the local extrema.
Spelling out the extremality conditions
\be
\left({\partial{\cal F}_m \over\partial D}\right)_\xi\ =\ 0,\quad
\left({\partial{\cal F}_m \over\partial \xi}\right)_D\ =\ 0,
\ee
we obtain
\begin{align}
{\pi^2 N_c\over 20\lambda M}\times{2{\cal F}_m(\xi;M_3/M_4)\over D^3}\
-\ N_c\lambda MM_2^2\times 2D\xi^2\ &
=\ {\bf F}
\label{Feq}\\
\intertext{and}
{\pi^2 N_c\over 20\lambda M}\times{1\over D^2}\times{\partial{\cal F}_m \over\partial\xi}\
+\ N_c\lambda MM_2^2\times 2D^2\xi\ &
=\ 0.
\label{XIeq}
\end{align}
The last equation here has two branches of solutions:
the {\it zigzag branch}
\be
\xi>0,\qquad D^4\
=\ {\pi^2\over 20(\lambda MM_2)^2}\times {-1\over2\xi}\,{\partial{\cal F}_m \over\partial\xi}\,,
\label{ZigzagBranch}
\ee
and the {\it straight-line branch}
\be
\xi=0,\qquad {\rm any}\ D>0
\label{SLBranch}
\ee
which follows from $\partial{\cal F}_m/\partial\xi=0$ for $\xi=0$.
Plugging these two solution branches into the compression force equation \eqref{Feq}, we arrive at
\begin{subequations}
\label{Forceq}
\begin{align}
{\bf F}(D)\ &
=\ N_c\sqrt{\tfrac{\pi}{\sqrt{20}}\lambda MM_2^3}\times
\dfrac{2{\cal F}_m(\xi=0)\,=\,4}{\left[ D\times\sqrt{\tfrac{\sqrt{20}}{\pi}\lambda MM_2}\right]^3} &
\qquad\text{for the straight-line branch,}\\
{\bf F}(\xi)\ &
=\ N_c\sqrt{\tfrac{\pi}{\sqrt{20}}\lambda MM_2^3}\times
\dfrac{\displaystyle{2{\cal F}_m\,+\,\xi\,{\partial{\cal F}_m\over\partial\xi}}}%
	{\displaystyle{\left[{-1\over2\xi}\,{\partial{\cal F}_m\over\partial\xi}\right]^{3/4}}} &
\text{for the zigzag branch.}
\end{align}
\end{subequations}
Between the two branches, for any particular value of the compression force $\bf F$
we may have 1, 2, or more solutions for the lattice moduli $(D,\xi)$.
Only one of these solutions is the global minimum of the chemical potential~\eqref{ChemPot}
--- the rest are merely local extrema --- and the simplest way to identify that global minimum
is to evaluate
\begin{subequations}
\label{CPE}
\begin{align}
\hat\mu(D)\ &
=\ N_cM_2\tfrac{\pi}{\sqrt{20}}\times
\dfrac{6}{\left[ D\times\sqrt{\tfrac{\sqrt{20}}{\pi}\lambda MM_2}\right]^2} &
\qquad\text{for the straight-line branch,}\\[10pt]
\hat\mu(\xi)\ &
=\ N_cM_2\tfrac{\pi}{\sqrt{20}}\times
\dfrac{\displaystyle{3{\cal F}_m\,+\,{\xi\over2}\,{\partial{\cal F}_m\over\partial\xi}}}%
	{\displaystyle{\left[{-1\over2\xi}\,{\partial{\cal F}_m\over\partial\xi}\right]^{1/2}}} &
\text{for the zigzag branch.}
\end{align}
\end{subequations}
for all the solutions of eqs.~\eqref{Forceq} and select the solution
with the lowest $\hat\mu$.

When we vary the compression force $\bf F$ or the $M_3/M_4$ ratio of the background, sometimes
the global minimum `jumps' from one solution to another;
physically, this corresponds to a phase transition between different lattice geometries.
To see how this works, let's pick an $M_3/M_4$ ratio --- say, $M_3/M_4=0.5$ ---
and plot the chemical potential $\hat\mu$ as a function of the compression force $\bf F$.
for all the solutions of eqs.~\eqref{Forceq}.
In practice, this means combining the parametric plots $({\bf F}(D),\hat\mu(D))$ for the straight-line
branch (the black curve)
and $({\bf F}(\xi),\hat\mu(\xi))$ for the zigzag branch (the red and green curves,
according to the zigzag's orientation phase), thus
\be
\psset{xunit=1.9cm,yunit=1cm,linewidth=1pt}
\begin{pspicture}[shift=-2](0,-0.2)(7.5,4.7)
\rput(0,0){
	\psline{->}(0,0)(3.1,0)
	\uput[r](3.1,0){$\bf F$}
	\psline{->}(0,0)(0,4.25)
	\uput[u](0,4.25){$\hat\mu$}
%
%
%
%
	\pscurve[linecolor=black](0,0)(0.035,0.2)(0.137,0.5)(0.39,1)(0.71,1.5)(1.1,2.0)%
		(1.53,2.5)(2.0,3.0)(2.31,3.29)
	\pscurve[linecolor=green](1.17,2.23)(1.3,2.32)(1.5,2.47)(1.75,2.64)(2.0,2.8)(2.5,3.15)(3.0,3.47)
	\pscurve[linecolor=green](2.68,3.58)(2.31,3.35)(2.0,3.1)(1.75,2.85)(1.5,2.6)(1.25,2.33)(1.17,2.23)
	\pscurve[linecolor=red](2.31,3.29)(2.4,3.37)(2.5,3.45)(2.6,3.53)(2.68,3.58)
	\psline[linestyle=dotted](1.5,-0.1)(1.5,2.47)
	\uput[d](1.5,-0.1){${\bf F}_c$}
	}
\rput(4,0){
	\psline{->}(0,0)(3.1,0)
	\uput[r](3.1,0){$\bf F$}
	\psline{->}(0,0)(0,4.25)
	\uput[u](0,4.25){$\hat\mu$}
	\psset{linewidth=1.5pt}
	\pscurve[linecolor=black](0,0)(0.035,0.2)(0.137,0.5)(0.39,1)(0.71,1.5)(1.1,2.0)(1.5,2.47)
	\pscurve[linecolor=green](1.5,2.47)(1.75,2.64)(2.0,2.8)(2.5,3.15)(3.0,3.47)
	\psline[linestyle=dotted](1.5,-0.1)(1.5,2.47)
	\uput[d](1.5,-0.1){${\bf F}_c$}
	}
\end{pspicture}
\label{FmuPlot}
\ee
The left plot here shows all the solutions while the right plot shows only the global minimum
of the chemical potential $\hat\mu$ for any compression force $\bf F$.
We see that at ${\bf F}={\bf F}_c$ the global minimum switches from one solution branch to another;
physically, this corresponds to a first-order phase transition from a straight line to a zigzag.

For other values of the $M_3/M_4$ ratio, we get more complicated $({\bf F},\hat\mu)$ plots than
\eqref{FmuPlot}
For example, for $M_3/M_4=0.73$ the zigzag branch goes back and forth several times --- although
this is very hard to see graphically without zooming on very small portions of the curve ---
so that for some values of the force $\bf F$ we  get up to six different solutions of eqs.~\eqref{Forceq}.
Consequently, when we vary $\bf F$ there is a {\it sequence of three phase transitions:}
First, a second-order transition from a straight line to a zigzag with the antiferromagnetic order
of orientations.
Second, a first order transition in which the zigzag amplitude drastically increases, the lattice
spacing shrinks about 20\%, and the orientation pattern changes from {\red AF} to the non-abelian
{\green NA2}.
Third, another first-order phase transition, this time with a smaller jump of the lattice geometry
while the orientation pattern changes from the {\green NA2} to another non-abelian phase {\blue NA1}.

Altogether, we have found 7 different transition sequences for different $M_3/M_4$ ratios!
But instead of spelling out all these sequences in gory detail, let us simply present
in figures~\ref{BigPhaseDiagramFM}(a), \ref{BigPhaseDiagramFM}(b), and \ref{BigPhaseDiagramRho}
the phase diagram of all the zigzag and straight-chain phases in three different planes:
the compression force $\bf F$ versus $M_3/M_4$,
the non-relativistic chemical potential $\hat\mu$ versus $M_3/M_4$,
and the linear instanton density $\rho=1/D$ versus $M_3/M_4$.
\begin{figure}[p]
\psset{yunit=70mm,xunit=25mm,linewidth=0.4pt,framearc=0,linearc=0}
$$
\begin{pspicture}[shift=-40mm](-0.4,-0.15)(5.5,1.0)
\psline{->}(0,0)(5.25,0)
\uput[r](5.25,0){$\bf F$}
\psline{->}(0,0)(0,1.05)
\uput[u](0,1.05){$M_3/M_4$}
\pscustom[linestyle=none,fillstyle=solid,fillcolor=pink]{%
	\pscurve(0.3,0)(0.5,0.1)(0.66,0.2)(0.88,0.3)(1.15,0.395)(1.5,0.5)(2.0,0.64)(2.22,0.7)(2.3,0.725)
	\psline(2.3,0.725)(2.3,1)(0,1)(0,0)(0.3,0)
	}
\pscustom[linestyle=none,fillstyle=solid,fillcolor=red]{%
	\pscurve(2.77,0.93)(2.6,0.84)(2.5,0.79)(2.38,0.735)
	\psline(2.77,0.93)(2.77,1)(2.3,1)(2.3,0.725)(2.38,0.735)
	}
\pscustom[linestyle=none,fillstyle=solid,fillcolor=yellow]{%
	\pscurve(2.77,0.93)(2.8,0.94)(3.03,0.95)(3.36,0.96)(3.8,0.97)(5,0.98)
	\psline(5,0.98)(5,1)(2.77,1)(2.77,0.93)
	}
\pscustom[linestyle=none,fillstyle=solid,fillcolor=blue]{%
	\pscurve(5,0.98)(3.8,0.97)(3.36,0.96)(3.03,0.95)(2.8,0.94)(2.77,0.93)
	\pscurve(2.77,0.93)(2.75,0.92)(2.73,0.91)(2.71,0.90)(2.6,0.84)(2.5,0.79)(2.38,0.735)
	\psecurve(2.3,0.77)(2.38,0.735)(2.5,0.70)(2.98,0.65)(4.15,0.6)(5.0,0.575)(6.0,0.57)
	\psline(5.0,0.575)(5.0,0.98)
	}
\pscustom[linestyle=none,fillstyle=solid,fillcolor=green]{%
	\pscurve(0.3,0)(0.5,0.1)(0.66,0.2)(0.88,0.3)(1.15,0.395)(1.5,0.5)(2.0,0.64)(2.22,0.7)(2.3,0.725)
	\psline(2.3,0.725)(2.38,0.735)
	\psecurve(2.3,0.77)(2.38,0.735)(2.5,0.70)(2.98,0.65)(4.15,0.6)(5.0,0.575)(6.0,0.57)
	\psline(5.0,0.575)(5.0,0)(0.3,0)
	}
\pscustom[linestyle=none,fillstyle=solid,fillcolor=blue]{%
	\pscurve(0.3,0)(0.5,0.1)(0.66,0.2)(0.88,0.3)(1.15,0.395)
	\pscurve(1.15,0.395)(1.3,0.39)(1.5,0.365)(1.8,0.305)(2.15,0.2)(2.385,0.1)(2.44,0.0)
	\psline(2.44,0)(0.3,0)
	}
\multido{\n=0.0+0.1}{11}{%
	\psline(-0.05,\n)(5,\n)
	\uput[l](-0.05,\n){\small\n}
	}
\multido{\n=0.0+0.5}{11}{%
	\psline(\n,-0.015)(\n,1.01)
	\uput[d](\n,-0.015){\small\n}
	}
\psset{linewidth=1pt}
\pscurve(0.3,0)(0.5,0.1)(0.66,0.2)(0.88,0.3)(1.15,0.395)(1.5,0.5)(2.0,0.64)(2.22,0.7)(2.3,0.725)
\psline(2.3,0.725)(2.38,0.735)
\pscurve(2.38,0.735)(2.5,0.79)(2.6,0.84)(2.77,0.93)
\psline(2.77,0.93)(2.77,1.0)
\psline[linecolor=white](2.3,0.725)(2.3,1.0)
\pscurve(2.77,0.93)(2.8,0.94)(3.03,0.95)(3.36,0.96)(3.8,0.97)(5.0,0.98)
\pscurve(2.38,0.735)(2.5,0.70)(2.98,0.65)(4.15,0.6)(5.0,0.575)
\pscurve[linecolor=white](1.15,0.395)(1.3,0.39)(1.5,0.365)(1.8,0.305)(2.15,0.2)(2.385,0.1)(2.44,0.0)
\rput*(1.0,0.7){\small straight}
\rput*(1.25,0.2){\scriptsize NA1}
\rput*(4.0,0.8){\scriptsize NA1}
\rput*(3.5,0.3){\scriptsize NA2}
\psset{framesep=1pt}
\rput*(2.5,0.92){\scriptsize AF}
\rput*(3.2,0.99){\scriptsize AB}
\end{pspicture}
\eqno\rm(a)
$$
$$
\begin{pspicture}[shift=-40mm](-0.4,-0.1)(5.5,1.1)
\psline{->}(0,0)(5.25,0)
\uput[r](5.25,0){$\hat\mu$}
\psline{->}(0,0)(0,1.05)
\uput[u](0,1.05){$M_3/M_4$}
\pscustom[linestyle=none,fillstyle=solid,fillcolor=pink]{%
	\pscurve(0.82,0)(1.0,0.04)(1.5,0.22)(2.07,0.395)(2.5,0.51)(3.0,0.645)(3.28,0.725)
	\psline(3.28,0.725)(3.28,1)(0,1)(0,0)(0.82,0)
	}
\pscustom[linestyle=none,fillstyle=solid,fillcolor=red]{%
	\pscurve(3.35,0.735)(3.5,0.805)(3.71,0.93)
	\psline(3.71,0.93)(3.71,1)(3.28,1)(3.28,0.725)(3.35,0.735)
	}
\pscustom[linestyle=none,fillstyle=solid,fillcolor=yellow]{%
	\pscurve(3.71,0.93)(3.73,0.94)(3.88,0.95)(4.1,0.96)(4.4,0.97)(5.0,0.98)
	\psline(5,0.98)(5,1)(3.71,1)(3.71,0.93)
	}
\pscustom[linestyle=none,fillstyle=solid,fillcolor=blue]{%
	\pscurve(5,0.98)(4.4,0.97)(4.1,0.96)(3.88,0.95)(3.73,0.94)(3.71,0.93)
	\pscurve(3.71,0.93)(3.695,0.92)(3.61,0.87)(3.5,0.805)(3.35,0.735)
	\pscurve(3.35,0.735)(3.41,0.70)(3.6,0.67)(4.1,0.62)(5.0,0.575)
	\psline(5.0,0.575)(5.0,0.98)
	}
\pscustom[linestyle=none,fillstyle=solid,fillcolor=green]{%
	\pscurve(0.82,0)(1.0,0.04)(1.5,0.22)(2.07,0.395)(2.5,0.51)(3.0,0.645)(3.28,0.725)
	\psline(3.28,0.725)(3.35,0.735)
	\pscurve(3.35,0.735)(3.41,0.70)(3.6,0.67)(4.1,0.62)(5.0,0.575)
	\psline(5.0,0.575)(5.0,0)(0.82,0)
	}
\pscustom[linestyle=none,fillstyle=solid,fillcolor=blue]{%
	\pscurve(0.82,0)(1.0,0.04)(1.5,0.22)(2.07,0.395)
	\pscurve(2.07,0.395)(2.2,0.39)(2.5,0.34)(2.8,0.23)(3.0,0.1)(3.02,0.0)
	\psline(3.02,0)(0.82,0)
	}
\multido{\n=0.1+0.1}{10}{%
	\psline(-0.05,\n)(5,\n)
	\uput[l](-0.05,\n){\n}
	}
\multido{\n=0.5+0.5}{10}{%
	\psline(\n,-0.01)(\n,1.01)
	\uput[d](\n,-0.01){\n}
	}
\psset{linewidth=1pt}
\pscurve(0.82,0)(1.0,0.04)(1.5,0.22)(2.07,0.395)(2.5,0.51)(3.0,0.645)(3.28,0.725)
\psline[linecolor=white](3.28,0.725)(3.28,1.0)
\psline(3.28,0.725)(3.35,0.735)
\pscurve(3.35,0.735)(3.5,0.805)(3.71,0.93)
\psline(3.71,0.93)(3.71,1.0)
\pscurve(3.71,0.93)(3.73,0.94)(3.88,0.95)(4.1,0.96)(4.4,0.97)(5.0,0.98)
\pscurve(3.35,0.735)(3.41,0.70)(3.6,0.67)(4.1,0.62)(5.0,0.575)
\pscurve[linecolor=white](2.07,0.395)(2.2,0.39)(2.5,0.34)(2.8,0.23)(3.0,0.1)(3.02,0.0)
\rput*(1.5,0.7){\small straight}
\rput*(2.0,0.2){\scriptsize NA1}
\rput*(4.2,0.8){\scriptsize NA1}
\rput*(3.8,0.3){\scriptsize NA2}
\psset{framesep=1pt}
\rput*(3.5,0.93){\scriptsize AF}
\rput*(4.0,0.99){\scriptsize AB}
\end{pspicture}
\eqno\rm(b)
$$
\caption{%
	Phase diagrams of the zigzag and straight-chain phases in the compression v.\ $M_3/M_3$ plane
	(top diagram (a), $\bf F$ in units of $N_c\sqrt{\lambda MM_2^3}$)
	and in the chemical potential v.\ $M_3/M_3$ plane
	(bottom diagram (b), the non-relativistic $\hat\mu$ in units of $N_cM_2$).
	The straight-chain phase is colored pink, while 4 other colors --- red, yellow, blue, and green ---
	denote zigzag phases with different instanton orientation patterns.
	The first-order transition between phases are indicated by black lines, the second-order
	transitions by white lines.
	}
\label{BigPhaseDiagramFM}
\end{figure}
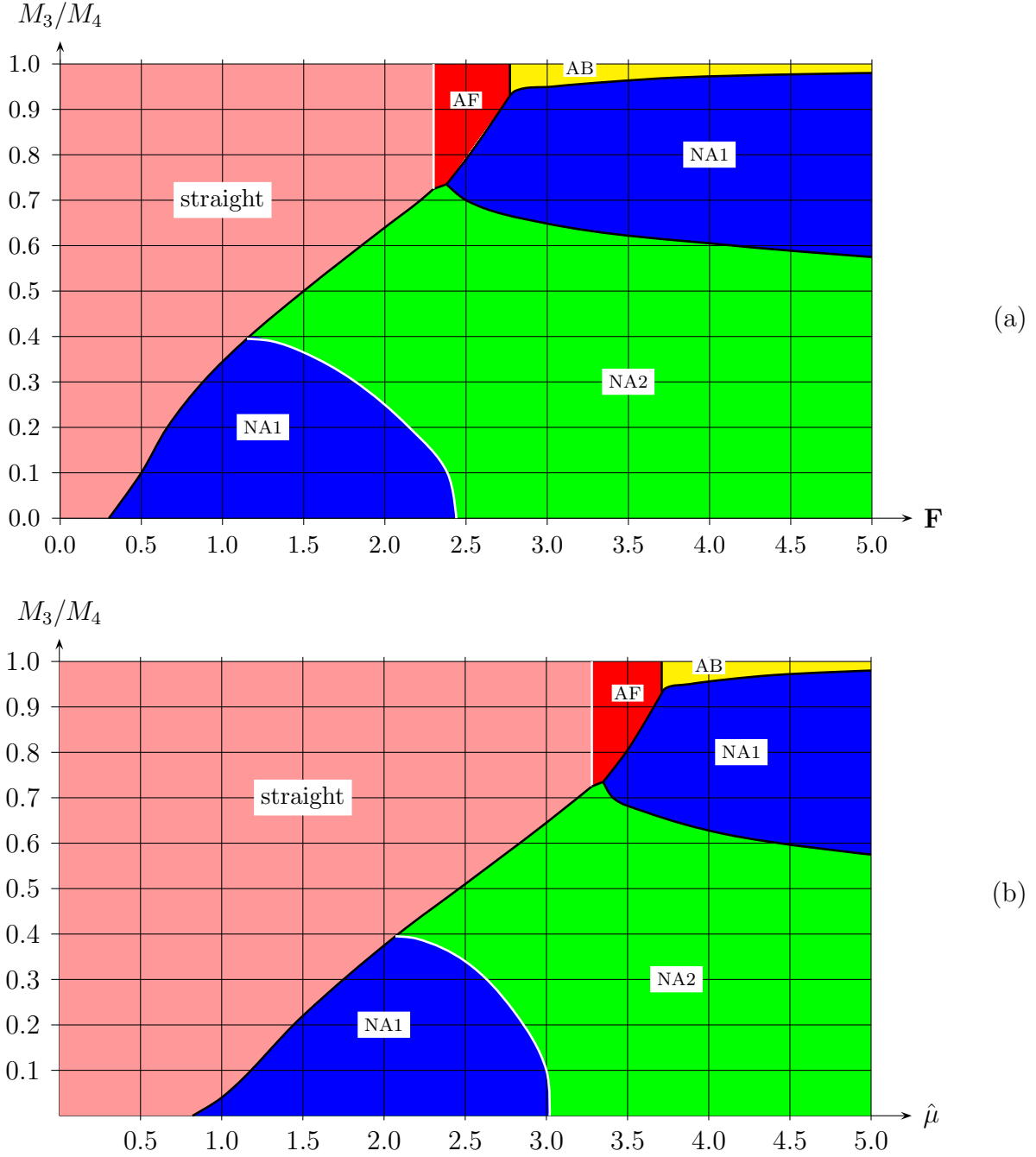
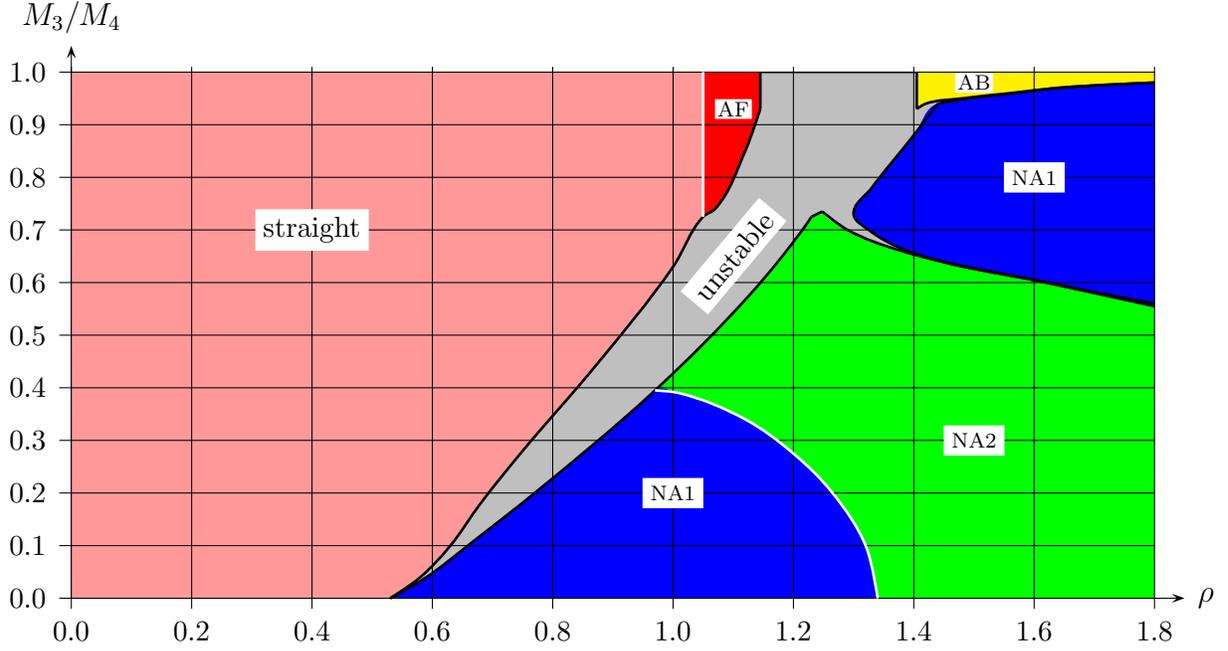
\begin{figure}[t]
\psset{yunit=70mm,xunit=80mm,linewidth=0.4pt,framearc=0,linearc=0}
$$
\begin{pspicture}[shift=-40mm](-0.15,-0.15)(1.9,1.0)
\psline{->}(0,0)(1.85,0)
\uput[r](1.85,0){$\rho$}
\psline{->}(0,0)(0,1.05)
\uput[u](0,1.05){$M_3/M_4$}
\psframe[fillcolor=lightgray,fillstyle=solid](0,0)(1.8,1)
\pscustom[linestyle=none,fillstyle=solid,fillcolor=pink]{%
	\pscurve(0.53,0)(0.58,0.04)(0.63,0.1)(0.68,0.18)(0.75,0.28)(0.87,0.44)(1.0,0.63)%
	(1.05,0.725)(1.07,0.74)(1.12,0.85)(1.145,0.93)
	\psline(1.145,0.93)(1.145,1)(0,1)(0,0)(0.53,0)
	}
\pscustom[linestyle=none,fillstyle=solid,fillcolor=red]{%
	\pscurve(1.05,0.725)(1.07,0.74)(1.12,0.85)(1.145,0.93)
	\psline(1.145,0.93)(1.145,1)(1.05,1)(1.05,0.725)
	}
\pscustom[linestyle=none,fillstyle=solid,fillcolor=yellow]{%
	\pscurve(1.405,0.93)(1.42,0.94)(1.48,0.95)(1.56,0.96)(1.64,0.97)(1.8,0.98)
	\psline(1.8,0.98)(1.8,1)(1.405,1)(1.405,0.93)
	}
\pscustom[linestyle=none,fillstyle=solid,fillcolor=blue]{%
	\pscurve(1.8,0.98)(1.64,0.97)(1.57,0.96)(1.49,0.95)(1.44,0.94)
	\pscurve(1.44,0.94)(1.4,0.88)(1.33,0.78)(1.3,0.73)(1.325,0.7)(1.39,0.66)(1.62,0.6)(1.8,0.56)
	\psline(1.8,0.56)(1.8,0.98)
	}
\pscustom[linestyle=none,fillstyle=solid,fillcolor=green]{%
	\pscurve(0.53,0)(0.59,0.04)(0.67,0.11)(0.77,0.2)(0.97,0.395)(1.13,0.58)(1.215,0.7)(1.23,0.725)
	\psline(1.23,0.725)(1.248,0.735)
	\pscurve(1.248,0.735)(1.29,0.70)(1.38,0.66)(1.62,0.6)(1.8,0.555)
	\psline(1.8,0.555)(1.8,0)(0.53,0)
	}
\pscustom[linestyle=none,fillstyle=solid,fillcolor=blue]{%
	\pscurve(0.53,0)(0.59,0.04)(0.67,0.11)(0.77,0.2)(0.97,0.395)
	\pscurve(0.97,0.395)(1.01,0.39)(1.12,0.34)(1.2,0.275)(1.265,0.2)(1.32,0.1)(1.34,0)
	\psline(1.34,0)(0.53,0)
	}
\multido{\n=0.0+0.1}{11}{%
	\psline(-0.02,\n)(1.8,\n)
	\uput[l](-0.02,\n){\small\n}
	}
\multido{\n=0.0+0.2}{10}{%
	\psline(\n,-0.02)(\n,1.01)
	\uput[d](\n,-0.02){\small\n}
	}
\psset{linewidth=1pt}
\pscurve(0.53,0)(0.58,0.04)(0.63,0.1)(0.68,0.18)(0.75,0.28)(0.87,0.44)(1.0,0.63)(1.05,0.725)%
	(1.07,0.74)(1.12,0.85)(1.145,0.93)
\psline(1.145,0.93)(1.145,1.0)
\psline[linecolor=white](1.05,0.725)(1.05,1)
\pscurve(0.53,0)(0.59,0.04)(0.67,0.11)(0.77,0.2)(0.97,0.395)(1.13,0.58)(1.215,0.7)(1.23,0.725)
\psline(1.23,0.725)(1.248,0.735)
\pscurve(1.248,0.735)(1.29,0.70)(1.38,0.66)(1.62,0.6)(1.8,0.555)
\pscurve[linecolor=white](0.97,0.395)(1.01,0.39)(1.12,0.34)(1.2,0.275)(1.265,0.2)(1.32,0.1)(1.34,0)
\psline(1.405,1.0)(1.405,0.93)
\pscurve(1.405,0.93)(1.42,0.94)(1.48,0.95)(1.56,0.96)(1.64,0.97)(1.8,0.98)
\pscurve(1.8,0.98)(1.64,0.97)(1.57,0.96)(1.49,0.95)(1.44,0.94)
\pscurve(1.44,0.94)(1.4,0.88)(1.33,0.78)(1.3,0.73)(1.325,0.7)(1.39,0.66)(1.62,0.6)(1.8,0.56)
\rput*{50}(1.1,0.65){\small unstable}
\rput*(0.4,0.7){\small straight}
\rput*(1.0,0.2){\scriptsize NA1}
\rput*(1.6,0.8){\scriptsize NA1}
\rput*(1.5,0.3){\scriptsize NA2}
\psset{framesep=1pt}
\rput*(1.1,0.93){\scriptsize AF}
\rput*(1.5,0.98){\scriptsize AB}
\end{pspicture}
$$
\caption{%
	Phase diagram of the zigzag and straight-chain phases in the linear density v.\ $M_3/M_3$ plane;
	the density $\rho=1/D$ is in units of $\sqrt{\lambda MM_2}$.
	The stable straight-chain phase is colored pink, while the stable zigzag phases are colored
	red, yellow, blue, or green according to the instanton orientation pattern.
	Finally, the gray color denotes densities at which a uniform zigzag or straight chain
	would be mechanically or thermodynamically unstable.
	}
\label{BigPhaseDiagramRho}
\end{figure}

\newpage

Here are a few particularly noteworthy features of these diagrams:
\begin{itemize}

\item
Since we assume $M_2\ll M_3$, the straight-chain phase
always has the anti-ferromagnetic order of the instantons' orientations.

\item
The very first transition from the straight chain to a zigzag
could be either first-order or second-order, depending on the $M_3/M_4$ ratio:
for $(M_3/M_4)<0.725$ the transition is first-order while for $(M_3/M_4)>0.725$ it's second-order.
This difference is due to different orientation phases of the zigzag immediately after the transition:
for $(M_3/M_4)>0.725$ the zigzag has the same antiferromagnetic order as the straight chain,
which allows a second-order transition;
but for $(M_3/M_4)<0.725$ the zigzag has a different orientation pattern
--- the non-abelian {\blue NA1} or {\green NA2} --- so the transition is first-order.

\item
The non-abelian phases {\blue NA1} and {\green NA2} of the zigzag cover much larger
areas of the phase diagrams \ref{BigPhaseDiagramFM}(a,b) and \ref{BigPhaseDiagramRho}
than the abelian phases {\red AF} and {\yellow AB}.
In particular, at larger compression forces $\bf F$ --- and hence larger chemical potentials
$\hat\mu$, larger densities, and larger zigzag amplitudes ---
the instanton orientations usually prefer the non-abelian patterns.
Only the backgrounds with $M_3\approx M_4$ --- such as the model we have analyzed in
\cite{Kaplunovsky:2012gb} --- favor the abelian orientations.

\item
Figure \ref{BigPhaseDiagramRho} has gray areas at which an instanton zigzag with a uniform
lattice spacing $D=1/\rho$ and a uniform amplitude $\epsilon$ (or a uniform straight chain for $\epsilon=0$)
would be unstable against instantons' motion along the $x^1$ axis (the long direction of the zigzag).
If we put $N\gg1$ instantons into a box of fixed length $L=N/\rho$ and let them seek the lowest-energy
configuration, they would organize themselves into domains of two different lattices with different
lattice spacings and different amplitudes.

\item
The {\blue NA1} phase of the zigzag occupies two separate regions of the phase diagram
separated by the region of the other non-abelian phase {\green NA2}.
The phase transition between the lower-left {\blue NA1} region and the {\green NA2} region
is second-order, while the transition between the upper-right {\blue NA1} region and the {\green NA2}
is weakly first-order: the lattice spacing and the zigzag amplitude are discontinuous across the transition,
but the discontinuity is very small and hard to see graphically on figure~\ref{BigPhaseDiagramRho}
or \eqref{FmuPlot}--like plots.

\item
Likewise, the transition between the upper-right region of the {\blue NA1} phase and the abelian
{\yellow AB} phases of the zigzag is weakly first-order.
On the other hand, the transitions between the antiferromagnetic phases of the straight chain or zigzag
and all the other zigzag phases is strongly first-order, with largish discontinuities of the lattice
spacing and even larger discontinuity of the zigzag amplitude.

\item
However, for small $M_3/M_4$ the discontinuity becomes small; for $M_3/M_4=0$ it vanishes altogether
and the phase transition between the straight chain and the {\blue NA1} phase of the zigzag becomes
second-order.
This is OK because for $M_3/M_4=0$ --- or rather for $M_3=M_2\ll M_4$ --- the instantons' orientations
in the straight-chain phase are no longer antiferromagnetic but form the $\rm period=4$ Klein-group
pattern similar to the {\blue NA1} pattern of the zigzag.

\end{itemize}

The phase diagrams on figures \ref{BigPhaseDiagramFM}(a,b) and \ref{BigPhaseDiagramRho}
completes out analysis of the instanton zigzag.
But the zigzag is only the first step in the transition sequence from a 1D to a 2D instanton lattice.
The subsequent steps --- from a zigzag to a 3-layer lattice, to 4-layer lattice, \etc\ ---
will be explored in our next paper \cite{OurNextPaper}.

\section{ Summary and outlook}
In~\cite{Kaplunovsky:2012gb}  we have investigated the crystalline structure
of the holographic nuclear matter.
The analysis done in that paper revealed the ``popcorn transition" of a one-dimensional infinite chain
of nucleons into a zigzag structure.
We have  also found out that the  phases between adjacent nucleons are abelian.
In this paper we continued the  exploration of 1D instanton systems and their popcorn transition
to the zigzag, but by generalizing the background to allow for  $M_3\neq M_4$,
we  found a much richer phase structure, including two distinct non-abelian phases.

In \S3 we proved that for low-density multi-instanton systems --- in which
the distances between the instantons are much larger that their radii ---
the interactions between the baryons are dominated by the two-body forces,
while the 3--body forces, {\it etc.,} are suppressed by powers of $\rm (radius/distance)^2$.
The two-body-force approximation makes for much easier calculations, so in \S4--5 we could
put the instanton chains and zigzags into richer backgrounds than we had in our previous paper.
In particular, we could use a less symmetric external potential for the baryons, by making
the 5D gauge coupling vary as
\be
{8\pi^2\over g_5^2(x)}\
=\ N_c\lambda M\times\Bigl( 1\ +\
	M_2^2\times x_2^2\,+\,M_3^2\times x_3^2\,+\,M_4^2\times x_4^2\
	O\bigl(M^4 x^4_{2,3,4}\bigr)\Bigr)
\tag{\ref{Ipot}}
\ee
with three independent curvature parameters $M_2,M_3,M_4$.

For the almost-holographic background $M_2,M_3\ll M_4$, we found that the 1D lattice of instantons
has a highly degenerate family of ground states with many different patterns of instantons'
orientations.
Among them are periodic patterns in which instanton orientations span  finite subgroups
of the $SU(2)/\Z_2$: the anti-ferromagnetic chains spanning the $\Z_2$ subgroup
(two alternating orientations related by a $180^\circ$ twist),
$\rm period=4$ chains spanning the Klein group $\Z_2\times \Z_2$, and ${\rm period}=2k$
chains spanning the prismatic groups $\Z_k\times\Z_2$ and the dihedral groups $D_{2k}$.
There are also many link-periodic patterns in which the relative orientations $y_n^\dagger y_{n+1}^{}$
of neighboring instantons are periodic but the orientations $y_n$ themselves are not periodic.
But the vast majority of the degenerate ground states are not periodic at all.
Raising the $M_2$ and $M_3$ parameters lifts the degeneracy.
For $M_2\ll M_3\sim M_4$, the ground state is the anti-ferromagnetic chain, while for
$M_2\sim M_3\sim M_4$ the ground state is periodic with a longer period or link-periodic,
depending on the $M_2/M_3$ ratio.
Specifically, for $M_2=M_3$ the instanton's orientation in the ground state span the Klein group,
for other rational $M_2^2/M_3^2$ they span a dihedral group, while the irrational
$M_2^2/M_3^2$ favor the link-periodic patterns.

For the zigzag-shaped chains of instantons we need $M_2\ll M_3\sim M_4$ to make sure
the transition from a straight chain to a zigzag happens for $\rm lattice\ spacing\ \gg\ instanton\
radius$, but we may vary the $M_3/M_4$ ratio.
Consequently, we found 5 distinct phases shown in figures~\ref{BigPhaseDiagramFM}
and~\ref{BigPhaseDiagramRho}:
The anti-ferromagnetic straight chain, the anti-ferromagnetic zigzag, the abelian
link-periodic zigzag ($\rm period=1$), and two different non-abelian link-periodic phases
($\rm period=2$).
Figures~\ref{BigPhaseDiagramFM} and~\ref{BigPhaseDiagramRho}
also show a rather complicated phase structure:
As we increase the compression force of the 1D lattice --- and hence increase the instanton
density and the chemical potential, ---
the zigzag amplitude and the orientation pattern
go through a sequence of 1, 2, or 3 phase transitions, but the number of transition and their
thermodynamic orders depend on the $M_3/M_4$ ratio.
In particular, the very first transition from a straight chain to a zigzag is
first-order for $M_3<0.725 M_4$ but second order for $M_3>0.725 M_4$.
The subsequent transition(s) may also be first-order or second-order,
depending on the $M_3/M_4$.

This paper is part of a program focused on exploring the holographic nuclear matter,
especially its low-temperature high-pressure phases.
Thus far 
we have worked out the 1D instanton lattices and their zigzag deformations
as a toy model for the transitions between 3D and 4D lattices.
But many interesting questions remain open, and we would like to address them in our future
papers; here is a short list:
\begin{itemize}

\item
The very next step in our program is to follow up the ``popcorn transitions'' from a 1D instanton chain
into the second dimension beyond the zigzag phases through multiple lines of instantons towards
a thick 2D lattice.
We should also check that that the very first popcorn transition indeed goes from the straight chain
to the zigzag rather than to some other configuration of the instantons.

\item
After that, we should set $M_2=0$ and study the infinite 2D lattices and their popcorn
transitions into into the third dimension for $M_3\ll M_4$.
The techniques developed in this paper --- using the two-body-forces approximation for $a\ll D$ ---
should work all the way up to infinite 3D lattices for $M_2=M_3=0$.
Alas, the ultimate popcorn transition from the 3D instanton crystal into the fourth space dimension
probably happens for $D\sim a$, so the two-body-forces approximation would not be accurate.
But we ought to to try it anyhow, just to get a qualitative picture of the transition.

\item
Our program (starting with \cite{Kaplunovsky:2012gb}) is based on the conjecture that the
popcorn transition from a 3D instanton crystal to a 4D crystal is holographically
dual to the transition of the large--$N_c$ nuclear matter from the baryonic phase to the ``quarkyonic
phase'' \cite{McLerran:2007qj}
(Fermi liquid of quarks, with baryon-like excitations near the Fermi surface).
Thus far, this conjecture was justified by qualitative arguments only, and we would like
to make a quantitative argument  by working out the details and the implication
of the transition.

\item
In the skyrmion model of large--$N_c$ nuclear matter,
Kugler and Shtrikman \cite{KuglerShtrikman} showed that at low pressures, the lowest-energy
configuration of is the FCC lattice of skyrmions with 4 different isospin orientations,
while the high-pressure configuration is the simple cubic lattice of half-skyrmions.
Later, Manton and Sutcliffe \cite{Manton:1994rf} described these lattices in terms of  instantons
on a $T^4$ torus.
We would like to re-interpret their results in terms of instanton lattices to see where the
Klebanov--Kugler--Shtrikman transition point (between whole-skyrmion and half-skyrmion lattices)
corresponds to the popcorn transition from a 3D to a 4D lattice.
Also, we would like to see how the chiral symmetry restoration in the high-density phase
works in terms of the instanton lattices.

\item
Finally, we would like to see if any of our 1D, 2D, 3D lattices --- or their generalizations ---
appear in condensed matter rather than in the holographic nuclear physics.
We imagine a crystalline lattice of some high-spin atoms / ions / whatever whose orientations may form
all kinds of complicated patterns --- perhaps non-abelian patterns --- depending on external
parameters such as temperature, pressure, or doping.
Hopefully, the methods developed in this article would be useful for studying such systems.
In this case, we would end up answering questions that have yet not been asked.

\end{itemize}

\section*{Acknowledgements}
We thank Dmitry Melnikov for many fruitful conversations.
This paper is based on research supported by the US National Science Foundation
(grant PHY--0969020) [VK] and
by the Israel Science Foundation (grant 1665/10).

\appendix
\section{Energy of the Instanton Zigzag}
In this Appendix we calculate the net  energy of a zigzag-shaped
chain (\ref{Zigzag}) of small instantons whose orientations
follow the ansatz~(\ref{ZigYAnsatz}) from section~5,
$$
y_n\ =\ \exp\bigl(i n\tfrac\phi2\,\tau_2\bigr)\times
\exp\Bigl(i\bigl(\tfrac\alpha2\,+\,(-1)^n\tfrac\beta2\bigr)\tau_1\Bigr).
\eqno(\ref{ZigYAnsatz}),
$$
and then we minimize that energy with respect to the orientation moduli
$\alpha$, $\beta$, and $\phi$.

The energy of a general 2D instanton configuration in a background with $M_2\ll M_3\sim M_4$
is spelled out in eq.~(\ref{E2D}):
\begin{align}
\E\ &
=\ N_c\lambda MM_2^2\times\sum_n (X_n^2)^2\
+\ {N_c\over 5\lambda M}\sum_{m\neq n}{Q_z(m,n)\over|X_m-X_n|^2}\\
\intertext{for}
Q_z(m,n)\ &
=\ \tfrac12\ +\ \tr^2\bigl(y_m^\dagger y_n^{}\bigr)\
+\ C_3\sum_{a=1,2}\tr^2\bigl(y_m^\dagger y_n^{}(-i\tau^a)\bigr)\nonumber\\
&\hskip 3em+\ (1-2C_3)\tr^2\bigl(y_m^\dagger y_n^{}(-i\vec\tau\cdot\vec N_{mn})\bigr),
\end{align}
where $C_3=M_3^2/(M_4^2+M_3^2)$, $0\le C_3\le\frac12$, and
$N^\mu_{mn}=(X_m^\mu-X_n^\mu)/|X_m-X_n|$.
For the zigzag arrangement of the instanton centers,
\begin{align}
\left.\begin{aligned}
	|X_m-X_n|^2\ &=\ (n-m^2)D^2,\\
	\vec\tau\cdot\vec N_{m,n}\ & =\ \pm\tau_1
	\end{aligned}\right\} &
\qquad{\rm for\ even}\  n-m,\\
\intertext{but}
\left.\begin{aligned}
	|N_m-X_n|^2\ &=\ (n-m^2)D^2\ +\ 4\epsilon^2,\\
	\vec\tau\cdot\vec N_{m,n}\ &
	=\ {(n-m)D\tau_1\,-\,(-1)^n(2\epsilon)\tau_2\over\sqrt{(n-m)^2D^2+4\epsilon^2}}
	\end{aligned}\right\} &
\qquad{\rm for\ odd}\ n-m.
\end{align}

Our first task is to calculate the orientation-dependent force coefficients
$Q_z(m,n)$ for the instantons oriented according to the ansatz~(\ref{ZigYAnsatz}).
For even $\ell=n-m$, we have
\begin{align}
y_m^\dagger y_n^{}\ &
=\ \exp\Bigl(-i\frac{\alpha\pm\beta}{2}\,\tau_1\Bigr)
\times\exp\Bigl(i\ell\frac\phi2\,\tau_2\Bigr)
\times \exp\Bigl(+i\frac{\alpha\pm\beta}{2}\,\tau_1\Bigr)\nonumber\\
&=\ \cos\frac{\ell\phi}{2}\
+\ \sin\frac{\ell\phi}{2}\cos(\alpha\pm\beta)\times(i\tau_2)\
+\ \sin\frac{\ell\phi}{2}\sin(\alpha\pm\beta)\times(i\tau_3)
\label{RelEven}\\
\intertext{where $\pm=(-1)^n=(-1)^m$, and consequently}
Q_z(m,n)\ &
=\ \tfrac12\ +\ 4\cos^2\frac{\ell\phi}{2}\
+\ C_3\times 4\sin^2\frac{\ell\phi}{2}\cos^2(\alpha\pm\beta).
\end{align}
Averaging between even and odd $n$ for fixed distance $\ell=n-m$,
we obtain
\be
\begin{split}
\vev{\cos^2(\alpha\pm\beta)}\ &
\equiv\ \frac12\times\cos^2(\alpha+\beta)\
+\ \frac12\times\cos^2(\alpha-\beta)\\
&=\ \cos^2\!\alpha\cos^2\!\beta\ + \sin^2\!\alpha\sin^2\!\beta
\end{split}
\ee
and hence
\be
\label{Qeven}
\begin{split}
\!\vev{Q_z(\ell=n-m)}\, &
=\ \frac12\ +\ 4\cos^2{\ell\phi\over2}\
+\ 4C_3\times
	(\cos^2\!\alpha\cos^2\!\beta+\sin^2\!\alpha\sin^2\!\beta)\times\sin^2{\ell\phi\over2}\\
&=\ \frac92\
+\ 2\Bigl(-1\,+\,C_3(\cos^2\!\alpha\cos^2\!\beta+\sin^2\!\alpha\sin^2\!\beta)\Bigr)
	\times(1-\cos(\ell\phi)).
\end{split}
\ee
Therefore, the average net energy of an instanton $\#m$ due to 2-body repulsive forces
from the other instantons $\#n$ at even distances $\ell=n-m$ comes up to
\begin{align}
\E_{\rm even}\ &
=\ {N_c\over5\lambda M}\sum_{{\rm even}\,\ell\neq0}{\vev{Q_z(\ell)}\over(\ell D)^2}\\[5pt]
&=\ {N_c\over5\lambda MD^2}\times\left(\begin{aligned}
	\frac92&
	\times\left(\sum_{{\rm even}\,\ell\neq0}{1\over\ell^2}\ =\ {\pi^2\over12}\right)\\
	&+\,2\Bigl(-1\,+\,C_3(\cos^2\!\alpha\cos^2\!\beta+\sin^2\!\alpha\sin^2\!\beta)\Bigr)
		\times{}\\
	&\qquad\times\left(\sum_{{\rm even}\,\ell\neq0}{1-\cos(\ell\phi)\over\ell^2}\
		=\ {\pi^2\over8}\,\Sigma_0(\phi)\right)
	\end{aligned}\right)\nonumber\\[5pt]
&=\ {\pi^2N_c\over20\lambda MD^2}\left(
	\frac32\,
	+\,\Bigl(-1\,+\,C_3(\cos^2\!\alpha\cos^2\!\beta+\sin^2\!\alpha\sin^2\!\beta)\Bigr)
		\times\Sigma_0(\phi)
	\right)
\end{align}
where
\be
\Sigma_0(\phi)\ =\ {4\phi(\pi-\phi)\over\pi^2}
\ee
{\it cf.}\ eq.~(\ref{Sigmas}a).

For instantons at odd distances from each other we have more complicated formulae.
For odd $\ell=n-m$ we have
\begin{align}
y_m^\dagger y_n^{}\ &
=\ \exp\Bigl({-i\alpha\pm i\beta\over2}\,\tau_1\Bigr)
\times\exp\Bigl(i\ell\frac\phi2\,\tau_2\Bigr)
\times \exp\Bigl({+i\alpha\pm i\beta\over2}\,\tau_1\Bigr)
\label{RelOdd}\\
&=\ \cos\frac{\ell\phi}{2}\times\exp\bigl(\pm i\beta\tau_1\bigr)
+\ \sin\frac{\ell\phi}{2}\times(i\tau_2)\times\exp\bigl(i\alpha\tau_1\bigr)
\nonumber\\
&=\ \cos\frac{\ell\phi}{2}\cos\beta\
\pm\ \cos\frac{\ell\phi}{2}\sin\beta\times(i\tau_1)\
+\ \sin\frac{\ell\phi}{2}\cos\alpha\times(i\tau_2)\
+\ \sin\frac{\ell\phi}{2}\sin\alpha\times(i\tau_3)\nonumber
\end{align}
where $\pm=(-1)^n=-(-1)^m$.
Consequently,
\be
\tr\bigl(y_m^\dagger y_n^{}(-i\vec\tau\cdot\vec N_{mn})\bigr)\
=\ {\pm2\over\sqrt{(\ell D)^2+4\epsilon^2}}\times\Bigl(
	\cos\frac{\ell\phi}{2}\sin\beta\times\ell D\,
	-\,\sin\frac{\ell\phi}{2}\cos\alpha\times 2\epsilon\Bigr)
\ee
and hence
\be
\label{Qodd}
\begin{split}
Q_z(m,n)\ &
=\ \frac12\ +\ 4\cos^2\frac{\ell\phi}{2}\cos^2\!\beta\
+\ C_3\Bigl( 4\cos^2\frac{\ell\phi}{2}\sin^2\!\beta\,
	+\,4\sin^2\frac{\ell\phi}{2}\cos^2\!\alpha\Bigr)\\
&\hskip 2.9em+\ {4(1-2C_3)\over \ell^2 D^2+4\epsilon^2}\times\Bigl(
	\cos\frac{\ell\phi}{2}\sin\beta\times\ell D\,
	-\,2\sin\frac{\ell\phi}{2}\cos\alpha\times 2\epsilon
	\Bigr)^2.
\end{split}
\ee
Note that this $Q_z(m,n)$ depends only on $\ell=n-m$, so there is no need to
average it over $m$.
Instead, it is convenient to expand the right hand side of this formula
and to re-group the terms according to their $\ell$ dependence, thus
\begin{align}
Q_z(\ell)\ =\ &
\bigl( \tfrac52-\sin^2\beta+\cos^2\alpha\bigr)\
+\ \bigl(2-\sin^2\beta-\cos^2\alpha\bigr)\times\cos(\ell\phi)\nonumber\\
&+\ (1-2C_3)\bigl(\sin^2\beta-\cos^2\alpha\bigr)\times
	{\ell^2D^2-4\epsilon^2\over\ell^2D^2+4\epsilon^2}\\
&+\ (1-2C_3)(\sin^2\beta+\cos^2\alpha\bigr)\times
	{\ell^2D^2-4\epsilon^2\over\ell^2D^2+4\epsilon^2}\,\cos(\ell\phi)\nonumber\\
&-\ 2(1-2C_3)\cos\alpha\sin\beta\times
	{4\ell D\epsilon\over\ell^2D^2+4\epsilon^2}\,\sin(\ell\phi).\nonumber
\end{align}
Therefore, the net energy per instanton due to repulsion from its odd-distance
neighbors
\be
\E_{\rm odd}\ =\ {N_c\over 5\lambda M}\sum_{{\rm odd}\,\ell}
{Q_z(\ell)\over\ell^2D^2+4\epsilon^2}
\ee
is a linear combination of five series,
\be
\label{Eodd}
\E_{\rm odd}\ =\ {\pi^2N_c\over 20\lambda MD^2}\left(\begin{aligned}
	\bigl( \tfrac52-\sin^2\beta+\cos^2\alpha\bigr)\times\Sigma_1\ &
	+\ \bigl(2-\sin^2\beta-\cos^2\alpha\bigr)\times\Sigma_2\\
	&+\ (1-2C_3)\bigl(\sin^2\beta-\cos^2\alpha\bigr)\times\Sigma_3\\
	&+\ (1-2C_3)(\sin^2\beta+\cos^2\alpha\bigr)\times\Sigma_4\\
	&-\ 2(1-2C_3)\cos\alpha\sin\beta\times\Sigma_5
	\end{aligned}\right)
\ee
where
\be
\def\jj#1{\Sigma_{#1}\ \equiv\ {4D^2\over\pi^2}\sum_{{\rm odd}\,\ell}}
\vcenter{\openup 1\jot \ialign{
	&$\displaystyle{#}$\hfil\cr
	\jj1 & {1\over \ell^2D^2+4\epsilon^2}\ &
	=\ {\tanh(\pi\epsilon/D)\over(\pi\epsilon/D)}\,,\cr
	\jj2 & {\cos\ell\phi\over \ell^2D^2+4\epsilon^2}\ &
	=\ {\sinh((\pi-2\phi)\epsilon/D)\over (\pi\epsilon/D)\cosh(\pi\epsilon/D)}\,,\cr
	\jj3 & {\ell^2D^2-4\epsilon^2\over [\ell^2D^2+4\epsilon^2]^2}\ &
	=\ {1\over\cosh^2(\pi\epsilon/D)}\,,\cr
	\jj4 & {(\ell^2D^2-4\epsilon^2)\times\cos\ell\phi\over [\ell^2D^2+4\epsilon^2]^2}\ &
	=\ {\cosh(2\phi\,\epsilon/D)\over\cosh^2(\pi\epsilon/D)}\
		-\ {2\phi\over\pi}\times{\cosh((\pi-2\phi)\epsilon/D)\over\cosh(\pi\epsilon/D)}\,,\cr
	\jj5 & {4\ell D\epsilon\times\sin\ell\phi\over [\ell^2D^2+4\epsilon^2]^2}\ &
	=\ {\sinh(2\phi\,\epsilon/D)\over\cosh^2(\pi\epsilon/D)}\
		+\ {2\phi\over\pi}\times{\sinh((\pi-2\phi)\epsilon/D)\over\cosh(\pi\epsilon/D)}\,,\cr
	}}
\ee
{\it cf.}\ eqs.~(\ref{Sigmas}.b--f).

Altogether, combining the interaction energies between even-distance
and odd-distance neighbors we obtain the net average energy of an instanton
in a zigzag,
\be
\label{Enet}
\E^{\rm interaction}_{\rm per\,instanton}\
=\ \E_{\rm even}\ +\ \E_{\rm odd}\
=\ {\pi^2N_c\over 20\lambda MD^2}\times{\cal F}(\phi,\alpha,\beta)
\ee
for
\be
\tag{\ref{Fdef}}
\begin{split}
{\cal F}\ =\ \frac32\ &
+\ \Bigl(-1\,+\,C_3(\cos^2\!\alpha\cos^2\!\beta+\sin^2\!\alpha\sin^2\!\beta)\Bigr)
	\times\Sigma_0(\phi)\\
&+\ \Bigl(\frac52+\cos^2\!\alpha-\sin^2\!\beta\Bigr)\times\Sigma_1(\epsilon/D)\\
&+\ \Bigl(2-\cos^2\!\alpha-\sin^2\!\beta\Bigr)\times\Sigma_2(\phi,\epsilon/D)\\
&+\ (1-2C_3)(\sin^2\!\beta-\cos^2\!\alpha)\times\Sigma_3(\epsilon/D)\\
&+\ (1-2C_3)(\sin^2\!\beta+\cos^2\!\alpha)\times\Sigma_4(\phi,\epsilon/D)\\
&-\,2(1-2C_3)\cos\alpha\sin\beta\times\Sigma_5(\phi,\epsilon/D).
\end{split}
\ee
Or rather, this is the net zigzag's energy (per instanton) due to 2-body interactions,
but there is also a 1-body potential energy
\be
\E_{\rm pot}\ =\ N_c\lambda MM_2^2\times\epsilon^2
\ee
due to instanton centers being at $X^2_n=\pm\epsilon\neq0$.
Adding this potential energy to eq.~(\ref{Enet}) we arrive at
\be
\E^{\rm zigzag}_{\rm per\,instanton}\
=\ {\pi^2N_c\over 20\lambda MD^2}\times{\cal F}(\phi,\alpha,\beta)\
+\ N_c\lambda MM_2^2\times\epsilon^2
\ee
and hence eq.~\eqref{ZigNetEnergy}.
{\it Quod erat demonstrandum.}

\bigskip\centerline{$\star\quad\star\quad\star$}\bigskip

In the second part of this Appendix we minimize the zigzag's energy with
respect to the orientation moduli $\phi,\alpha,\beta$.
Specifically, we are going to minimize ${\cal F}(\phi,\alpha,\beta)$ in two stages:
At first we hold $\phi$ fixed and minimize WRT $\alpha$ and $\beta$,
and then we allow $\phi$ to vary and seek the overall minimum.

For the first stage it is convenient to change variables
\be
\alpha\ \to\ a\ =\ \cos\alpha\quad{\rm and}\quad \beta\ \to\ b\ =\ \sin\beta,\quad
-1\ \le\ a,b\ \le\ +1.
\label{ABvars}
\ee
Then in terms of $(\phi,a,b)$ the $\cal F$ function becomes a polynomial in $a$
and $b$ with $\phi$-dependent coefficients,
\be
{\cal F}(\phi,a,b)\
=\ A(\phi)\ +\ B(\phi)\times a^2\ +\ C(\phi)\times b^2\
-\ 2J(\phi)\times ab\ -\ K(\phi)\times a^2b^2
\label{Fpoly}
\ee
where
\be
\label{FGHJKdef}
\begin{split}
A(\phi)\ &=\ \frac32\ -\ \Sigma_0(\phi)\ +\ \frac52\Sigma_1\ +\ 2\Sigma_2(\phi),\\
B(\phi)\ &=\ C_3\Sigma_0(\phi)\ + \Sigma_1\ -\ \Sigma_2(\phi)\
	-\ (1-2C_3)\Sigma_3\ +\ (1-2C_3)\Sigma_4(\phi),\\
C(\phi)\ &=\ C_3\Sigma_0(\phi)\ - \Sigma_1\ -\ \Sigma_2(\phi)\
	+\ (1-2C_3)\Sigma_3\ +\ (1-2C_3)\Sigma_4(\phi),\\
J(\phi)\ &=\ (1-2C_3)\Sigma_5(\phi),\\
K(\phi)\ &=\ 2C_3\Sigma_0(\phi).
\end{split}
\ee
Minimizing the polynomial \eqref{Fpoly} WRT $a$ and $b$ works differently
for $\phi=0$ or $\phi=\pi$ and for generic $\phi$'s, so let us start with the special cases.
For $\phi=\pi$ we have $\Sigma_0=\Sigma_5=0$ while $\Sigma_2=-\Sigma_1$
and $\Sigma_4=-\Sigma_3$, hence
\be
K\ =\ J\ =\ C\ =\ 0\quad{\rm while}\quad B\ >\ 0.
\ee
Consequently, $\cal F$ becomes independent of $b$, but that's OK because the $\beta$
angle does not affect the instantons' relative orientations $y_m^\dagger y_n^{}$
for $\phi=\pi$, {\it cf.}\ eqs.\ \eqref{RelEven} and \eqref{RelOdd}.
In other words, $\beta$ or $b$ becomes an irrelevant variable at $\phi=\pi$
--- just like the geographic longitude becomes an irrelevant coordinate at
latitudes $\pm90^\circ$.
As to the $a$ variable, $B>0$ means that the minimum is at $a=0$,
$i.\,e.,\ \alpha=\pm\frac\pi2$.
Altogether, we have
\be
y_n^\dagger y_{n+1}^{}\ =\ \pm i\tau_3
\label{AFphase}
\ee
which means the anti-ferromagnetic phase {\red AF} of the instanton orientations.
The energy of this phase corresponds to
\be
{\cal F}({\rm\red AF})\ =\ A(\phi=\pi)\
=\ \frac32\ +\ \frac12\Sigma_1\,.
\label{FofAF}
\ee
Similarly, for $\phi=0$ we have $\Sigma_0=\Sigma_5=0$ while $\Sigma_2=+\Sigma_1$
and $\Sigma_4=+\Sigma_3$, hence
\be
K\ =\ J\ =\ B\ =\ 0\quad{\rm while}\quad C\ <\ 0.
\ee
This time, $\cal F$ becomes independent of $a$, but that's OK since the $\alpha$
angle becomes irrelevant to the instanton's relative orientations $y_m^\dagger y_n^{}$
for $\phi=0$, {\it cf.}\ eqs.\ \eqref{RelEven} and \eqref{RelOdd}.
As to the $b$ variable, $C<0$ calls for maximal $b^2=\sin^2\beta$, thus
$b=\pm1$ and $\beta=\pm\frac\pi2$.
Altogether we obtain
\be
y_n^\dagger y_{n+1}^{}\ =\ \pm i\tau_1
\label{AFprime}
\ee
and hence a new anti-ferromagnetic phase $\rm AF'$.
However, this new anti-ferromagnetic phase $\rm AF'$ has a higher energy than the
old anti-ferromagnetic phase {\red AF} ---
\be
{\cal F}({\rm AF'})\ =\ A(\phi=0)\ +\ C(\phi=0)\
=\ \frac32\ +\ \frac52\Sigma_1\ +\ 2(1-2C_3)\Sigma_3\
>\ {\cal F}({\rm\red AF})
\ee
--- so it is not going to survive the minimization of energy WRT $\phi$.

Now consider generic values of $\phi$, $0<\phi<\pi$.
This time we have
\be
K\ >\ 0,\quad J\ \ge\ 0,\quad B\,-\,C\ >0,
\ee
although the signs of $B$ and $C$ themselves depend on $\phi$
and the parameters $\epsilon/D$ and $M_3/M_4$.
Thanks to $K>0$, the polynomial~\eqref{Fpoly} cannot have any minima at generic
values of $a$ and $b$.
More precisely, $a,b$ are limited to the square $-1\le a,b\le +1$, and
$\cal F$ cannot have any local minima in the interior of this square
($i.\,e.$ for $-1<a,b<+1$) except maybe at the center $a=b=0$.

To see this, consider any straight line through the center of the square.
Along such a line the ratio $a/b$ is fixed, so $\cal F$ becomes
a quadratic polynomial of ${\rm radius}^2=a^2+b^2$ with a non-positive coefficient
of the $\rm(radius^2)^2$ term.
As a function of an un-bounded real variable such a polynomial does not have
any minima at all.
But since the $\rm radius^2$ variable is bounded --- between zero at the center and
the maximum at the edge of the square --- $\cal F$ may have a minimum (or minima)
at the center or/and at the edge.
Applying this argument to all lines through the center, we find that the minimum
or minima of $\cal F$ may lie at the square's boundaries
$a=\pm1$ or $b=\pm1$, or at the center $a=b=0$, but not at any other interior point.

Let us consider such possible minima in more detail:
\begin{itemize}

\item[$\yellow\bullet$]
The minimum at $a=b=0$ --- $i.\,e.\ \alpha=\pm\frac\pi2,\ \beta=0$ --- leads to
\be
y_n^\dagger y_{n+1}^{}\ =\ \cos\frac\phi2\
\pm\ \sin\frac\phi2\times(i\tau_3)
\ee
and hence the abelian {\yellow AB} phase of the instanton orientations.
Local stability of this minimum requires $C>0$ and $BC>J^2$, and its energy
corresponds to
\be
{\cal F}({\rm\yellow AB})\ =\ A(\phi).
\label{FofAB}
\ee

\item[$\green\bullet$]
Minimum at the boundary $b=\pm1$ (but generic $-1<a<+1$) --- $i.\,e.\ \beta=\pm\frac\pi2$
and generic $\alpha$ --- leads to
\be
y_n^\dagger y_{n+1}^{}\ =\ (-1)^n\cos\frac\phi2\times(\pm i\tau_1)\
+\ \sin\frac\phi2\times\Bigl(\cos\alpha\times(i\tau_2)\,+\,\sin\alpha\times(i\tau_3)\Bigr)
\ee
which corresponds to the non-abelian orientation phase {\green NA2}.
The specific location of such a minimum is
\be
b\ =\ \pm1,\quad a\ =\ \pm{J\over B-K}
\ee
and its local stability at $|a|<1$ requires
\be
B\ -\ K\ >\ J\quad{\rm but}\quad
\left({J\over B-K}\right)^2\ >\ {C\over B}\,.
\ee
The energy of this minimum is
\be
{\cal F}({\rm\green NA2})\ =\
A(\phi)\ +\ B(\phi)\ -\ {J^2(\phi)\over B(\phi)-K(\phi)}\,.
\label{FofNA2}
\ee

\item[$\circ$]
In principle there could be a similar minimum at the other boundary $a=\pm1$
(but generic $b$).
Specifically, such a minimum would happen at
\be
a\ =\ \pm1,\quad b\ =\ \pm{J\over C-K}\
\ee
and its local stability at $|b|<1$ would require
\be
C\ -\ K\ >\ J\quad{\rm but}\quad
\left({J\over C-K}\right)^2\ >\ {B\over C}\,.
\ee
However, thanks to $B>C$ these two requirements contradict each other,
so this type of minimum does not happen.

\item[$\blue\bullet$]
Finally, we may have  minima at the corners of the $(a,b)$ square, at $a,b=\pm1$
$i.\,e.$, $\alpha=0$ or $\pi$ while $\beta=\pm\frac\pi2$.
Such minima lead to
\be
y_n^\dagger y_{n+1}^{}\ =\ (-1)^n\cos\frac\phi2\times(\pm i\tau_1)\
+\ \sin\frac\phi2(\pm i\tau_2)
\ee
and hence the non-abelian {\blue NA1} phase of the instanton orientations.
Local stability of this phase requires $B\ge J+K$ while its energy corresponds to
\be
\label{FofNA1}
{\cal F}({\rm\blue NA1})\
=\ A(\phi)\ +\ B(\phi)\ +\ C(\phi)\ -\ 2J(\phi)\ -\ K(\phi).
\ee
\end{itemize}

Altogether, for each generic $\phi\neq0,\pi$ we have one or more local energy minima
with respect to $\alpha$ and $\beta$ moduli, and we have analytic formulae \eqref{FofAB},
\eqref{FofNA2}, \eqref{FofNA1} for their energies (or rather $\cal F$ functions)
${\cal F}[{\yellow AB}](\phi)$, ${\cal F}[{\green NA2}](\phi)$,  ${\cal F}[{\blue NA1}](\phi)$
as functions of the $\phi$ modulus.
The next step is to minimize these functions with respect to $\phi$, but unfortunately
the zero-derivative equations
\be
{\partial{\cal F}[{\yellow AB}]\over\partial\phi}\ =\ 0,\quad
{\partial{\cal F}[{\green NA2}]\over\partial\phi}\ =\ 0,\quad
{\partial{\cal F}[{\blue NA1}]\over\partial\phi}\ =\ 0,
\ee
are transcendental and do not have analytic solutions.
In principle, we could solve these equations numerically and then seek the global minimum
among the solutions, but since we do not seek high precision we have used a simpler numeric method:
Covered the $0<\phi<\pi$ interval with a 1800-point grid, evaluated the
${\cal F}[{\yellow AB}]$, ${\cal F}[{\green NA2}]$, and ${\cal F}[{\blue NA1}]$
for all grid points --- and also the ${\cal F}[{\red AF}]$ for $\phi=\pi$, --- and looked for
the lowest value among these $3\times 1800+1$ data points.

The phase diagram on figure \ref{ZigzagDiagram} (on page \pageref{ZigzagDiagram})
was obtained by repeating this numerical minimization procedure for $100\times 100$
combinations of the $\epsilon/D$ and $M_3/M_4$ parameters.


\end{document}